\documentclass[aps,prd,preprint,superscriptaddress,nofootinbib]{revtex4-1}

\usepackage{latexsym}
\usepackage{amsmath}
\usepackage{amssymb}
\usepackage{amsfonts}
\usepackage{bm}

\usepackage{color}
\definecolor{purple}{rgb}{0.5,0,0.5}
\definecolor{blue}{rgb}{0.0,0,0.9}
\definecolor{prdblue}{rgb}{0.133,0.118,0.498}
\usepackage[colorlinks=true, pdfstartview=FitV, linkcolor=prdblue, citecolor= prdblue, urlcolor=prdblue]{hyperref}

\usepackage{supertabular} 
\usepackage{placeins}
\usepackage{epsfig}
\usepackage{graphicx}

\newcommand{\RNum}[1]{\uppercase\expandafter{\romannumeral #1\relax}}

\begin{document}

%\preprint{}

\title{Higher Order Corrections to the Effective Field Theory of Low-energy Axions}


\author{Bryan Cordero-Patino}
\email[]{bryan.cordero@epn.edu.ec}
\affiliation{Departamento de Física, Escuela Politécnica Nacional, Quito 170143, Ecuador}

\author{\'Alvaro Duenas-Vidal}
\email[]{alvaro.duenas@epn.edu.ec}
\affiliation{Departamento de Física, Escuela Politécnica Nacional, Quito 170143, Ecuador}

\author{Jorge Segovia}
\email[]{jsegovia@upo.es}
\affiliation{Departamento de Sistemas F\'isicos, Qu\'imicos y Naturales, \\ Universidad Pablo de Olavide, E-41013 Sevilla, Spain}


\date{\today}

\begin{abstract}
Dark matter (DM) can be composed of a collection of axions, or axion-like particles (ALPs), whose existence is due to the spontaneous breaking of the Peccei-Quinn $U(1)$ symmetry which is the most compelling solution of the strong $CP$-problem of Quantum Chromodynamics (QCD).
Axions must be spin-$0$ particles with very small masses and extremely weak interactions with themselves as well as with the particles that constitute the Standard Model. In general, the physics of axions is detailed by a quantum field theory of a real scalar field, $\phi$. Nevertheless, it is more convenient to implement a non-relativistic effective field theory with a complex scalar field, $\psi$, to characterize the mentioned axions in the low-energy regime. A possible application of this equivalent description is to study the collapse of cold dark matter into more complex structures.
There have been a few derivations of effective Lagrangians for the complex field $\psi$; resulting to be all equivalent after a nonlocal-space transformation between $\phi$ and $\psi$ was found, and some other corrections were introduced. Our contribution herein is to further provide higher order corrections, in particular, we compute the effective field theory Lagrangian up to order $(\psi^\ast\psi)^5$, incorporating also the fast-oscillating field fluctuations into the dominant slowly-varying non-relativistic field.
\end{abstract}

%\pacs{}
%\keywords{}

\maketitle

%%%%%%%%%%%%%%%%%%%%%%%%%%%%%%%%%%%%%%%%%%%%%%%%%%%%%%%%%%%%%%%%%%%%%%%%%%%%%%%%
%%%%%%%%%%%%%%%%%%%%%%%%%%%%%%%%%%%%%%%%%%%%%%%%%%%%%%%%%%%%%%%%%%%%%%%%%%%%%%%%

\section{Introduction}
\label{sec:intro}

Despite the broad consistency of the Standard Model (SM) of particle physics and the success of General Relativity (GR) to describe, respectively, the short and large scales of the Universe, there are still unsolved problems in modern physics. In particular, the nature of dark matter is one of the most intriguing enigmas as it is linked to both frameworks. Experimental observations indicate that DM should be abundant throughout the Universe, contributing about five times more to the Universe's energy density than ordinary (baryonic) matter. Additionally, the vast majority of DM should be cold and collisionless~\cite{ParticleDataGroup:2022pth}. In order to fully resolve the puzzle of dark matter, one needs, in one hand, to identify a plausible dark-matter candidate within realistic models of nuclear and particle physics and, in the other hand, to develop an accurate, systematically improvable, theoretical approach suitable for low-energy phenomena associated with cold DM.

Axions, and axion-like particles, are one of the most strongly motivated possibilities for the constituents that make up the dark matter of the Universe~\cite{Kim:2008hd}. The hypothetical particle was originally introduced as the pseudo-Goldstone boson associated with the spontaneously broken Peccei-Quinn $U(1)$ symmetry in order to address the strong CP problem of QCD~\cite{Peccei:1977hh, Weinberg:1977ma, Wilczek:1977pj}. Axions can be produced in the early universe with an abundance that is compatible with the observed dark matter density by a combination of the cosmic string decay~\cite{Davis:1986xc, Harari:1987ht} and the vacuum misalignment mechanisms~\cite{Preskill:1982cy, Abbott:1982af, Dine:1982ah}. The axions from the cosmic string decay mechanism are incoherent, while the ones from the vacuum misalignment mechanism are coherent. An important feature is also that both produce axions that are extremely nonrelativistic. Finally, Refs.~\cite{Sikivie:2009qn, Erken:2011dz, Saikawa:2012uk, Davidson:2013aba, Noumi:2013zga, Davidson:2014hfa} conclude that gravitational self-interactions may bring the axions in the early universe into thermal equilibrium, producing Bose-Einstein condensates.

A Bose-Einstein condensate of spin-$0$ bosons that fulfill certain properties of axion-like particles can be described by a quantum field theory for a real scalar field $\phi(x)$ whose interactions are depicted by a potential $V(\phi)$ that is a periodic function of $\phi$. With non-relativistic applications in mind, one can introduce a convenient field redefinition that relates $\phi$ with a complex scalar field, $\psi$, that is more suitable to describe the underlying theory in its low-energy limit. The resulting Effective Field Theory (EFT) for the complex field $\psi$ was first constructed in Refs.~\cite{Braaten:2016kzc, Braaten:2016dlp}. However, it is worth mentioning that there have been other articles whose goal was to propose a formalism to describe low-energy axions. The authors of Ref.~\cite{Mukaida:2016hwd}, by integrating out relativistic fluctuations of $\phi$, developed a non-relativistic effective Lagrangian for $\psi$ whose expression is completely different than that given in Ref.~\cite{Braaten:2016kzc}. An exact relation between the real scalar field $\phi$ and the complex scalar field $\psi$ was given in Ref.~\cite{Namjoo:2017nia} which, correcting some errors in both~\cite{Braaten:2016kzc} and~\cite{Mukaida:2016hwd}, demonstrates that the two proposed non-relativistic effective Lagrangians were equivalent. And, lastly, Ref.~\cite{Eby:2017teq} developed a method that gives a sequence of improvements to equations for a complex field $\psi$ with harmonic time dependence. 

Having at hand what we could call a unique effective field theory for low-energy axions, the main purpose of this work is the computation of higher order corrections, up to $(\psi^\ast\psi)^5$. This is motivated by the fact that the axion dynamics is primarily governed by the attractive gravitational interaction and scalar-scalar contact forces. Therefore, a Bose-Einstein condensate can be formed but, instead of being a state with cosmologically long-range correlations and located in the high density/occupancy regime, the mentioned interactions may lead to localized Bose clumps that only exhibit short range correlations with either scarce occupancy or low coherence. In these conditions the system may be poorly approximated by the classical field theory since other behavior would be possible and higher order corrections important. Another relevant aspect to consider when constructing EFTs for low-energy axions are field fluctuations that oscillate rapidly in time scales $m_a^{-1}$~\cite{Namjoo:2017nia}. Typically, in the non-relativistic limit, these fluctuations would be averaged to $0$ on time scales much larger than the inverse of the axion mass. However, as we shall see later, non-linear self-couplings of axions can induce a back-reaction of the fast-oscillating terms on the dominant slowly-varying non-relativistic field and thus they must be treated with caution.

The manuscript is arranged as follows. Section~\ref{sec:R-FT} simply defines the relativistic field theory of a real scalar field. Afterwards, we present in Sec.~\ref{sec:NR-FT} the general procedure by means of which the relativistic Lagrangian, described by the real scalar field $\phi$, transforms into the non-relativistic Lagrangian, described by the complex scalar field $\psi$. Next, Section~\ref{sec:HO-FT} is devoted to present our computation of the $(\psi^\ast\psi)^5$ higher-order corrections taking into account the effects of fast-oscillating fields into the dominant slowly-varying non-relativistic one due to axion's non-linear self-interactions. Finally, some conclusions and possible future steps are given in Sec.~\ref{sec:summary}.

%%%%%%%%%%%%%%%%%%%%%%%%%%%%%%%%%%%%%%%%%%%%%%%%%%%%%%%%%%%%%%%%%%%%%%%%%%%%%%%%
%%%%%%%%%%%%%%%%%%%%%%%%%%%%%%%%%%%%%%%%%%%%%%%%%%%%%%%%%%%%%%%%%%%%%%%%%%%%%%%%

\section{Axion's Relativistic Field Theory}
\label{sec:R-FT} 

For momentum scales much smaller than the axion's decay constant, $f_a$, the axion field can be represented by an elementary quantum field $\phi(x)$ that is a real Lorentz scalar. Since $\phi$ ultimately represents an angle, $\phi$ must obey the shift symmetry $\phi \to \phi + 2 \pi f_a$. At even smaller momenta, \emph{i.e.} below the QCD confinement scale ($\approx\! 1\,\text{GeV}$), the self-interactions of axions can be described by a relativistic potential $V(\phi)$, known as the axion potential. This EFT is often regarded as Relativistic Axion Field Theory (see Ref.~\cite{Braaten:2019knj} for a nice review). The Lagrangian density of the field $\phi$ can be written as follows:\footnote{Note here that some authors define $V(\phi)$ without including the mass term $\frac{1}{2} m_a^2 \phi^2$ and, thus, it is directly depicted in the Lagrangian.} 
\begin{equation}
\mathcal{L} = \frac{1}{2} \partial_u \phi \partial^u \phi - V(\phi) \,,
\label{L_eff_rlt}
\end{equation}
where $V(\phi)$ is a periodic function of $\phi$ with period $2\pi f_a$ as a result of the aforementioned shift symmetry. It is worth mentioning that $\phi$ has also a $Z_2$ symmetry $\phi(x) \to -\phi(x)$. Consequently, $V(\phi)$ is an even function of $\phi$, and for $\phi/f_a \ll 1$  it can be expanded in powers of $\phi^2$:
\begin{equation}
V(\phi) = \frac{1}{2} m_a^2 \phi^2 + m_a^2 f_a^2 \sum_{n=2}^{\infty} \frac{\lambda_{2n}}{(2n)!} \left( \frac{\phi^2}{f_a^2} \right)^n \,,
\label{potential_relativistic}
\end{equation}
where $\lambda_{2n}$ are dimensionless coupling constants. In reasonable axion models the constants $\lambda_{2n}$ have natural values of order $1$. The Feynman rule for the ($2n$)-axion vertex is $-i \lambda_{2n} m_a^2 / f_a^{2n-2}$ while the quantum-loop suppression factor for the QCD axion in the post-inflation scenario is approximately $m_a^2 / f_a^2 \simeq 10^{-48 \pm 4}$.\footnote{The $\pm$ does not symbolize uncertainty but the possible range of magnitude of the quantum-loop suppression factor as a direct consequence of the allowed regions of $m_a$ and $f_a$. In the post-inflationary scenario $10^{9}\,\text{GeV} \lesssim f_a \lesssim 10^{12}\,\text{GeV}$ and $10^{-6}\,\text{eV} \lesssim m_a \lesssim 10^{-3}\,\text{eV}$~\cite{Sikivie:2009qn}.} Consequently, for most purposes, the effects of quantum loops can be omitted, \emph{i.e.} the relativistic axion field theory can be treated as a classical field theory. It is important to mention that in fundamental quantum field theory the series of the effective potential must end at the power $\phi^4$ so the system can be renormalizable~\cite{peskin}. However, this requirement is not necessary for either a classical field theory or an effective field theory \cite{Braaten:2019knj}.

In the literature, one can find two alternatives for the relativistic axion potential. The first one is known as the instanton potential and it is given by the following expression:
\begin{equation}
V(\phi) = m_a^2 f_a^2 \left[ 1-\cos \left( \frac{\phi}{f_a} \right) \right].
\label{instaton}
\end{equation}
This potential was first derived by Peccei and Quinn~\cite{Peccei:1977hh, Peccei:1977ur} using a dilute instanton gas approximation and thus there is no known way to improve it. Therefore, the instanton potential should be considered only as a good phenomenological approximation, subsequently implemented in most phenomenological studies regarding axion physics~\cite{Marsh:2015xka, GrillidiCortona:2015jxo}. If one compares the Taylor series expansion of Eq.~\eqref{instaton} with~\eqref{potential_relativistic}, the dimensionless coupling constants $\lambda_{2n}$ for axion self-interactions are defined by $\lambda_{2n}=(-1)^{n+1}$. It is clear that $\lambda_4 = -1$ which implies that axion pair interactions are attractive.

The second option for the axion's relativistic potential is known as chiral potential because it is derived from the leading order chiral effective field theory considering the light pseudoscalar mesons of QCD and the axion field:
\begin{equation}
V(\phi) = m_\pi^2 f_\pi^2 \left[ 1 - \sqrt{1-\frac{4z}{(1+z)^2} \sin^2 \left( \frac{\phi}{2f_a} \right)} \right] \,,
\label{chiral}
\end{equation}
where $m_\pi$ is the pion's mass, $f_\pi$ is the pion's decay constant, and $z=m_u /m_d$ is the ratio of the up and down quark masses. This potential was first derived by Di Vecchia and Veneziano~\cite{DiVecchia:1980yfw}, continued by diverse studies such as those in Refs.~\cite{GrillidiCortona:2015jxo, Braaten:2019knj, DiLuzio:2020wdo}. 
Note also that the instanton potential can be derived from the chiral potential in Eq.~\eqref{chiral}. To do so, one should apply the binomial expansion to the square root, then truncate it after the $\sin^2(\phi/2f_a)$ term, and finally use a trigonometric identity.

According to the chiral potential, the dimensionless coupling constant for the 4-axion vertex is,
\begin{equation}
\lambda_4 = - \frac{1-z+z^2}{(1+z^2)} \,.
\end{equation}
Since $z>0$, it is easy to see that $\lambda_4 < 0$ and, consequently, the axion pair interactions are, once again, attractive~\cite{Braaten:2016kzc, Braaten:2019knj}.

%%%%%%%%%%%%%%%%%%%%%%%%%%%%%%%%%%%%%%%%%%%%%%%%%%%%%%%%%%%%%%%%%%%%%%%%%%%%%%%%
%%%%%%%%%%%%%%%%%%%%%%%%%%%%%%%%%%%%%%%%%%%%%%%%%%%%%%%%%%%%%%%%%%%%%%%%%%%%%%%%

\section{Non-relativistic Effective Field Theory}
\label{sec:NR-FT}

The predominant method to build a Non-Relativistic Effective Field Theory for axions is to transform the real scalar field $\phi$ into the complex scalar field $\psi$. This may seem inconsistent because the complex scalar field is formed by two real scalar fields. However, the aim of the procedure is to obtain a non-relativistic Lagrangian from the original relativistic one while maintaining the number of degrees of freedom. The relativistic Lagrangian is quadratic in time derivatives of $\phi$ while the non-relativistic one has only linear terms in time derivatives of $\psi$. Furthermore, the number of propagating degrees of freedom in a Lagrangian containing up to quadratic terms in time derivatives is equal to the number of real fields in it and, on the other hand, the number of propagating degrees of freedom for a Lagrangian containing only first derivatives in time is equal to half the number of real fields in it. Therefore, the number of degrees of freedom in both Lagrangians remains equal.

Although both Refs.~\cite{Namjoo:2017nia} and~\cite{Braaten:2018lmj} have implemented different methods in order to construct an EFT for non-relativistic axions, the final effective Lagrangian has been demonstrated to be the same. Therefore, it becomes compelling to take the results of these investigations as a basis for future developments and applications. Specifically, the procedure first detailed in  Ref.~\cite{Namjoo:2017nia} is implemented herein to compute later the higher order corrections.

%%%%%%%%%%%%%%%%%%%%%%%%%%%%%%%%%%%%%%%%%%%%%%%%%%%%%%%%%%%%%%%%%%%%%%%%%%%%%%%%

\subsection{The non-relativistic EFT Lagrangian density}

The analysis of Ref.~\cite{Namjoo:2017nia} begins from the relativistic Lagrangian of $\phi$ where only one self-interaction term is utilized in the derivation. In this work four self-interaction terms are taken into account,
\begin{align}
\mathcal{L} =& \, \frac{1}{2} \partial_u \phi \partial^u \phi -\frac{1}{2} m_a^2 \phi^2 - m_a^2 f_a^2 \sum_{n=2}^{5} \frac{\lambda_{2n}}{(2n)!} \frac{\phi^{2n}}{f_a^{2n}} \,, 
\end{align}
\noindent
where the constants $m_a^2 \lambda_{2n}/f_a^{2n-2}$ indicate the strength of the self-interaction terms. To simplify this notation is convenient to define
\begin{align}
\bar{\lambda}_{2n} &= \frac{m_a^2}{f_a^{2n-2}} \lambda_{2n} \,,
\label{constants_lambda}
\end{align}
and thus the relativistic Lagrangian becomes
\begin{align}
\mathcal{L} =& \frac{1}{2} \partial_u \phi \partial^u \phi -\frac{1}{2} m_a^2 \phi^2 - \sum_{n=2}^{5} \bar{\lambda}_{2n} \frac{\phi^{2n}}{(2n)!} \,.
\end{align}
Consequently, the Hamiltonian density of the system is
\begin{align}
\mathcal{H} =& \, \pi(t,x) \dot{\phi}(t,x) - \mathcal{L} \nonumber \\
=& \, \frac{1}{2} \pi^2 - \frac{1}{2} (\bigtriangledown \phi)^2 +\frac{1}{2} m_a^2 \phi^2 +  \sum_{n=2}^{5} \bar{\lambda}_{2n} \frac{\phi^{2n}}{(2n)!} \,,
\end{align}
where $\pi=\partial \cal{L}/\partial \dot{\phi}=\dot{\phi}$ is the canonical momentum field of $\phi$. Finally, the system's equations of motion are
\begin{align}
\dot{\phi} = + \frac{\delta \mathcal{H}}{\delta \pi} =& \, \pi \,, \label{orig_motion_1} \\
\dot{\pi} = - \frac{\delta \mathcal{H}}{\delta \phi} =& \,(\bigtriangledown^2 - m_a^2) \phi - \sum_{n=2}^{5} \bar{\lambda}_{2n} \frac{\phi^{2n-1}}{(2n-1)!} \,.
\label{orig_motion_2}
\end{align}

To obtain the non-relativistic EFT for low-energy axions, one should typically relate $\phi(t,x)$ and $\pi(t,x)$ with $\psi(t,x)$ as follows:
\begin{align}
\phi &= \frac{1}{\sqrt{2m_a}} \left( e^{-im_at} \psi + e^{im_at} \psi^* \right) \,, \label{typical-transf1} \\
\pi &= -i\sqrt{\frac{m_a}{2}} \left( e^{-im_at} \psi - e^{im_at} \psi^* \right) \,. \label{typical-transf2}
\end{align}
However, in order to obtain relativistic corrections in the EFT, the authors in Ref.~\cite{Namjoo:2017nia} proposed the following non-local canonical field transformation:
\begin{align}
\phi &= \frac{1}{\sqrt{2m_a}} \mathcal{P}^{-\frac{1}{2}} \left( e^{-im_at} \psi + e^{im_at} \psi^* \right) \,, \label{transf_Namjoo1} \\
\pi &= -i\sqrt{\frac{m_a}{2}} \mathcal{P}^{\frac{1}{2}} \left( e^{-im_at} \psi - e^{im_at} \psi^* \right) \,, \label{transf_Namjoo2}
\end{align}
where,
\begin{equation}
\mathcal{P} \equiv \sqrt{ 1 - \frac{\bigtriangledown^2}{m_a^2}} \,,
\label{def_P}
\end{equation}
is an operator which will be later expanded in powers of $\bigtriangledown^2 / m_a^2$. The operator $\mathcal{P}$ is defined so that the equation of motion of $\psi$ does not contain rapidly oscillating terms in the free field theory. It is also important to mention that $m_a \mathcal{P}$ corresponds to the total energy of a free relativistic particle. 

The transformations in Eqs.~\eqref{transf_Namjoo1} and~\eqref{transf_Namjoo2} uniquely define both $\psi$ and $\psi^\ast$; that is, an expression of $\psi$ in terms of $\phi$ and $\pi$ can be derived
\begin{equation}
\psi = \sqrt{\frac{m_a}{2}} e^{im_at} \mathcal{P}^{\frac{1}{2}} \left[\phi + \frac{i}{m_a} \mathcal{P}^{-1} \pi \right] \,. \label{transf_Namjoo3}
\end{equation}
Additionally, it is possible to obtain the equation of motion of $\psi$. First, we derive with respect time Eq.~\eqref{transf_Namjoo3} and obtain
\begin{align}
\dot{\psi} = \, i m_a \sqrt{\frac{m_a}{2}} e^{im_at} \mathcal{P}^{\frac{1}{2}} \left[\phi + \frac{i}{m_a} \mathcal{P}^{-1} \pi \right] 
+ \sqrt{\frac{m_a}{2}} e^{im_at} \mathcal{P}^{\frac{1}{2}} \left[\dot{\phi} + \frac{i}{m_a} \mathcal{P}^{-1} \dot{\pi} \right] \,. \label{derv_transf_Namjoo3_1}
\end{align}
Afterwards, by replacing Eq.~\eqref{transf_Namjoo3} in~\eqref{derv_transf_Namjoo3_1}, and then multiplying by $i$, the following expression
\begin{equation}
i\dot{\psi} = -m_a\psi +  \sqrt{\frac{m_a}{2}} e^{im_at} \mathcal{P}^{\frac{1}{2}} \left[i\dot{\phi} - \frac{1}{m_a} \mathcal{P}^{-1} \dot{\pi} \right] \,, \label{derv_transf_Namjoo3_2}
\end{equation}
is obtained. On the other hand, from Eq.~\eqref{def_P}, 
\begin{equation}
\bigtriangledown^2 - m_a^2 = -m_a^2 \mathcal{P}^2 \,,
\end{equation}
and therefore Eq.~\eqref{orig_motion_2} can be written as
\begin{equation}
\dot{\pi} = -m_a^2 \mathcal{P}^2 \phi -  \sum_{n=2}^{5} \Bar{\lambda}_{2n} \frac{\phi^{2n-1}}{(2n-1)!} \,,
\label{orig_motion_2_P}
\end{equation}
in such a way that Eq.~\eqref{derv_transf_Namjoo3_2} is given by
\begin{align}
i\dot{\psi} =& -m_a\psi + m_a \mathcal{P} \left[ \sqrt{\frac{m_a}{2}} e^{im_at} \mathcal{P}^{\frac{1}{2}} \left( \phi + \frac{i}{m_a} \mathcal{P}^{-1}\pi \right) \right] \nonumber \\
&
+ \frac{1}{\sqrt{2m_a}} \sum_{n=2}^{5} e^{im_at} \mathcal{P}^{-\frac{1}{2}} \left( \bar{\lambda}_{2n} \frac{\phi^{2n-1}}{(2n-1)!} \right) \,, \label{derv_transf_Namjoo3_4}
\end{align}
when using Eqs.~\eqref{orig_motion_1} and~\eqref{orig_motion_2_P}. Lastly, by replacing Eqs.~\eqref{transf_Namjoo1} and~\eqref{transf_Namjoo3}, one obtains the equation of motion for the field $\psi$:
\begin{align}
i\dot{\psi} = \, m_a(\mathcal{P}-1)\psi + \sum_{n=2}^{5} \frac{n \, \Bar{\lambda}_{2n} }{2^{n-1} (2n)! \, m_a^n } e^{im_at} \mathcal{P}^{-\frac{1}{2}}  \mathcal{A}^{2n-1} \,, \label{eq_motion_psi}
\end{align}
where
\begin{equation}
\mathcal{A} = e^{-im_at}\mathcal{P}^{-\frac{1}{2}}\psi + e^{im_at}\mathcal{P}^{-\frac{1}{2}} \psi^*.
\label{def_A}
\end{equation}
As a side note, the  Lagrangian that describes the dynamics of  $\psi$ can be found by requiring that its equation of motion matches the one above:
\begin{align}
\mathcal{L}_{\text{eff}} = \, \frac{i}{2}(\dot{\psi}\psi^* - \psi\dot{\psi}^*) - m_a \psi^* (\mathcal{P}-1) \psi 
- \sum_{n=2}^{5} \Bar{\lambda}_{2n} \left( \frac{1}{2 \, m_a} \right)^{n}  \frac{\mathcal{A}^{2n}}{(2n)!}  \,.
\end{align}

%%%%%%%%%%%%%%%%%%%%%%%%%%%%%%%%%%%%%%%%%%%%%%%%%%%%%%%%%%%%%%%%%%%%%%%%%%%%%%%%

\subsection{Importance of rapidly field fluctuations}

Once derived the equation of motion of $\psi$ it becomes tempting to expand $\mathcal{P}$ in powers of $\bigtriangledown^2 / m_a^2$ to obtain an expression that describes the behavior of low-energy axions. However, this would not incorporate how the fast-oscillation terms inside $\psi$ affect the behavior of the slowly-varying field. In the original work of Ref.~\cite{Namjoo:2017nia}, to show the importance of this effect, all expressions of $\mathcal{P}$ were approximated as
\begin{equation}
\mathcal{P} - 1 \approx -\frac{\bigtriangledown^2}{2m_a^2} \label{P-1} \,, \quad \quad \quad
\mathcal{P}^{-\frac{1}{2}} \approx 1 \,.
\end{equation}
This is also replicated in this paper, but only with two self-interaction terms in the effective potential so as not to overload the text with more mathematical expressions than are necessary to illustrate. By replacing Eq.~\eqref{P-1} into~\eqref{eq_motion_psi} one obtains
\begin{align}
i\dot{\psi} \approx &-\frac{1}{2m_a} \bigtriangledown^2 \psi + \frac{\Bar{\lambda}_4}{8 m_a^2} |\psi|^2 \psi + \frac{\Bar{\lambda}_6}{4 \cdot 4! m_a^3} |\psi|^4 \psi  + \frac{\Bar{\lambda}_4}{4! m_a^2} \big( e^{-2im_at}\psi^3 + 3 e^{2im_at}|\psi|^2 \psi^* \nonumber \\
&+ e^{4im_at}{\psi^*}^3 \big)  + \frac{\Bar{\lambda}_6}{8 \cdot 5! m_a^3} \big( e^{-4im_at}\psi^5 + 5 e^{-2im_at}|\psi|^2 \psi^3 + 10 e^{2im_at}|\psi|^4 \psi^* \nonumber \\ &+ 5 e^{4im_at}|\psi|^2{\psi^*}^3 + e^{6im_at}{\psi^*}^5 \big) \,. \label{eq_motion_psi_fast_terms}
\end{align}
All terms inside parenthesis are fast-oscillating factors. Typically, they would be approximated to $0$ in the limit of large $m_a$, \emph{i.e.} when averaging over time scales $\Delta t \gg m_a^{-1}$. Nevertheless, to analyze how the equation behaves, let us expand $\psi$ as the following \emph{ans\"atze} 
\begin{equation}
\psi = \psi_s + e^{2im_at}\, \delta\psi,
\label{ansatz_fast_terms}
\end{equation}
where $\psi_s$ is slow-varying and $e^{2im_at}\,\delta\psi$ is clearly not. Then, by inserting Eq.~\eqref{ansatz_fast_terms} into~\eqref{eq_motion_psi_fast_terms} and only keeping terms up to linear order in $\delta \psi$, which also vary in time more slowly than $e^{\pm im_at}$, one obtains 
\begin{align}
i\dot{\psi}_s \approx &-\frac{1}{2m_a} \bigtriangledown^2 \psi_s 
+ \frac{\Bar{\lambda}_4}{8 m_a^2} \, |\psi_s|^2 \psi_s + \frac{\Bar{\lambda}_6}{4 \cdot 4! m_a^3} \, |\psi_s|^4 \psi_s \nonumber \\
&
+ \frac{\Bar{\lambda}_4}{8 m_a^2} \big(\psi_s^2 \, \delta \psi + 2 \, \psi_s^2 \, \delta \psi^* \big) \nonumber \\
&
+ \frac{\Bar{\lambda}_6}{4 \cdot 4! m_a^3} \big(2 \, |\psi_s|^2 \psi_s^2 \, \delta\psi + 3 \, |\psi_s|^4 \, \delta\psi^* \big) \,.
    \label{eq_motion_psi_fast_terms2}
\end{align}
The last two terms on the right hand side (RHS) of Eq.~\eqref{eq_motion_psi_fast_terms2} clearly show that the dynamics of the slow-varying field are affected by the fast-oscillating factors in non-trivial ways for each self-interaction included in the calculation. 

%%%%%%%%%%%%%%%%%%%%%%%%%%%%%%%%%%%%%%%%%%%%%%%%%%%%%%%%%%%%%%%%%%%%%%%%%%%%%%%%

\subsection{Defining the iterative process}

To study the effect of fast-oscillating factors, $\psi$ must be expanded as an infinite series of harmonics
\begin{equation}
\psi(t,x) = \sum_{\nu = -\infty}^{+\infty} \psi_\nu(t,x) e^{i\nu m_at} \,,
\label{psi_expansion}
\end{equation}
where each $\psi_\nu$ varies slowly on a time scale $m_a^{-1}$. Additionally, the mode with $\nu=0$ is designated as the lowest slowly-varying contribution of the field and it is denoted as $\psi_{\nu=0} \equiv \psi_s $. In the non-relativistic limit it is assumed that $|\psi_{\nu}-\psi_s| \ll |\psi_s|$, \emph{i.e.} $\psi_s$ is the dominant term of $\psi$. Next, it is convenient to define
\begin{align}
    \Psi_\nu &\equiv  \mathcal{P}^{-\frac{1}{2}} \psi_\nu \label{psi_mayus} \,,
\end{align}
and,
\begin{align}
\widetilde{G} &\equiv e^{im_at} \mathcal{P}^{-\frac{1}{2}}  \mathcal{A}^3 \,, \qquad \widetilde{F} \equiv e^{im_at} \mathcal{P}^{-\frac{1}{2}} \mathcal{A}^5 \nonumber  \,, \\
\widetilde{D} &\equiv e^{im_at} \mathcal{P}^{-\frac{1}{2}} \mathcal{A}^7 \,, \qquad  \widetilde{R} \equiv e^{im_at} \mathcal{P}^{-\frac{1}{2}} \mathcal{A}^9  \label{GFDR-exp1} \,,
\end{align}
in such a way that, by replacing Eqs.~\eqref{psi_expansion} and ~\eqref{psi_mayus} into Eqs.~\eqref{GFDR-exp1}, it is possible to expand $\widetilde{G}$, $\widetilde{F}$, $\widetilde{D}$, and $\widetilde{R}$ as an infinite series of harmonics:
\begin{align}
\widetilde{G}(t,x) &= \sum_{\nu= - \infty}^{+\infty} \widetilde{G}_\nu(t,x) e^{\nu im_at} \,, \qquad
\widetilde{F}(t,x) = \sum_{\nu= - \infty}^{+\infty} \widetilde{F}_\nu(t,x) e^{\nu im_at} \,, \nonumber \\
\widetilde{D}(t,x) &= \sum_{\nu= - \infty}^{+\infty} \widetilde{D}_\nu(t,x) e^{\nu im_at} \,, \qquad
\widetilde{R}(t,x) = \sum_{\nu= - \infty}^{+\infty} \widetilde{R}_\nu(t,x) e^{\nu im_at} \,, \label{GFDR_exp2}
\end{align}
where, for shortness,  the mode functions $\widetilde{G}_\nu$, $\widetilde{F}_\nu$, $\widetilde{D}_\nu$ and $\widetilde{R}_\nu$ can be found in the Appendix~\ref{App-A}. Replacing Eq.~\eqref{psi_mayus} and  Eqs.~\eqref{G_exp3} to \eqref{R_exp3} into Eq.~\eqref{eq_motion_psi} results in the following equation of motion for each mode $\nu$:
\begin{align}
i \dot{\psi}_\nu - \nu m_a \psi_\nu =& m_a (\mathcal{P}-1) \psi_\nu + \frac{\Bar{\lambda}_4}{4! \, m_a^2} \widetilde{G}_\nu  \nonumber \\
&+ \frac{3 \, \Bar{\lambda}_6}{4 \cdot 6! \, m_a^3} \widetilde{F}_\nu + \frac{\Bar{\lambda}_8}{2 \cdot 8! \, m_a^4} \widetilde{D}_\nu \nonumber \\
&+ \frac{5 \, \Bar{\lambda}_{10}}{16 \cdot 10! \, m_a^5} \widetilde{R}_\nu \,.
\label{eq_motion_psi2}
\end{align}
Then, by multiplying equation (\ref{eq_motion_psi2}) by $\mathcal{P}^{-\frac{1}{2}}$ and rearranging terms, it is possible to obtain
\begin{equation}
\Psi_\nu = - \frac{i}{m_a} \Gamma_\nu \dot{\Psi}_\nu + \Bar{\lambda}_4 G_\nu + \Bar{\lambda}_6 F_\nu + \Bar{\lambda}_8 D_\nu + \Bar{\lambda}_{10} R_\nu \,,
\label{eq_motion_psi3}
\end{equation}
where,
\begin{equation}
\Gamma_\nu = (1-\nu-\mathcal{P})^{-1} \,,
\label{def_Gamma}
\end{equation}
and,
\begin{align}
G_\nu &= \frac{\Gamma_\nu \mathcal{P}^{-\frac{1}{2}}}{4! \, m_a^3} \widetilde{G}_\nu \,, \qquad
F_\nu = \frac{3 \, \Gamma_\nu \mathcal{P}^{-\frac{1}{2}}}{4 \cdot 6! \, m_a^4} \widetilde{F}_\nu \,, \nonumber \\
D_\nu &= \frac{\Gamma_\nu \mathcal{P}^{-\frac{1}{2}}}{2 \cdot 8! \, m_a^5} \widetilde{D}_\nu \,, \qquad
R_\nu = \frac{5 \, \Gamma_\nu \mathcal{P}^{-\frac{1}{2}}}{16 \cdot 10! \, m_a^6} \widetilde{R}_\nu \,. \label{GFDR_exp4}
\end{align}
Now, given a physical object $\mathcal{F}(t,x)$, the spatial and temporal variations are compared to $m_a^{-1}$ and the self-interactions are mediated by the constants $\Bar{\lambda}_{2n}$. It is useful to parameterize the magnitude of spatial and temporal variations as
\begin{equation}
\frac{\nabla^2 \mathcal{F}}{m_a^2} \sim \epsilon_x \mathcal{F} \,, \quad\quad  \frac{\dot{\mathcal{F}}}{m_a} \sim \epsilon_t \mathcal{F}.
\end{equation}
It is appropriate to assume $\epsilon_x$, $\epsilon_t$, $\Bar{\lambda}_{2n} \ll 1$ for weakly interacting systems in the non-relativistic limit. Then, Eq.~\eqref{eq_motion_psi3} holds for all orders of $\epsilon_x,\epsilon_t$, and $\Bar{\lambda}_{2n}$. Additionally, the first term of the RHS of Eq.~\eqref{eq_motion_psi3} is suppressed relative to $\Psi_\nu$ by $\epsilon_t$, the second term is suppressed relative to $G_\nu$ by $\Bar{\lambda}_4$, the third term is suppressed relative to $F_\nu$ by $\Bar{\lambda}_6$, the fourth term is suppressed relative to $D_\nu$ by $\Bar{\lambda}_8$, and the fifth term is suppressed relative to $R_\nu$ by $\Bar{\lambda}_{10}$. Therefore, the RHS of~\eqref{eq_motion_psi3} can be interpreted as a perturbative source for $\Psi_\nu$.

As long as the functions $\psi_\nu$ are constructed so that each mode satisfies Eq.~\eqref{eq_motion_psi2}, then the full series will satisfy the equation of motion~(\ref{eq_motion_psi}). Taking this into account, the authors of Ref.~\cite{Namjoo:2017nia} proposed the following iterative process to find $\psi$. The $0th$ order approximation of $\Psi_\nu$ is
\begin{equation}
\Psi^{(0)}_\nu(t,x) = 
\begin{cases}
\Psi_s(t,x) & \text{if } \nu=0 \,, \\
0 & \text{if } \nu \neq 0 \,,
\end{cases}
\label{Psi_iter_0}
\end{equation}
For $\nu \neq 0$, the $nth$ correction to $\Psi_\nu$ at the $nth$ iteration is denoted as $\Psi_\nu^{(n)}$. Mathematically, this results in $\Psi_\nu$ being expressed as
\begin{equation}
\Psi_\nu(t,x) = \sum_{n=0}^{\infty} \Psi_\nu^{(n)}(t,x) \,,
\label{Psi_iterat_n}
\end{equation}
where $\Psi_\nu^{(0)}=0$ because the way of defining the $0th$ order approximation. Additionally, $\Psi_0^{(n)}$ is set to $0$ for $n>0$. As mentioned before, the perturbative source of $\Psi_\nu$ is Eq.~\eqref{eq_motion_psi3}, \emph{i.e.} $\Psi_\nu^{(n)}$ will always be proportional to the $n$ powers of $\epsilon_t$ and $\Bar{\lambda}_{2n}$. Additionally, for all $\nu$, $G_\nu$, $F_\nu$, $D_\nu$, and $R_\nu$ are expanded as
\begin{align}
&
G_\nu(t,x) = \sum_{n=0}^{\infty} G_\nu^{(n)}(t,x) \,, 
 \qquad F_\nu(t,x) = \sum_{n=0}^{\infty} F_\nu^{(n)}(t,x) \,, 
\nonumber \\
&
D_\nu(t,x) = \sum_{n=0}^{\infty} D_\nu^{(n)}(t,x) \,, 
 \qquad R_\nu(t,x) = \sum_{n=0}^{\infty} R_\nu^{(n)}(t,x) \,, 
\label{GFDR_iterat_n}
\end{align}
where $G_\nu^{(n)}$, $F_\nu^{(n)}$, $D_\nu^{(n)}$, and $R_\nu^{(n)}$ contain all the terms proportional to the $nth$ power of $\epsilon_t$ and $\Bar{\lambda}_{2n}$ in Eqs~\eqref{GFDR_exp4}, respectively. Next, by replacing Eq.~\eqref{Psi_iterat_n} and  Eqs.~\eqref{GFDR_iterat_n} into Eq.~\eqref{eq_motion_psi3}, and collecting terms with the same power of $\epsilon_t$ and $\Bar{\lambda}_{2n}$, one obtains the following expression:
\begin{align}
\Psi_\nu^{(n)} =& - \frac{i}{m_a} \Gamma_\nu \dot{\Psi}_\nu^{(n-1)} + \Bar{\lambda}_4 G_\nu^{(n-1)} + \Bar{\lambda}_6 F_\nu^{(n-1)} + \Bar{\lambda}_8 D_\nu^{(n-1)} + \Bar{\lambda}_{10} R_\nu^{(n-1)} \,.
\label{eq_motion_psi_iterat_n}
\end{align}
It is important to note that the above equation only works for $\nu \neq 0$. By replacing Eqs.~\eqref{GFDR_exp4} into~\eqref{eq_motion_psi2} for $\nu=0$, it is possible to derive the following
\begin{align}
i \Dot{\psi}_s  =& \, m_a (\mathcal{P}-1) \psi_s + m_a \Gamma_0^{-1} \mathcal{P}^{\frac{1}{2}} \left[ \Bar{\lambda}_4 G_0 + \Bar{\lambda}_6 F_0 \right. + \left. \Bar{\lambda}_8 D_0 + \Bar{\lambda}_{10} R_0 \right] \,.
\label{eq_motion_psi_s}
\end{align}
Therefore, one comes across the following way of proceeding. First, obtain an expression for $G_0$, $F_0$, $D_0$, and $R_0$ in terms of $\Psi_s$ up to the $n$ power of $\epsilon_t$ and $\Bar{\lambda}_{2n}$. This results in the last term in the RHS of Eq.~\eqref{eq_motion_psi_s} only containing $\Psi_s$ terms proportional up to the $(n+1)$ power of $\epsilon_t$ and $\bar{\lambda}_{2n}$. Next, expand all expressions of $\mathcal{P}$ in Eq.~\eqref{eq_motion_psi_s} up to the $(n+1)$ power of $\epsilon_x$. However, only the terms proportional up to the $(n+1)$ power of $\epsilon_x$, $\epsilon_t$ and $\Bar{\lambda}_{2n}$ are kept. 

One apparent problem with this process is that new degrees of freedom are added from the first term in the RHS of Eq.~\eqref{eq_motion_psi_iterat_n}. How exactly this dilemma is solved is more effectively shown in the next section. Nevertheless, the basic idea is to represent all expressions of $\Dot{\Psi}_\nu^{(m)}$, where $m \leq n$, in terms of $\Psi_s$ and powers of $\epsilon_t$. These terms are maintained as such thought the whole process up until the effective Lagrangian is calculated. A final step is added at this point: all terms that contain $\epsilon_t$, other than $\frac{i}{2}(\dot{\psi}_s\psi_s^* - \psi_s\dot{\psi}_s^*)$, are expressed in terms of $\psi_s$ by replacing the $0th$ order approximation equation of motion. The final expression is the $nth$ approximation of the non-relativistic effective Lagrangian of $\psi$ when $\psi_s$ is the dominant term. 

The above procedure is iterative because $G_\nu^{(n)}$, $F_\nu^{(n)}$, $D_\nu^{(n)}$, and $R_\nu^{(n)}$ are expressed in terms of $\Psi_\nu^{(m)}$ where $m\leq n$, and, from Eq. (\ref{eq_motion_psi_iterat_n}), these terms are expressed in terms of $G_\nu^{(m-1)}$, $F_\nu^{(m-1)}$, $D_\nu^{(m-1)}$, and $R_\nu^{(m-1)}$. In other words, in order to obtain $G_\nu^{(n)}$, $F_\nu^{(n)}$, $D_\nu^{(n)}$, and $R_\nu^{(n)}$ it is necessary to already know an expression for all $G_\nu^{(m)}$, $F_\nu^{(m)}$, $D_\nu^{(m)}$, and $R_\nu^{(m)}$ where $m\leq n$.

%%%%%%%%%%%%%%%%%%%%%%%%%%%%%%%%%%%%%%%%%%%%%%%%%%%%%%%%%%%%%%%%%%%%%%%%%%%%%%%%
%%%%%%%%%%%%%%%%%%%%%%%%%%%%%%%%%%%%%%%%%%%%%%%%%%%%%%%%%%%%%%%%%%%%%%%%%%%%%%%%

\section{Higher Order Corrections}
\label{sec:HO-FT} 

Before implementing the iterative process just described in the previous Section, some other considerations need to be applied for practical purposes. First, every new iteration will not take into account the highest order of $\Bar{\lambda}_{2n}$ of the previous iteration. Consequently, the $0th$ order iteration takes into account all four interaction terms, meanwhile, the $1st$ order iteration only takes into account the three lowest order interaction terms and so on. This is done in order to reduce the amount of expressions. However, as a result of this methodology, the expansion of the $\mathcal{P}$ operator must be done with care. Secondly, the length of $G_\nu^{(n)}$, $F_\nu^{(n)}$, $D_\nu^{(n)}$, and $R_\nu^{(n)}$ increases drastically with each iteration. As a result, it is necessary to introduce new notation to divide these expressions into pieces. How exactly this is done is better shown in the subsequent subsections. Finally, to reduce the numerical constants associated with some expressions, it is convenient to define
\begin{align}
    \alpha(n) = 2^n (n!)^2 \,. \label{def-const-alpha}
\end{align}

%%%%%%%%%%%%%%%%%%%%%%%%%%%%%%%%%%%%%%%%%%%%%%%%%%%%%%%%%%%%%%%%%%%%%%%%%%%%%%%%

\subsection{0\textit{th} order iteration}

Given that the $0th$ order approximation of $\Psi_\nu$ is Eq.~\eqref{Psi_iter_0}, Eqs.~\eqref{G_exp_0th} to \eqref{R_exp_0th} can be derived from the expressions in Eq.~\eqref{GFDR_exp4}:
\begin{align}
    G_\nu^{(0)} =& \, \frac{\Gamma_\nu \mathcal{P}^{-1}}{4! \, m_a^3} \bigg[ \Psi_s^3 \delta_{\nu,-2} + 3 \, \Psi_s |\Psi_s|^2 \delta_{\nu,0} + 3 \, \Psi_s^* |\Psi_s|^2 \delta_{\nu,2} + {\Psi_s^*}^3 \delta_{\nu,4} \bigg] \,, \label{G_exp_0th}
\end{align}
\begin{align}
    F_\nu^{(0)} = \, \frac{3 \, \Gamma_\nu \mathcal{P}^{-1}}{4 \cdot 6! \, m_a^4} \bigg[& \Psi_s^5 \delta_{\nu,-4} + 5 \, \Psi_s^3 |\Psi_s|^2 \delta_{\nu,-2} + 10 \, \Psi_s |\Psi_s|^4 \delta_{\nu,0} \nonumber \\
    &+ 10 \, \Psi_s^* |\Psi_s|^4 \delta_{\nu,2} + 5 \, {\Psi_s^*}^3 |\Psi_s|^2 \delta_{\nu,4} + {\Psi_s^*}^5 \delta_{\nu,6} \bigg] \,, \label{F_exp_0th}
\end{align}
\begin{align}
    D_\nu^{(0)} = \, \frac{\Gamma_\nu \mathcal{P}^{-1}}{2 \cdot 8! \, m_a^5} \bigg[ &\Psi_s^7 \delta_{\nu,-6} + 7 \, \Psi_s^5 |\Psi_s|^2 \delta_{\nu,-4} + 21 \, \Psi_s^3 |\Psi_s|^4 \delta_{\nu,-2}\nonumber \\
    &+ 35 \, \Psi_s |\Psi_s|^6 \delta_{\nu,0} +  35 \, \Psi_s^* |\Psi_s|^6  \delta_{\nu,2} + 21 \, {\Psi_s^*}^3 |\Psi_s|^4 \delta_{\nu,4} \nonumber \\
    &+ 7 \, {\Psi_s^*}^5 |\Psi_s|^2 \delta_{\nu,6} + {\Psi_s^*}^7 \delta_{\nu,8} \bigg] \,, \label{D_exp_0th}
\end{align}
\begin{align}
    R_\nu^{(0)} = \, \frac{5 \, \Gamma_\nu \mathcal{P}^{-1}}{16 \cdot 10! \, m_a^6} \bigg[ &\Psi_s^9 \delta_{\nu,-8} + 9 \, \Psi_s^7 |\Psi_s|^2 \delta_{\nu,-6} + 36 \, \Psi_s^5 |\Psi_s|^4 \delta_{\nu,-4} + 84 \, \Psi_s^3 |\Psi_s|^6 \delta_{\nu,-2}\nonumber \\
    &+ 126 \, \Psi_s |\Psi_s|^8 \delta_{\nu,0} + 126 \, \Psi_s^* |\Psi_s|^8  \delta_{\nu,2} +  84 \, {\Psi_s^*}^3 |\Psi_s|^6  \delta_{\nu,4} \nonumber \\
    &+ 36 \, {\Psi_s^*}^5 |\Psi_s|^4 \delta_{\nu,6} + 9 \, {\Psi_s^*}^7 |\Psi_s|^2 \delta_{\nu,8} + {\Psi_s^*}^9 \delta_{\nu,10} \bigg] \,, \label{R_exp_0th}
\end{align}
For $\nu = 0$ these expressions take the following form
\begin{align}
    G_0^{(0)} =& \, \frac{2 \, \Gamma_0 \mathcal{P}^{-1}}{\alpha(2) \, m_a^3} \left( \Psi_s |\Psi_s|^2 \right) \,, \qquad  F_0^{(0)} = \, \frac{3 \, \Gamma_0 \mathcal{P}^{-1}}{\alpha(3) \, m_a^4} \left( \Psi_s |\Psi_s|^4 \right) \,, \nonumber \\
    D_0^{(0)} =& \, \frac{4 \, \Gamma_0 \mathcal{P}^{-1}}{\alpha(4) \, m_a^5} \left( \Psi_s |\Psi_s|^6 \right) \,, \qquad  R_0^{(0)} =\, \frac{5 \, \Gamma_0 \mathcal{P}^{-1}}{\alpha(5) \, m_a^6} \left( \Psi_s |\Psi_s|^8 \right) \,. \label{GFDR_exp_0th_nu_0}    
\end{align}

%%%%%%%%%%%%%%%%%%%%%%%%%%%%%%%%%%%%%%%%%%%%%%%%%%%%%%%%%%%%%%%%%%%%%%%%%%%%%%%%

\subsection{1\textit{st} order iteration}

In this iteration $\Psi_\nu = \Psi_\nu^{(1)}$. With this in mind, Eqs.~\eqref{G_exp_1st} to \eqref{D_exp_1st} can be derived again from expressions in Eq.~\eqref{GFDR_exp4}
\begin{align}
    G_\nu^{(1)} = \, \frac{3 \, \Gamma_\nu \mathcal{P}^{-1}}{4! \, m_a^3} \bigg[& \Psi_s^2 \left( \Psi_{2 + \nu}^{(1)} + {\Psi_{-\nu}^{(1)*}} \right) + 2 \, |\Psi_s|^2 \left( {\Psi_{\nu}^{(1)}} + \Psi_{2-\nu}^{(1)*} \right)\nonumber \\
    &+ {\Psi_s^*}^2 \left( \Psi_{-2 + \nu}^{(1)} + {\Psi_{4-\nu}^{(1)*}} \right) \bigg] \label{G_exp_1st} \,,
\end{align}
\begin{align}
    F_\nu^{(1)} = \, \frac{15 \, \Gamma_\nu \mathcal{P}^{-1}}{4 \cdot 6! \, m_a^4} \bigg[& \Psi_s^4 \left( \Psi_{4+\nu}^{(1)} + \Psi_{-2-\nu}^{(1)*} \right) + 4 \, \Psi_s^2 |\Psi_s|^2 \left( \Psi_{2+\nu}^{(1)} + \Psi_{-\nu}^{(1)*} \right) \nonumber \\
    &+ 6 \, |\Psi_s|^4 \left( \Psi_{\nu}^{(1)} + \Psi_{2-\nu}^{(1)*} \right) + 4 \, {\Psi_s^*}^2 |\Psi_s|^2 \left( \Psi_{-2+\nu}^{(1)} + \Psi_{4-\nu}^{(1)*} \right) \nonumber \\
    &+ {\Psi_s^*}^4 \left( \Psi_{-4+\nu}^{(1)} + \Psi_{6-\nu}^{(1)*} \right) \bigg] \label{F_exp_1st} \,,
\end{align}
\begin{align}
    D_\nu^{(1)} = \, \frac{7 \, \Gamma_\nu \mathcal{P}^{-1}}{2 \cdot 8! \, m_a^5} \bigg[ &\Psi_s^6 \left( \Psi_{6+\nu}^{(1)} + \Psi_{-4-\nu}^{(1)*} \right) + 6 \, \Psi_s^4 |\Psi_s|^2 \left( \Psi_{4+\nu}^{(1)} + \Psi_{-2-\nu}^{(1)*} \right) \nonumber \\
    &+ 15 \, \Psi_s^2 |\Psi_s|^4 \left( \Psi_{2+\nu}^{(1)} + \Psi_{-\nu}^{(1)*} \right) + 20  \, |\Psi_s|^6 \left( \Psi_{\nu}^{(1)} + \Psi_{2-\nu}^{(1)*} \right) \nonumber \\
    &+ 15 \,  {\Psi_s^*}^2 |\Psi_s|^4 \left( \Psi_{-2+\nu}^{(1)} + \Psi_{4-\nu}^{(1)*} \right) + 6 \, {\Psi_s^*}^4 |\Psi_s|^2 \left( \Psi_{-4+\nu}^{(1)} + \Psi_{6-\nu}^{(1)*} \right) \nonumber \\ 
    &+ {\Psi_s^*}^6 \left( \Psi_{-6+\nu}^{(1)} + \Psi_{8-\nu}^{(1)*} \right) \bigg] \label{D_exp_1st} \,.
\end{align}
For $\nu = 0$, these expressions take the following form:
\begin{align}
    G_0^{(1)} =& \, \frac{3 \, \Gamma_0 \mathcal{P}^{-1}}{4! \, m_a^3} \bigg[ \Psi_s^2 \Psi_{2}^{(1)}  + 2 \, |\Psi_s|^2 \Psi_{2}^{(1)*} + {\Psi_s^*}^2 \left( {\Psi_{4}^{(1)*}} + \Psi_{-2}^{(1)} \right) \bigg] \label{G_exp_1st_nu_0} \,,
\end{align}
\begin{align}
    F_0^{(1)} = \, \frac{15 \, \Gamma_0 \mathcal{P}^{-1}}{4 \cdot 6! \, m_a^4} \bigg[& \Psi_s^4 \left( \Psi_{4}^{(1)} + \Psi_{-2}^{(1)*} \right) + 4 \, \Psi_s^2 |\Psi_s|^2 \Psi_{2}^{(1)} + 6 \, |\Psi_s|^4 \Psi_{2}^{(1)*} \nonumber \\
    &+ 4 \, {\Psi_s^*}^2 |\Psi_s|^2 \left( \Psi_{4}^{(1)*} + \Psi_{-2}^{(1)} \right)+ {\Psi_s^*}^4 \left( \Psi_{6}^{(1)*} + \Psi_{-4}^{(1)} \right) \bigg] \label{F_exp_1st_nu_0} \,,
\end{align}
\begin{align}
    D_0^{(1)} = \, \frac{7 \, \Gamma_0 \mathcal{P}^{-1}}{2 \cdot 8! \, m_a^5} \bigg[ &\Psi_s^6 \left( \Psi_{6}^{(1)} + \Psi_{-4}^{(1)*} \right) + 6 \, \Psi_s^4 |\Psi_s|^2 \left( \Psi_{4}^{(1)} + \Psi_{-2}^{(1)*} \right) + 15 \, \Psi_s^2 |\Psi_s|^4 \Psi_{2}^{(1)}\nonumber \\
    &+ 20 \, |\Psi_s|^6 \Psi_{2}^{(1)*} + 15 \, {\Psi_s^*}^2 |\Psi_s|^4 \left( \Psi_{4}^{(1)*} + \Psi_{-2}^{(1)} \right) \nonumber \\
    &+ 6 \,{\Psi_s^*}^4 |\Psi_s|^2 \left( \Psi_{6}^{(1)*} + \Psi_{-4}^{(1)} \right) + {\Psi_s^*}^6 \left( \Psi_{8}^{(1)*} + \Psi_{-6}^{(1)} \right)  \bigg] \label{D_exp_1st_nu_0} \,,
\end{align}

Afterwards, it is necessary to express all $\Psi_\nu^{(1)}$ in terms of $\Psi_s$. This is computed by taking $\Bar{\lambda}_{10}=0$ and $n=1$ in Eq.~\eqref{eq_motion_psi_iterat_n},
\begin{align}
\Psi_\nu^{(1)} =& \, \Bar{\lambda}_4 G_\nu^{(0)} + \Bar{\lambda}_6 F_\nu^{(0)} + \Bar{\lambda}_8 D_\nu^{(0)} \,,
\label{eq_motion_psi_iterat_1}
\end{align}
where the first term in the RHS of Eq.~\eqref{eq_motion_psi_iterat_n} becomes zero because $\Psi_\nu^{(0)}=0$. By inspecting Eqs.~\eqref{G_exp_0th} to \eqref{D_exp_0th}, it becomes clear that the only non-zero modes are $\nu =-6$, $-4$, $-2$, $2$, $4$, $6$ and $8$, when utilizing only the first three self-interaction terms. For example, $\Psi_{2}^{(1)}$ equals
\begin{align}
    \Psi_{2}^{(1)} =& \, \Bar{\lambda}_4 \, \frac{3 \, \Gamma_{2} \mathcal{P}^{-1}}{4! \, m_a^3} \left( \Psi_s^* |\Psi_s|^2 \right) + \Bar{\lambda}_6 \, \frac{15 \, \Gamma_{2} \mathcal{P}^{-1}}{2 \cdot 6! \, m_a^4} \left( \Psi_s^* |\Psi_s|^4 \right) + \Bar{\lambda}_8 \, \frac{35 \, \Gamma_{2} \mathcal{P}^{-1}}{2 \cdot 8! \, m_a^5} \left( \Psi_s^* |\Psi_s|^6  \right) \,.
\end{align}

Replacing all the required modes in Eqs.~\eqref{G_exp_1st_nu_0} to \eqref{D_exp_1st_nu_0}, they result in 
\begin{align}
     G_0^{(1)} =& \, \Bar{\lambda}_4 \, \frac{3 \, \Gamma_0 \mathcal{P}^{-\frac{1}{2}}}{2 \, \alpha(3) \, m_a^6} \mathrm{G} {\scriptstyle (4)}_{0,(1)}^{(1)} + \Bar{\lambda}_6 \, \frac{6 \, \Gamma_0 \mathcal{P}^{-\frac{1}{2}}}{\alpha(4) \, m_a^7} \mathrm{G} {\scriptstyle (6)}_{0,(1)}^{(1)} + \Bar{\lambda}_8 \, \frac{5 \, \Gamma_0 \mathcal{P}^{-\frac{1}{2}}}{\alpha(5) \, m_a^8} \mathrm{G} {\scriptstyle (8)}_{0,(1)}^{(1)} \,, \nonumber \\
     F_0^{(1)} =& \, \Bar{\lambda}_4 \, \frac{2 \, \Gamma_0 \mathcal{P}^{-\frac{1}{2}}}{\alpha(4) \, m_a^7} \mathrm{F} {\scriptstyle (4)}_{0,(1)}^{(1)} + \Bar{\lambda}_6 \, \frac{5 \, \Gamma_0 \mathcal{P}^{-\frac{1}{2}}}{2 \, \alpha(5) \, m_a^8} \mathrm{F} {\scriptstyle (6)}_{0,(1)}^{(1)}+ \Bar{\lambda}_8 \, \frac{15 \, \Gamma_0 \mathcal{P}^{-\frac{1}{2}}}{\alpha(6) \, m_a^9} \mathrm{F} {\scriptstyle (8)}_{0,(1)}^{(1)} \,, \nonumber \\
     D_0^{(1)} =& \, \Bar{\lambda}_4 \, \frac{5 \, \Gamma_0 \mathcal{P}^{-\frac{1}{2}}}{\alpha(5) \, m_a^8} \mathrm{D} {\scriptstyle (4)}_{0,(1)}^{(1)} + \Bar{\lambda}_6 \, \frac{3 \, \Gamma_0 \mathcal{P}^{-\frac{1}{2}}}{\alpha(6) \, m_a^9} \mathrm{D} {\scriptstyle (6)}_{0,(1)}^{(1)} + \Bar{\lambda}_8 \, \frac{7 \, \Gamma_0 \mathcal{P}^{-\frac{1}{2}}}{2 \, \alpha(7) \, m_a^{10}} \mathrm{D} {\scriptstyle (8)}_{0,(1)}^{(1)} \,, \label{GFD_exp_1st_nu_0-complete}
\end{align}
where a new notation is introduced, \emph{e.g.} $\mathrm{G} {\scriptstyle (4)}_{0,(1)}^{(1)}$ corresponds to the term related to $\Bar{\lambda}_4$ inside $G_0^{(1)}$. The extra sub-index $(1)$ indicates that this term arises from the part of $G_0^{(1)}$ that only includes $\Psi_\nu^{(1)}$. This seems redundant as $G_0^{(1)}$ is completely expressed in terms of $\Psi_\nu^{(1)}$; however, this detail shall become important in the next iterations. The exact form of all the terms in Eq.~\eqref{GFD_exp_1st_nu_0-complete} is presented in Appendix~\ref{App-B}.

%%%%%%%%%%%%%%%%%%%%%%%%%%%%%%%%%%%%%%%%%%%%%%%%%%%%%%%%%%%%%%%%%%%%%%%%%%%%%%%%

\subsection{2\textit{nd} order iteration}

In this iteration $G_{\nu}^{(2)}$ and $F_{\nu}^{(2)}$ need to be computed. These terms are divided into two in order to reduce the length of their expressions:
\begin{align}
G_{\nu}^{(2)} =& \, G_{\nu,(1)}^{(2)} + G_{\nu,(2)}^{(2)}  \,, \qquad F_{\nu}^{(2)} = \, F_{\nu,(1)}^{(2)} + F_{\nu,(2)}^{(2)} \label{GF_exp_2nd} \,,
\end{align}
where $G_{\nu,(1)}^{(2)}$ and $F_{\nu,(1)}^{(2)}$ contain only terms proportional to $\Psi_{\mu}^{(1)} \Psi_{\sigma}^{(1)}$. On the other hand, $G_{\nu,(2)}^{(2)}$ and $F_{\nu,(2)}^{(2)}$ contain only terms proportional to $\Psi_{\mu}^{(2)}$. The four terms shown in Eq.~\eqref{GF_exp_2nd} can be derived from those expressions in Eq.~\eqref{GFDR_exp4}:
\begin{align}
    G_{\nu,(1)}^{(2)} =& \, \frac{3 \, \Gamma_\nu \mathcal{P}^{-1}}{4! \, m_a^3} \sum_{\nu_1} \bigg[ \Psi_s \Big( \Psi_{\nu_1}^{(1)} \Psi_{2+\nu-\nu_1}^{(1)} + 2 \, \Psi_{\nu_1}^{(1)} \Psi_{-\nu + \nu_1}^{(1)*} + \Psi_{\nu_1}^{(1)*} \Psi_{2-\nu-\nu_1}^{(1)*} \Big) \nonumber \\
    &+ \Psi_s^* \Big( \Psi_{\nu_1}^{(1)} \Psi_{\nu-\nu_1}^{(1)} + 2 \, \Psi_{\nu_1}^{(1)*} \Psi_{-2+\nu + \nu_1}^{(1)} +\Psi_{\nu_1}^{(1)*} \Psi_{4-\nu-\nu_1}^{(1)*} \Big) \bigg] \label{G_exp_2nd_(1)} \,,
\end{align}
\begin{align}
    F_{\nu,(1)}^{(2)}  =& \, \frac{15 \, \Gamma_\nu \mathcal{P}^{-1}}{2 \cdot 6! \, m_a^4} \sum_{\nu_1} \bigg[ \Psi_s^3 \Big( \Psi_{\nu_1}^{(1)} \Psi_{4+\nu-\nu_1}^{(1)} + 2 \, \Psi_{\nu_1}^{(1)} \Psi_{-2-\nu+\nu_1}^{(1)*} + \Psi_{\nu_1}^{(1)*} \Psi_{-\nu-\nu_1}^{(1)*} \Big) \nonumber \\
    &+ 3\, \Psi_s |\Psi_s|^2 \Big( \Psi_{\nu_1}^{(1)} \Psi_{2+\nu-\nu_1}^{(1)} + 2 \, \Psi_{\nu_1}^{(1)} \Psi_{-\nu+\nu_1}^{(1)*} + \Psi_{\nu_1}^{(1)*} \Psi_{2-\nu-\nu_1}^{(1)*} \Big) \nonumber \\
    &+ 3 \, \Psi_s^* |\Psi_s|^2 \Big( \Psi_{\nu_1}^{(1)} \Psi_{\nu-\nu_1}^{(1)} + 2 \, \Psi_{\nu_1}^{(1)*} \Psi_{-2+\nu+\nu_1}^{(1)} + \Psi_{\nu_1}^{(1)*} \Psi_{4-\nu-\nu_1}^{(1)*} \Big) \nonumber \\
    &+ {\Psi_s^*}^3 \Big( \Psi_{\nu_1}^{(1)} \Psi_{-2+\nu-\nu_1}^{(1)} + 2 \, \Psi_{\nu_1}^{(1)*} \Psi_{-4+\nu+\nu_1}^{(1)} + \Psi_{\nu_1}^{(1)*} \Psi_{6-\nu-\nu_1}^{(1)*} \Big) \bigg] \label{F_exp_2nd_(1)} \,.
\end{align}
Its important to note that the sums inside Eqs.~\eqref{G_exp_2nd_(1)} and \eqref{F_exp_2nd_(1)} are finite since $\Psi_\nu^{(1)}$ only exists for $\nu= -4$, $-2$, $2$, $4$ and $6$, when only the first two self-interaction terms are considered. For $\nu=0$ these expressions take the following form:
\begin{align}
    G_{0,(1)}^{(2)} =& \, \frac{3 \, \Gamma_0 \mathcal{P}^{-1}}{4! \, m_a^3} \sum_{\nu_1} \bigg[ \Psi_s \Big( \Psi_{\nu_1}^{(1)} \Psi_{2-\nu_1}^{(1)} + 2 \, \Psi_{\nu_1}^{(1)} \Psi_{\nu_1}^{(1)*} + \Psi_{\nu_1}^{(1)*} \Psi_{2-\nu_1}^{(1)*} \Big) \nonumber \\
    &+ \Psi_s^* \Big( \Psi_{\nu_1}^{(1)} \Psi_{-\nu_1}^{(1)} + 2\, \Psi_{\nu_1}^{(1)*} \Psi_{-2+ \nu_1}^{(1)} +\Psi_{\nu_1}^{(1)*} \Psi_{4-\nu_1}^{(1)*} \Big) \bigg] \label{G_exp_2nd_(1)_nu_0} \,,
\end{align}
\begin{align}
    F_{0,(1)}^{(2)} =& \, \frac{15 \, \Gamma_0 \mathcal{P}^{-1}}{2 \cdot 6! \, m_a^4} \sum_{\nu_1} \bigg[ \Psi_s^3 \Big( \Psi_{\nu_1}^{(1)} \Psi_{4-\nu_1}^{(1)} + 2 \, \Psi_{\nu_1}^{(1)} \Psi_{-2+\nu_1}^{(1)*} + \Psi_{\nu_1}^{(1)*} \Psi_{-\nu_1}^{(1)*} \Big) \nonumber \\
    &+ 3\, \Psi_s |\Psi_s|^2 \Big( \Psi_{\nu_1}^{(1)} \Psi_{2-\nu_1}^{(1)} + 2 \, \Psi_{\nu_1}^{(1)} \Psi_{\nu_1}^{(1)*} + \Psi_{\nu_1}^{(1)*} \Psi_{2-\nu_1}^{(1)*} \Big) \nonumber \\
    &+ 3 \, \Psi_s^* |\Psi_s|^2 \Big( \Psi_{\nu_1}^{(1)} \Psi_{-\nu_1}^{(1)} + 2 \, \Psi_{\nu_1}^{(1)*} \Psi_{-2+\nu_1}^{(1)} + \Psi_{\nu_1}^{(1)*} \Psi_{4-\nu_1}^{(1)*} \Big) \nonumber \\
    &+ {\Psi_s^*}^3 \Big( \Psi_{\nu_1}^{(1)} \Psi_{-2-\nu_1}^{(1)} + 2 \, \Psi_{\nu_1}^{(1)*} \Psi_{-4+\nu_1}^{(1)} + \Psi_{\nu_1}^{(1)*} \Psi_{6-\nu_1}^{(1)*} \Big) \bigg] \label{F_exp_2nd_(1)_nu_0} \,.
\end{align}
Meanwhile,
\begin{align}
    G_{\nu,(2)}^{(2)} = \, \frac{3 \, \Gamma_\nu \mathcal{P}^{-1}}{4! \, m_a^3} \bigg[& \Psi_s^2 \left( \Psi_{2 + \nu}^{(2)} + {\Psi_{-\nu}^{(2)*}} \right) + 2  \, |\Psi_s|^2 \left( {\Psi_{\nu}^{(2)}} + \Psi_{2-\nu}^{(2)*} \right) \nonumber\\&
    + {\Psi_s^*}^2 \left( \Psi_{-2+\nu}^{(2)} + {\Psi_{4-\nu}^{(2)*}} \right) \bigg] \label{G_exp_2nd_(2)} \,,
\end{align}
\begin{align}
    F_{\nu,(2)}^{(2)} = \, \frac{15}{4} \, \frac{\Gamma_\nu \mathcal{P}^{-1}}{6! \, m_a^4} \bigg[ &\Psi_s^4 \left( \Psi_{4+\nu}^{(2)} + \Psi_{-2-\nu}^{(2)*} \right) + 4 \, \Psi_s^2 |\Psi_s|^2 \left( \Psi_{2+\nu}^{(2)} + \Psi_{-\nu}^{(2)*} \right) \nonumber \\
    &+ 6 \, |\Psi_s|^4 \left( \Psi_{\nu}^{(2)} + \Psi_{2-\nu}^{(2)*} \right) + 4 \, {\Psi_s^*}^2 |\Psi_s|^2 \left( \Psi_{-2+\nu}^{(2)} + \Psi_{4-\nu}^{(2)*} \right) \nonumber \\
    & + {\Psi_s^*}^4 \left( \Psi_{-4+\nu}^{(2)} + \Psi_{6-\nu}^{(2)*} \right) \bigg] \label{F_exp_2nd_(2)} \,.
\end{align}
where, for $\nu=0$, they become in
\begin{align}
    G_{0,(2)}^{(2)} &= \, \frac{3 \, \Gamma_0 \mathcal{P}^{-1}}{4! \, m_a^3} \bigg[ \Psi_s^2 \Psi_{2}^{(2)}  + 2 \, |\Psi_s|^2 \Psi_{2}^{(2)*} + {\Psi_s^*}^2 \left( {\Psi_{4}^{(2)*}} + \Psi_{-2}^{(2)} \right) \bigg] \label{G_exp_2nd_(2)_nu_0} \,,
\end{align}
\begin{align}
    F_{0,(2)}^{(2)} = \, \frac{15}{4} \, \frac{\Gamma_0 \mathcal{P}^{-1}}{6! \, m_a^4} \bigg[& \Psi_s^4 \left( \Psi_{4}^{(2)} + \Psi_{-2}^{(2)*} \right) + 4 \, \Psi_s^2 |\Psi_s|^2 \Psi_{2}^{(2)} + 6 \, |\Psi_s|^4 \Psi_{2}^{(2)*} \nonumber \\
    &+ 4 \,  {\Psi_s^*}^2 |\Psi_s|^2 \left( \Psi_{4}^{(2)*} + \Psi_{-2}^{(2)} \right)+ {\Psi_s^*}^4 \left( \Psi_{6}^{(2)*} + \Psi_{-4}^{(2)} \right) \bigg] \label{F_exp_2nd_(2)_nu_0} \,.
\end{align}
Next, all $\Psi_\nu^{(1)}$ in Eqs.~\eqref{G_exp_2nd_(1)_nu_0} and \eqref{F_exp_2nd_(1)_nu_0} are expressed in terms of $\Psi_s$ by implementing Eq.~\eqref{eq_motion_psi_iterat_1} but taking into account that $\Bar{\lambda}_8 = \Bar{\lambda}_{10} = 0$ in this iteration. This results in 
\begin{align}
    G_{0,(1)}^{(2)} =& \, \Bar{\lambda}_4^2 \, \frac{2 \, \Gamma_0 \mathcal{P}^{-\frac{1}{2}}}{\alpha(4) \, m_a^{9}}  \mathrm{G} {\scriptstyle (4,4)}_{0,(1)}^{(2)} + \Bar{\lambda}_4 \Bar{\lambda}_6 \, \frac{5 \, \Gamma_0 \mathcal{P}^{-\frac{1}{2}}}{\alpha(5) \, m_a^{10}}  \mathrm{G} {\scriptstyle (4,6)}_{0,(1)}^{(2)} + \Bar{\lambda}_6^2 \, \frac{9 \, \Gamma_0 \mathcal{P}^{-\frac{1}{2}}}{\alpha(6) \, m_a^{11}} \mathrm{G} {\scriptstyle (6,6)}_{0,(1)}^{(2)} \label{G_exp_2nd_(1)_nu_0-complete} \,,
\end{align}
\begin{align}
    F_{0,(1)}^{(2)} =& \, \Bar{\lambda}_4^2 \, \frac{25 \, \Gamma_0 \mathcal{P}^{-\frac{1}{2}}}{\alpha(5) \, m_a^{10}} \mathrm{F} {\scriptstyle (4,4)}_{0,(1)}^{(2)} + \Bar{\lambda}_4 \Bar{\lambda}_6 \, \frac{30 \, \Gamma_0 \mathcal{P}^{-\frac{1}{2}}}{\alpha(6) \, m_a^{11}} \mathrm{F} {\scriptstyle (4,6)}_{0,(1)}^{(2)} + \Bar{\lambda}_6^2 \, \frac{147 \, \Gamma_0 \mathcal{P}^{-\frac{1}{2}}}{2 \, \alpha(7) \, m_a^{12}} \mathrm{F} {\scriptstyle (6,6)}_{0,(1)}^{(2)} \label{F_exp_2nd_(1)_nu_0-complete} \,.
\end{align}
Similarly, it is necessary to express all $\Psi_\nu^{(2)}$ in Eqs.~\eqref{G_exp_2nd_(2)_nu_0} and \eqref{F_exp_2nd_(2)_nu_0} in terms of $\Psi_s$. These expressions are computed by taking $n=2$ in Eq.~\eqref{eq_motion_psi_iterat_n}:

\begin{align}
\Psi_\nu^{(2)} =& - \frac{i}{m_a} \Gamma_\nu \dot{\Psi}_\nu^{(1)} + \Bar{\lambda}_4 G_\nu^{(1)} + \Bar{\lambda}_6 F_\nu^{(1)} \,,
\label{eq_motion_psi_iterat_2}
\end{align}

\noindent
where $\dot{\Psi}_\nu^{(1)}$ is calculated by deriving with respect time Eq.~\eqref{eq_motion_psi_iterat_1}, \emph{e.g.}, for $\nu=2$, this results in
\begin{align}
    - \frac{i}{m_a} \Gamma_{2} \Dot{\Psi}_{2}^{(1)} =& - i \, \Bar{\lambda}_4 \, \frac{6 \, (\Gamma_{2})^2 \mathcal{P}^{-1}}{4! \, m_a^4} \left( |\Psi_s|^2 \Dot{\Psi}_s^* \right) - i \, \Bar{\lambda}_4 \, \frac{3 \, (\Gamma_{2})^2 \mathcal{P}^{-1}}{4! \, m_a^4} \left( {\Psi_s^*}^2 \Dot{\Psi}_s \right) \nonumber \\ 
    &- i \, \Bar{\lambda}_6 \, \frac{15 \, (\Gamma_{2})^2 \mathcal{P}^{-1}}{4 \cdot 6! \, m_a^5} \left( 6 \, |\Psi_s|^4 \Dot{\Psi}_s^* \right. + \left. 4 \, {\Psi_s^*}^2 |\Psi_s|^2 \Dot{\Psi}_s  \right) \,.
\end{align}

As mentioned before, the only non-zero modes are $\nu =-6$, $-4$, $-2$, $2$, $4$, $6$ and $8$, when only the first two self-interaction terms are utilized. As an example of calculation, $\Psi_2^{(2)}$ equals
\begin{align}
   & \Psi_{2}^{(2)} = - i \, \Bar{\lambda}_4 \, \frac{3 \, (\Gamma_{2})^2 \mathcal{P}^{-1}}{4! \, m_a^4} \left( 2 \, |\Psi_s|^2 \Dot{\Psi}_s^* + {\Psi_s^*}^2 \Dot{\Psi}_s \right) - i \, \Bar{\lambda}_6 \, \frac{15 \, (\Gamma_{2})^2 \mathcal{P}^{-1}}{4 \cdot 6! \, m_a^5} \left( 6 \, |\Psi_s|^4 \Dot{\Psi}_s^* + 4 \, {\Psi_s^*}^2 |\Psi_s|^2 \Dot{\Psi}_s \right) \nonumber \\  
    &+ \Bar{\lambda}_4^2 \, \frac{3 \, \Gamma_{2} \mathcal{P}^{-1}}{(4!)^2 \, m_a^{6}} \bigg[ \Psi_s^2 (\Gamma_{4} + \Gamma_{-2}) \mathcal{P}^{-1} \left( {\Psi_s^*}^3 \right) + 6 \, |\Psi_s|^2 \Gamma_{2} \mathcal{P}^{-1} \left( \Psi_s^* |\Psi_s|^2  \right) + 3 \, {\Psi_s^*}^2 \Gamma_{2} \mathcal{P}^{-1} \left( \Psi_s |\Psi_s|^2  \right) \bigg] \nonumber \\
    &+ \Bar{\lambda}_4 \Bar{\lambda}_6 \, \frac{\Gamma_{2} \mathcal{P}^{-1}}{40 \cdot (4!)^2 \, m_a^{7}} \bigg[ 15 \, \Psi_s^2 (\Gamma_{4} + \Gamma_{-2}) \mathcal{P}^{-1} \left( {\Psi_s^*}^3 |\Psi_s|^2   \right) + 60 \, |\Psi_s|^2 \Gamma_{2} \mathcal{P}^{-1} \left( \Psi_s^* |\Psi_s|^4  \right) \nonumber \\ 
    &+ 30 \, {\Psi_s^*}^2 \Gamma_{2} \mathcal{P}^{-1} \left( \Psi_s |\Psi_s|^4  \right) + 20 \, \Psi_s^2 |\Psi_s|^2 (\Gamma_{4} + \Gamma_{-2}) \mathcal{P}^{-1} \left( {\Psi_s^*}^3   \right) + 90 \, |\Psi_s|^4 \Gamma_{2} \mathcal{P}^{-1} \left( \Psi_s^* |\Psi_s|^2  \right) \nonumber \\ 
    &+ 60 \, {\Psi_s^*}^2 |\Psi_s|^2 \Gamma_{2} \mathcal{P}^{-1} \left( \Psi_s |\Psi_s|^2  \right) + 5 \, {\Psi_s^*}^4 (\Gamma_{4} + \Gamma_{-2}) \mathcal{P}^{-1} \left( \Psi_s^3   \right) \bigg] \nonumber \\
    &+ \Bar{\lambda}_6^2 \, \frac{45 \, \Gamma_{2} \mathcal{P}^{-1}}{16 \cdot (6!)^2 \, m_a^{8}} \bigg[ \Psi_s^4 (\Gamma_{6} + \Gamma_{-4}) \mathcal{P}^{-1} \left( {\Psi_s^*}^5  \right) + 20 \, \Psi_s^2 |\Psi_s|^2 (\Gamma_{4} + \Gamma_{-2}) \mathcal{P}^{-1} \left( {\Psi_s^*}^3 |\Psi_s|^2  \right) \nonumber \\ 
    &+ 60 \, |\Psi_s|^4 \Gamma_{2} \mathcal{P}^{-1} \left( \Psi_s^* |\Psi_s|^4  \right) + 40 \, {\Psi_s^*}^2 |\Psi_s|^2 \Gamma_{2} \mathcal{P}^{-1} \left( \Psi_s |\Psi_s|^4  \right) \nonumber \\
    &+ 5 \, {\Psi_s^*}^4 (\Gamma_{4} + \Gamma_{-2}) \mathcal{P}^{-1} \left( \Psi_s^3 |\Psi_s|^2   \right) \bigg] \,.
\end{align} 
Replacing all the required modes in Eqs.~\eqref{G_exp_2nd_(2)_nu_0} and \eqref{F_exp_2nd_(2)_nu_0} results in 
\begin{align}
    G_{0,(2)}^{(2)} =& \, i \, \Bar{\lambda}_4 \, \frac{9 \, \Gamma_0 \mathcal{P}^{-\frac{1}{2}}}{2 \, \alpha(3) \, m_a^{7}} \mathrm{G} {\scriptstyle (4, \textbf{\RNum{1}})}_{0,(2)}^{(2)} + i \, \Bar{\lambda}_6 \, \frac{6 \, \Gamma_0 \mathcal{P}^{-\frac{1}{2}}}{\alpha(4) \, m_a^{8}} \mathrm{G} {\scriptstyle (6, \textbf{\RNum{1}})}_{0,(2)}^{(2)} + \Bar{\lambda}_4^2 \, \frac{6 \, \Gamma_0 \mathcal{P}^{-\frac{1}{2}}}{\alpha(4) \, m_a^{9}} \mathrm{G} {\scriptstyle (4,4)}_{0,(2)}^{(2)}\nonumber \\
    &+ \Bar{\lambda}_4 \Bar{\lambda}_6 \, \frac{5 \, \Gamma_0 \mathcal{P}^{-\frac{1}{2}}}{2 \, \alpha(5) \, m_a^{10}} \mathrm{G} {\scriptstyle (4,6)}_{0,(2)}^{(2)} + \Bar{\lambda}_6^2 \, \frac{45 \, \Gamma_0 \mathcal{P}^{-\frac{1}{2}}}{2 \, \alpha(6) \, m_a^{11}} \mathrm{G} {\scriptstyle (6,6)}_{0,(2)}^{(2)} \label{G_exp_2nd_(2)_nu_0-complete} \,,
\end{align}
\begin{align}
    F_{0,(2)}^{(2)} =& \, i \, \Bar{\lambda}_4 \, \frac{6 \, \Gamma_0 \mathcal{P}^{-\frac{1}{2}}}{\alpha(4) \, m_a^{8}} \mathrm{F} {\scriptstyle (4, \textbf{\RNum{1}})}_{0,(2)}^{(2)} + i \, \Bar{\lambda}_6 \, \frac{25 \, \Gamma_0 \mathcal{P}^{-\frac{1}{2}}}{2 \, \alpha(5) \, m_a^{9}} \mathrm{F} {\scriptstyle (6, \textbf{\RNum{1}})}_{0,(2)}^{(2)} + \Bar{\lambda}_4^2 \, \frac{25 \, \Gamma_0 \mathcal{P}^{-\frac{1}{2}}}{2 \, \alpha(5) \, m_a^{10}} \mathrm{F} {\scriptstyle (4,4)}_{0,(2)}^{(2)} \nonumber \\
    &+ \Bar{\lambda}_4 \Bar{\lambda}_6 \, \frac{15 \, \Gamma_0 \mathcal{P}^{-\frac{1}{2}}}{2 \, \alpha(6) \, m_a^{11}} \mathrm{F} {\scriptstyle (4,6)}_{0,(2)}^{(2)} + \Bar{\lambda}_6^2 \, \frac{735 \, \Gamma_0 \mathcal{P}^{-\frac{1}{2}}}{4 \, \alpha(7) \, m_a^{12}} \mathrm{F} {\scriptstyle (6,6)}_{0,(2)}^{(2)} \label{F_exp_2nd_(2)_nu_0-complete} \,,
\end{align}
where a new feature has been added to the notation. The roman numerals inside a term indicate that its complete expression is proportional to $\epsilon_t$. Furthermore, the number depicted in roman numerals is equal to the order of $\epsilon_t$, \emph{e.g.} $\mathrm{G} {\scriptstyle (4, \textbf{\RNum{1}})}_{0,(2)}^{(2)}$ is proportional to $\Dot{\Psi}_s$ and its complex conjugate. 

Replacing Eqs.~\eqref{G_exp_2nd_(1)_nu_0-complete}, \eqref{F_exp_2nd_(1)_nu_0-complete}, \eqref{G_exp_2nd_(2)_nu_0-complete} and \eqref{F_exp_2nd_(2)_nu_0-complete} into~\eqref{GF_exp_2nd}, results in
\begin{align}
    G_0^{(2)} =& \, i \,  \Bar{\lambda}_4 \, \frac{9 \, \Gamma_0 \mathcal{P}^{-\frac{1}{2}}}{2 \, \alpha(3) \, m_a^{7}} \mathrm{G} {\scriptstyle (4, \textbf{\RNum{1}})}_{0,(2)}^{(2)} + i \, \Bar{\lambda}_6 \, \frac{6 \, \Gamma_0 \mathcal{P}^{-\frac{1}{2}}}{\alpha(4) \, m_a^{8}} \mathrm{G} {\scriptstyle (6, \textbf{\RNum{1}})}_{0,(2)}^{(2)} \nonumber \\ 
    &+ \Bar{\lambda}_4^2 \, \frac{2 \,  \Gamma_0 \mathcal{P}^{-\frac{1}{2}}}{\alpha(4) \, m_a^{9}} \Big( \mathrm{G} {\scriptstyle (4,4)}_{0,(1)}^{(2)} + 3 \, \mathrm{G} {\scriptstyle (4,4)}_{0,(2)}^{(2)} \Big)\nonumber \\
    &+ \Bar{\lambda}_4 \Bar{\lambda}_6 \, \frac{5 \,  \Gamma_0 \mathcal{P}^{-\frac{1}{2}}}{2 \, \alpha(5) \, m_a^{10}} \Big( 2 \, \mathrm{G} {\scriptstyle (4,6)}_{0,(1)}^{(2)} + \mathrm{G} {\scriptstyle (4,6)}_{0,(2)}^{(2)} \Big) \nonumber \\
    &+ \Bar{\lambda}_6^2 \, \frac{9 \, \Gamma_0 \mathcal{P}^{-\frac{1}{2}}}{2 \, \alpha(6) \, m_a^{11}} \Big( 2 \, \mathrm{G} {\scriptstyle (6,6)}_{0,(1)}^{(2)} + 5 \, \mathrm{G} {\scriptstyle (6,6)}_{0,(2)}^{(2)} \Big) \label{G_exp_2nd_nu_0-complete} \,,
\end{align}
\begin{align}
    F_0^{(2)} =& \, i \, \Bar{\lambda}_4 \, \frac{6 \, \Gamma_0 \mathcal{P}^{-\frac{1}{2}}}{\alpha(4) \, m_a^{8}} \mathrm{F} {\scriptstyle (4, \textbf{\RNum{1}})}_{0,(2)}^{(2)} + i \, \Bar{\lambda}_6 \, \frac{25 \, \Gamma_0 \mathcal{P}^{-\frac{1}{2}}}{2 \, \alpha(5) \, m_a^{9}} \mathrm{F} {\scriptstyle (6, \textbf{\RNum{1}})}_{0,(2)}^{(2)}\nonumber \\
    &+ \Bar{\lambda}_4^2 \, \frac{25 \,  \Gamma_0 \mathcal{P}^{-\frac{1}{2}}}{2 \, \alpha(5) \, m_a^{10}} \Big( 2 \, \mathrm{F} {\scriptstyle (4,4)}_{0,(1)}^{(2)} + \mathrm{F} {\scriptstyle (4,4)}_{0,(2)}^{(2)} \Big) \nonumber \\
    &+ \Bar{\lambda}_4 \Bar{\lambda}_6 \, \frac{15 \,  \Gamma_0 \mathcal{P}^{-\frac{1}{2}}}{2 \, \alpha(6) \, m_a^{11}} \Big( 4 \, \mathrm{F} {\scriptstyle (4,6)}_{0,(1)}^{(2)} + \mathrm{F} {\scriptstyle (4,6)}_{0,(2)}^{(2)} \Big) \nonumber \\
    &+ \Bar{\lambda}_6^2 \, \frac{147 \, \Gamma_0 \mathcal{P}^{-\frac{1}{2}}}{4 \, \alpha(7) \, m_a^{12}} \Big( 2 \, \mathrm{F} {\scriptstyle (6,6)}_{0,(1)}^{(2)} + 5 \, \mathrm{F} {\scriptstyle (6,6)}_{0,(2)}^{(2)} \Big) \label{F_exp_2nd_nu_0-complete} \,.
\end{align}
The detailed expressions of all the terms inside Eqs.~\eqref{G_exp_2nd_nu_0-complete} and \eqref{F_exp_2nd_nu_0-complete} are presented in Appendix~\ref{App-C}.

\subsection{3\textit{rd} order iteration}

Similarly to the previous iteration, $G_{\nu}^{(3)}$ is divided in three terms in order to reduce its length and improve its readability:
\begin{align}
G_{\nu}^{(3)} =& \, G_{\nu,(1)}^{(3)} + G_{\nu,(1)(2)}^{(3)} + G_{\nu,(3)}^{(3)} \label{G_exp_3rd} \,,
\end{align}
where $G_{\nu,(1)}^{(3)}$ contains only terms proportional to $\Psi_{\mu}^{(1)} \Psi_{\sigma}^{(1)} \Psi_{\rho}^{(1)}$ . On the other hand, $G_{\nu,(1)(2)}^{(3)}$ contains only terms proportional to $\Psi_{\mu}^{(1)} \Psi_{\sigma}^{(2)}$. Lastly,  $G_{\nu,(3)}^{(3)}$ contains only terms proportional to $\Psi_{\mu}^{(3)}$. The three terms depicted in Eq.~\eqref{G_exp_3rd} can be derived from the $G$'s expression in Eq.~\eqref{GFDR_exp4}:
\begin{align}
    G_{\nu,(1)}^{(3)} = \frac{\Gamma_\nu \mathcal{P}^{-1}}{4! \, m_a^3} \sum_{\nu_1,\nu_2} \bigg[& \Psi_{\nu_1}^{(1)} \Psi_{\nu_2}^{(1)} \bigg( \Psi_{2 + \nu - \nu_1 - \nu_2}^{(1)} + 3 \, \Psi_{-\nu + \nu_1 + \nu_2}^{(1)*} \bigg) \nonumber \\
    &+ \Psi_{\nu_1}^{(1)*} \Psi_{\nu_2}^{(1)*} \bigg( 3 \,  \Psi_{-2 + \nu + \nu_1 + \nu_2}^{(1)} + \Psi_{4-\nu-\nu_1 -\nu_2}^{(1)*} \bigg) \bigg] \label{G_exp_3rd_(1)} \,,
\end{align}
\begin{align}
    G_{\nu,(1)(2)}^{(3)} = \frac{6 \, \Gamma_\nu \mathcal{P}^{-1}}{4! \, m_a^3} \sum_{\nu_1} \bigg[& \Psi_s \bigg( \Psi_{2 + \nu - \nu_1}^{(2)} \Psi_{\nu_1}^{(1)} + \Psi_{-\nu + \nu_1}^{(2)*} \Psi_{\nu_1}^{(1)} + \Psi_{\nu_1}^{(2)} \Psi_{-\nu + \nu_1}^{(1)*} + \Psi_{2-\nu-\nu_1}^{(2)*} \Psi_{\nu_1}^{(1)*} \bigg) \nonumber \\
    &+ \Psi_s^* \bigg( \Psi_{\nu - \nu_1}^{(2)} \Psi_{\nu_1}^{(1)} + \Psi_{-2 + \nu + \nu_1}^{(2)} \Psi_{\nu_1}^{(1)*} + \Psi_{\nu_1}^{(2)*} \Psi_{-2 + \nu + \nu_1}^{(1)} \nonumber \\
    &\qquad\quad+ \Psi_{4-\nu-\nu_1}^{(2)*} \Psi_{\nu_1}^{(1)*} \bigg) \bigg] \label{G_exp_3rd_(1)(2)} \,,
\end{align}
\begin{align}
    G_{\nu,(3)}^{(3)} = \, \frac{3 \, \Gamma_\nu \mathcal{P}^{-1}}{4! \, m_a^3} \bigg[& \Psi_s^2 \left( \Psi_{2 + \nu}^{(3)} + {\Psi_{-\nu}^{(3)*}} \right) + 2  \, |\Psi_s|^2 \left( {\Psi_{\nu}^{(3)}} + \Psi_{2-\nu}^{(3)*} \right) \nonumber \\
    &+ {\Psi_s^*}^2 \left( \Psi_{-2+\nu}^{(3)} + {\Psi_{4-\nu}^{(3)*}} \right) \bigg] \label{G_exp_3rd_(3)} \,.
\end{align}

Once again, it is noteworthy to mention that the sums inside Eqs.~\eqref{G_exp_3rd_(1)} and \eqref{G_exp_3rd_(1)(2)} are finite since $\Psi_\nu^{(1)}$a and $\Psi_\nu^{(2)}$ only exists for $\nu=-2$, $2$ and $4$, when considering only the first self-interaction term. For $\nu=0$, these expressions take the following form
\begin{align}
    G_{0,(1)}^{(3)} = \frac{\Gamma_0 \mathcal{P}^{-1}}{4! \, m_a^3} \sum_{\nu_1,\nu_2} \bigg[& \Psi_{\nu_1}^{(1)} \Psi_{\nu_2}^{(1)} \bigg( \Psi_{2 - \nu_1 - \nu_2}^{(1)} + 3 \, \Psi_{\nu_1 + \nu_2}^{(1)*} \bigg) \nonumber \\
    &+ \Psi_{\nu_1}^{(1)*} \Psi_{\nu_2}^{(1)*} \bigg( 3 \,  \Psi_{-2 + \nu_1 + \nu_2}^{(1)} + \Psi_{4-\nu_1 -\nu_2}^{(1)*} \bigg) \bigg] \label{G_exp_3rd_(1)_nu_0} \,,
\end{align}
\begin{align}
    G_{0,(1)(2)}^{(3)} = \frac{6 \, \Gamma_0 \mathcal{P}^{-1}}{4! \, m_a^3} \sum_{\nu_1} \bigg[& \Psi_s \bigg( \Psi_{2 - \nu_1}^{(2)} \Psi_{\nu_1}^{(1)} + \Psi_{ \nu_1}^{(2)*} \Psi_{\nu_1}^{(1)} + \Psi_{\nu_1}^{(2)} \Psi_{ \nu_1}^{(1)*} + \Psi_{2-\nu_1}^{(2)*} \Psi_{\nu_1}^{(1)*} \bigg) \nonumber \\
    &+ \Psi_s^* \bigg( \Psi_{- \nu_1}^{(2)} \Psi_{\nu_1}^{(1)} + \Psi_{-2 + \nu_1}^{(2)} \Psi_{\nu_1}^{(1)*} + \Psi_{\nu_1}^{(2)*} \Psi_{-2 + \nu_1}^{(1)} \nonumber\\
    &\qquad \quad+ \Psi_{4-\nu_1}^{(2)*} \Psi_{\nu_1}^{(1)*} \bigg) \bigg] \label{G_exp_3rd_(1)(2)_nu_0} \,,
\end{align}
\begin{align}
    G_{0,(3)}^{(3)} =& \, \frac{3 \, \Gamma_0 \mathcal{P}^{-1}}{4! \, m_a^3} \bigg[ \Psi_s^2 \Psi_{2 }^{(3)} + 2  \, |\Psi_s|^2 \Psi_{2}^{(3)*} + {\Psi_s^*}^2 \left( {\Psi_{4}^{(3)*}} + \Psi_{-2}^{(3)} \right) \bigg] \label{G_exp_3rd_(3)_nu_0} \,.
\end{align}
Next, all $\Psi_\nu^{(1)}$ and $\Psi_\nu^{(2)}$ in Eqs.~\eqref{G_exp_3rd_(1)_nu_0} and \eqref{G_exp_3rd_(1)(2)_nu_0} are expressed in terms of $\Psi_s$ by implementing Eqs.~\eqref{eq_motion_psi_iterat_1} and \eqref{eq_motion_psi_iterat_2} but taking into account that $\Bar{\lambda}_6 = \Bar{\lambda}_8 = \Bar{\lambda}_{10} = 0$ in this iteration. This results in
\begin{align}
    G_{0,(1)}^{(3)} =& \, \Bar{\lambda}_4^3 \frac{25 \, \Gamma_0 \mathcal{P}^{-\frac{1}{2}}}{2 \, \alpha(5) \, m_a^{12}} \mathrm{G} {\scriptstyle (4,4,4)}_{0,(1)}^{(3)} \label{G_exp_3rd_(1)_nu_0-complete} \,,
\end{align}
\begin{align}
    G_{0,(1)(2)}^{(3)} =& \, i \Bar{\lambda}_4^2 \, \frac{12 \, \Gamma_0 \mathcal{P}^{-\frac{1}{2}}}{\alpha(4) \, m_a^{10}} \mathrm{G} {\scriptstyle (4,4, \textbf{\RNum{1}})}_{0,(1)(2)}^{(3)} + \Bar{\lambda}_4^3 \frac{25 \, \Gamma_0 \mathcal{P}^{-\frac{1}{2}}}{\alpha(5) \, m_a^{12}} \mathrm{G} {\scriptstyle (4,4,4)}_{0,(1)(2)}^{(3)} \label{G_exp_3rd_(1)(2)_nu_0-complete} \,.
\end{align}
Afterwards, it is necessary to express all $\Psi_\nu^{(3)}$ in Eq.~\eqref{G_exp_3rd_(3)_nu_0} in terms of $\Psi_s$. This is done by taking $n=3$ in Eq.~\eqref{eq_motion_psi_iterat_n}:
\begin{align}
\Psi_\nu^{(3)} =& - \frac{i}{m_a} \Gamma_\nu \dot{\Psi}_\nu^{(2)} + \Bar{\lambda}_4 G_\nu^{(2)} \,,
\label{eq_motion_psi_iterat_3}
\end{align}
where $\dot{\Psi}_\nu^{(2)}$ is calculated by deriving with respect time Eq.~\eqref{eq_motion_psi_iterat_2}, \emph{e.g.}, for $\nu=2$, this results in
\begin{align}
    &- \frac{i}{m_a} \Gamma_{2} \Dot{\Psi}_{2}^{(2)} = - \Bar{\lambda}_4 \, \frac{3 \, (\Gamma_{2})^3 \mathcal{P}^{-1}}{4! \, m_a^5} \left( 2 \, |\Psi_s|^2 \Ddot{\Psi}_s^* + {\Psi_s^*}^2 \Ddot{\Psi}_s \right) - \Bar{\lambda}_4 \, \frac{6 \, (\Gamma_{2})^3 \mathcal{P}^{-1}}{4! \, m_a^5} \left( \Psi_s \left( \Dot{\Psi}_s^* \right)^2 + 2 \, \Psi_s^* \left| \Dot{\Psi}_s \right|^2 \right) \nonumber \\
    &- i \, \Bar{\lambda}_4^2 \, \frac{3 \, (\Gamma_{2})^2 \mathcal{P}^{-1}}{(4!)^2 \, m_a^{7}} \bigg[ 2 \, \Psi_s \Dot{\Psi}_s (\Gamma_{4} + \Gamma_{-2}) \mathcal{P}^{-1} \left( {\Psi_s^*}^3 \right) + 3 \, \Psi_s^2 (\Gamma_{4} + \Gamma_{-2}) \mathcal{P}^{-1} \left( {\Psi_s^*}^2 \Dot{\Psi}_s^* \right) \nonumber \\
    &+ 6 \, \Psi_s \Dot{\Psi}_s^* \Gamma_{2} \mathcal{P}^{-1} \left( \Psi_s^* |\Psi_s|^2  \right) + 6 \, \Psi_s^* \Dot{\Psi}_s \Gamma_{2} \mathcal{P}^{-1} \left( \Psi_s^* |\Psi_s|^2  \right) + 12 \, |\Psi_s|^2 \Gamma_{2} \mathcal{P}^{-1} \left( \Dot{\Psi}_s^* |\Psi_s|^2  \right) \nonumber \\
    &+ 6 \, |\Psi_s|^2 \Gamma_{2} \mathcal{P}^{-1} \left( {\Psi_s^*}^2 \Dot{\Psi}_s  \right) + 6 \, \Psi_s^* \Dot{\Psi}_s^* \Gamma_{2} \mathcal{P}^{-1} \left( \Psi_s |\Psi_s|^2  \right) + 6 \, {\Psi_s^*}^2 \Gamma_{2} \mathcal{P}^{-1} \left( |\Psi_s|^2 \Dot{\Psi}_s \right) \nonumber \\ 
    &+ 3 \, {\Psi_s^*}^2 \Gamma_{2} \mathcal{P}^{-1} \left( \Psi_s^2 \Dot{\Psi}_s^*  \right) \bigg] \,.
\end{align}
As mentioned before, the only non-zero modes are $\nu =-2$, $2$ and $4$, when only the first two self-interaction terms are utilized. As an example of calculation, $\Psi_3^{(2)}$ equals
\begin{align}
    \Psi_{2}^{(3)} =& - \Bar{\lambda}_4 \, \frac{3 \, (\Gamma_{2})^3 \mathcal{P}^{-1}}{4! \, m_a^5} \left( 2 \, |\Psi_s|^2 \Ddot{\Psi}_s^* + {\Psi_s^*}^2 \Ddot{\Psi}_s \right) - \Bar{\lambda}_4 \, \frac{6 \, (\Gamma_{2})^3 \mathcal{P}^{-1}}{4! \, m_a^5} \left( \Psi_s \left( \Dot{\Psi}_s^* \right)^2 + 2 \, \Psi_s^* \left| \Dot{\Psi}_s \right|^2 \right) \nonumber \\
    &- i \, \Bar{\lambda}_4^2 \, \frac{3}{(4!)^2 \, m_a^{7}} \bigg[ (\Gamma_{2})^2 \mathcal{P}^{-1} \bigg( 2 \, \Psi_s \Dot{\Psi}_s (\Gamma_{4} + \Gamma_{-2}) \mathcal{P}^{-1} \left( {\Psi_s^*}^3 \right) \nonumber \\
    &+ 3 \, \Psi_s^2 (\Gamma_{4} + \Gamma_{-2}) \mathcal{P}^{-1} \left( {\Psi_s^*}^2 \Dot{\Psi}_s^* \right) + 6 \, \Psi_s \Dot{\Psi}_s^* \Gamma_{2} \mathcal{P}^{-1} \left( \Psi_s^* |\Psi_s|^2  \right) + 6 \, \Psi_s^* \Dot{\Psi}_s \Gamma_{2} \mathcal{P}^{-1} \left( \Psi_s^* |\Psi_s|^2  \right) \nonumber \\
    &+ 12 \, |\Psi_s|^2 \Gamma_{2} \mathcal{P}^{-1} \left( \Dot{\Psi}_s^* |\Psi_s|^2  \right) + 6 \, |\Psi_s|^2 \Gamma_{2} \mathcal{P}^{-1} \left( {\Psi_s^*}^2 \Dot{\Psi}_s  \right) + 6 \, \Psi_s^* \Dot{\Psi}_s^* \Gamma_{2} \mathcal{P}^{-1} \left( \Psi_s |\Psi_s|^2  \right)\nonumber \\
    &+ 6 \, {\Psi_s^*}^2 \Gamma_{2} \mathcal{P}^{-1} \left( |\Psi_s|^2 \Dot{\Psi}_s \right) + 3 \, {\Psi_s^*}^2 \Gamma_{2} \mathcal{P}^{-1} \left( \Psi_s^2 \Dot{\Psi}_s^*  \right) \bigg)\nonumber \\
    &+ 3 \, \Gamma_2 \mathcal{P}^{-1} \bigg( \Psi_s^2 ((\Gamma_{4})^2 - (\Gamma_{-2})^2) \mathcal{P}^{-1} \left( {\Psi_s^*}^2 \Dot{\Psi}_s^* \right) \nonumber \\ 
    &+ 2  \, |\Psi_s|^2 (\Gamma_{2})^2 \mathcal{P}^{-1} \left( 2 \, |\Psi_s|^2 \Dot{\Psi}_s^* + {\Psi_s^*}^2 \Dot{\Psi}_s \right) - {\Psi_s^*}^2 (\Gamma_{2})^2 \mathcal{P}^{-1} \left( 2 \, |\Psi_s|^2 \Dot{\Psi}_s + \Psi_s^2 \Dot{\Psi}_s^* \right) \bigg) \bigg] \nonumber \\
    &+ \Bar{\lambda}_4^3 \, \frac{3 \, \Gamma_2 \mathcal{P}^{-1}}{(4!)^3 \, m_a^9} \bigg[ 3 \, \Psi_s \Big( 3 \left( \Gamma_{2} \mathcal{P}^{-1} \left( \Psi_s^* |\Psi_s|^2 \right) \right)^2 \nonumber \\
    &+ 2 \, \Gamma_{2} \mathcal{P}^{-1} \left( \Psi_s |\Psi_s|^2 \right)  (\Gamma_{4} + \Gamma_{-2}) \mathcal{P}^{-1} \left( {\Psi_s^*}^3 \right) \Big) \nonumber \\
    &+ 2 \, \Psi_s^* \Big( 9 \left| \Gamma_{2} \mathcal{P}^{-1} \left( \Psi_s |\Psi_s|^2 \right)  \right|^2 + \left| (\Gamma_{4} + \Gamma_{-2}) \mathcal{P}^{-1} \left( \Psi_s^3 \right) \right|^2 \Big) \nonumber
\end{align}
\begin{align}
    &+ 3 \, \Psi_s^2 (\Gamma_{4} + \Gamma_{-2}) \mathcal{P}^{-1} \bigg( 2 \, |\Psi_s|^2 (\Gamma_{4} + \Gamma_{-2}) \mathcal{P}^{-1} \left( {\Psi_s^*}^3 \right) + 3 \, {\Psi_s^*}^2 \Gamma_{2} \mathcal{P}^{-1} \left( \Psi_s^* |\Psi_s|^2  \right) \bigg) \nonumber \\
    &+ 6  \, |\Psi_s|^2 \Gamma_{2} \mathcal{P}^{-1} \bigg( \Psi_s^2 (\Gamma_{4} + \Gamma_{-2}) \mathcal{P}^{-1} \left( {\Psi_s^*}^3 \right) + 6 \, |\Psi_s|^2 \Gamma_{2} \mathcal{P}^{-1} \left( \Psi_s^* |\Psi_s|^2  \right) \nonumber \\
    &+ 3 \, {\Psi_s^*}^2 \Gamma_{2} \mathcal{P}^{-1} \left( \Psi_s |\Psi_s|^2  \right) \bigg) + 3 \, {\Psi_s^*}^2 \Gamma_{2} \mathcal{P}^{-1} \bigg( {\Psi_s^*}^2 (\Gamma_{4} + \Gamma_{-2}) \mathcal{P}^{-1} \left( \Psi_s^3 \right)\nonumber \\
    &+ 6 \, |\Psi_s|^2 \Gamma_{2} \mathcal{P}^{-1} \left( \Psi_s |\Psi_s|^2  \right) + 3 \, \Psi_s^2 \Gamma_{2} \mathcal{P}^{-1} \left( \Psi_s^* |\Psi_s|^2  \right) \bigg) \bigg] \,.
\end{align}

Replacing all the required modes in Eq.~\eqref{G_exp_3rd_(3)_nu_0} results in
\begin{align}
    G_{0,(3)}^{(3)} =& - \Bar{\lambda}_4 \, \frac{9 \, \Gamma_0 \mathcal{P}^{-\frac{1}{2}}}{2 \, \alpha(3) \, m_a^8} \mathrm{G} {\scriptstyle (4, \textbf{\RNum{2}})}_{0,(3)}^{(3)} - \Bar{\lambda}_4 \, \frac{18 \, \Gamma_0 \mathcal{P}^{-\frac{1}{2}}}{2 \, \alpha(3) \, m_a^8} \mathrm{G} {\scriptstyle (4, \textbf{\RNum{1}}, \textbf{\RNum{1}})}_{0,(3)}^{(3)} \nonumber \\
    &+ i \, \Bar{\lambda}_4^2 \, \frac{6 \, \Gamma_0 \mathcal{P}^{-\frac{1}{2}}}{\alpha(4) \, m_a^{10}} \mathrm{G} {\scriptstyle (4,4, \textbf{\RNum{1}})}_{0,(3)}^{(3)} + \Bar{\lambda}_4^3 \, \frac{25 \, \Gamma_0 \mathcal{P}^{-\frac{1}{2}}}{2 \, \alpha(5) \, m_a^{12}} \mathrm{G} {\scriptstyle (4,4,4)}_{0,(3)}^{(3)} \label{G_exp_3rd_(3)_nu_0-complete} \,,
\end{align}
and substituting Eqs.~\eqref{G_exp_3rd_(1)_nu_0-complete}, \eqref{G_exp_3rd_(1)(2)_nu_0-complete}, and \eqref{G_exp_3rd_(3)_nu_0-complete} into~\eqref{G_exp_3rd} terminates in
\begin{align}
    G_0^{(3)} =& - \Bar{\lambda}_4 \, \frac{9 \, \Gamma_0 \mathcal{P}^{-\frac{1}{2}}}{2 \, \alpha(3) \, m_a^8} \left( \mathrm{G} {\scriptstyle (4, \textbf{\RNum{2}})}_{0,(3)}^{(3)} + 2 \, \mathrm{G} {\scriptstyle (4, \textbf{\RNum{1}}, \textbf{\RNum{1}})}_{0,(3)}^{(3)} \right)  \nonumber \\
    &+ i \, \Bar{\lambda}_4^2 \, \frac{6 \, \Gamma_0 \mathcal{P}^{-\frac{1}{2}}}{\alpha(4) \, m_a^{10}} \left( 2 \, \mathrm{G} {\scriptstyle (4,4, \textbf{\RNum{1}})}_{0,(1)(2)}^{(3)} + \mathrm{G} {\scriptstyle (4,4, \textbf{\RNum{1}})}_{0,(3)}^{(3)} \right) \nonumber \\ 
    &+ \Bar{\lambda}_4^3 \, \frac{25 \, \Gamma_0 \mathcal{P}^{-\frac{1}{2}}}{2 \, \alpha(5) \, m_a^{12}} \left( \mathrm{G} {\scriptstyle (4,4,4)}_{0,(1)}^{(3)} + 2 \, \mathrm{G} {\scriptstyle (4,4,4)}_{0,(1)(2)}^{(3)} + \mathrm{G} {\scriptstyle (4,4,4)}_{0,(3)}^{(3)} \right) \label{G_exp_3rd_nu_0-complete} \,.
\end{align}
The complete expressions for all the terms inside Eq.~\eqref{G_exp_3rd_nu_0-complete} are presented in Appendix~\ref{App-D}.

%%%%%%%%%%%%%%%%%%%%%%%%%%%%%%%%%%%%%%%%%%%%%%%%%%%%%%%%%%%%%%%%%%%%%%%%%%%%%%%%

\subsection{Expansion of the $\mathcal{P}$ operator}

Replacing Eqs.~\eqref{GFDR_exp_0th_nu_0}, \eqref{GFD_exp_1st_nu_0-complete}, \eqref{G_exp_2nd_nu_0-complete}, \eqref{F_exp_2nd_nu_0-complete}, and \eqref{G_exp_3rd_nu_0-complete} into Eq.~\eqref{eq_motion_psi_s} results in
\begin{align}
    i \Dot{\psi}_s  =& \, m_a (\mathcal{P}-1) \psi_s + \Bar{\lambda}_4 \, \frac{2 \, \mathcal{P}^{-\frac{1}{2}}}{\alpha(2) \, m_a^2} \left( \Psi_s |\Psi_s|^2 \right) + \Bar{\lambda}_6 \, \frac{3 \, \mathcal{P}^{-\frac{1}{2}}}{\alpha(3) \, m_a^3} \left( \Psi_s |\Psi_s|^4 \right) \nonumber \\
    &+ \Bar{\lambda}_4^2 \, \frac{3}{2 \, \alpha(3) \, m_a^5} \mathrm{G} {\scriptstyle (4)}_{0,(1)}^{(1)} + i \,  \Bar{\lambda}_4^2 \, \frac{9}{2 \, \alpha(3) \, m_a^6} \mathrm{G} {\scriptstyle (4, \textbf{\RNum{1}})}_{0,(2)}^{(2)} \nonumber \\
    &- \Bar{\lambda}_4^2 \, \frac{9}{2 \, \alpha(3) \, m_a^7} \left( \mathrm{G} {\scriptstyle (4, \textbf{\RNum{2}})}_{0,(3)}^{(3)} + 2 \, \mathrm{G} {\scriptstyle (4, \textbf{\RNum{1}}, \textbf{\RNum{1}})}_{0,(3)}^{(3)} \right) + \Bar{\lambda}_8 \, \frac{4 \, \mathcal{P}^{-\frac{1}{2}}}{\alpha(4) \, m_a^4} \left( \Psi_s |\Psi_s|^6 \right) \nonumber \\
    &+ \Bar{\lambda}_4 \Bar{\lambda}_6 \, \frac{2}{\alpha(4) \, m_a^6} \left( 3 \, \mathrm{G} {\scriptstyle (6)}_{0,(1)}^{(1)} + \mathrm{F} {\scriptstyle (4)}_{0,(1)}^{(1)} \right) + i \, \Bar{\lambda}_4 \Bar{\lambda}_6 \, \frac{6}{\alpha(4) \, m_a^7} \left( \mathrm{G} {\scriptstyle (6, \textbf{\RNum{1}})}_{0,(2)}^{(2)} + \mathrm{F} {\scriptstyle (4, \textbf{\RNum{1}})}_{0,(2)}^{(2)} \right) \nonumber \\
    &+ \Bar{\lambda}_4^3 \, \frac{2}{\alpha(4) \, m_a^{8}} \Big( \mathrm{G} {\scriptstyle (4,4)}_{0,(1)}^{(2)} + 3 \, \mathrm{G} {\scriptstyle (4,4)}_{0,(2)}^{(2)} \Big) + i \, \Bar{\lambda}_4^3 \, \frac{6}{\alpha(4) \, m_a^{9}} \left( 2 \, \mathrm{G} {\scriptstyle (4,4, \textbf{\RNum{1}})}_{0,(1)(2)}^{(3)} + \mathrm{G} {\scriptstyle (4,4, \textbf{\RNum{1}})}_{0,(3)}^{(3)} \right) \nonumber \\
    &+ \Bar{\lambda}_{10} \, \frac{5 \, \mathcal{P}^{-\frac{1}{2}}}{\alpha(5) \, m_a^5} \left( \Psi_s |\Psi_s|^8 \right) + \Bar{\lambda}_4 \Bar{\lambda}_8 \, \frac{5}{\alpha(5) \, m_a^7} \left( \mathrm{G} {\scriptstyle (8)}_{0,(1)}^{(1)} + \mathrm{D} {\scriptstyle (4)}_{0,(1)}^{(1)} \right) \nonumber
\end{align}
\begin{align}
    &+ \Bar{\lambda}_6^2 \, \frac{5}{2 \, \alpha(5) \, m_a^7} \mathrm{F} {\scriptstyle (6)}_{0,(1)}^{(1)} + i \, \Bar{\lambda}_6^2 \, \frac{25}{2 \, \alpha(5) \, m_a^{8}} \mathrm{F} {\scriptstyle (6, \textbf{\RNum{1}})}_{0,(2)}^{(2)} \nonumber \\
    &+ \Bar{\lambda}_4^2 \Bar{\lambda}_6 \, \frac{5}{2 \, \alpha(5) \, m_a^{9}} \Big( 2 \, \mathrm{G} {\scriptstyle (4,6)}_{0,(1)}^{(2)} + \mathrm{G} {\scriptstyle (4,6)}_{0,(2)}^{(2)} + 10 \, \mathrm{F} {\scriptstyle (4,4)}_{0,(1)}^{(2)} + 5 \, \mathrm{F} {\scriptstyle (4,4)}_{0,(2)}^{(2)} \Big) \nonumber \\
    &+ \Bar{\lambda}_4^4 \, \frac{25}{2 \, \alpha(5) \, m_a^{11}} \left( \mathrm{G} {\scriptstyle (4,4,4)}_{0,(1)}^{(3)} + 2 \, \mathrm{G} {\scriptstyle (4,4,4)}_{0,(1)(2)}^{(3)} + \mathrm{G} {\scriptstyle (4,4,4)}_{0,(3)}^{(3)} \right) \nonumber \\
    &+ \Bar{\lambda}_6 \Bar{\lambda}_8 \, \frac{3}{\alpha(6) \, m_a^8} \left( 5 \, \mathrm{F} {\scriptstyle (8)}_{0,(1)}^{(1)} + \mathrm{D} {\scriptstyle (6)}_{0,(1)}^{(1)} \right) \nonumber \\
    &+ \Bar{\lambda}_4 \Bar{\lambda}_6^2 \, \frac{3}{2 \, \alpha(6) \, m_a^{10}} \Big( 6 \, \mathrm{G} {\scriptstyle (6,6)}_{0,(1)}^{(2)} + 15 \, \mathrm{G} {\scriptstyle (6,6)}_{0,(2)}^{(2)} + 20 \, \mathrm{F} {\scriptstyle (4,6)}_{0,(1)}^{(2)} + 5 \, \mathrm{F} {\scriptstyle (4,6)}_{0,(2)}^{(2)} \Big) \nonumber \\
    &+ \Bar{\lambda}_8^2 \, \frac{7}{2 \, \alpha(7) \, m_a^9} \mathrm{D} {\scriptstyle (8)}_{0,(1)}^{(1)} \nonumber \\
    &+ \Bar{\lambda}_6^3 \, \frac{147}{4 \, \alpha(7) \, m_a^{11}} \Big( 2 \, \mathrm{F} {\scriptstyle (6,6)}_{0,(1)}^{(2)} + 5 \, \mathrm{F} {\scriptstyle (6,6)}_{0,(2)}^{(2)} \Big) \label{eq_motion_4th_no_exp_P} \,.
\end{align}

All instances of $\mathcal{P}$ in Eq.~\eqref{eq_motion_4th_no_exp_P} need to be expressed in terms of $\epsilon_x$. However, the order of this expansion is not clear since with each new iteration an additional self-interaction term was neglected. Therefore, according to $\Bar{\lambda}_6$ the iterative process ended at the $2nd$ iteration, but according to $\Bar{\lambda}_8$ the iterative process concluded at the $1st$ iteration. Consequently, each term inside Eq.~\eqref{eq_motion_4th_no_exp_P} is treated differently based on the constant that multiplies it. First, all terms proportional to $\Bar{\lambda}_{10}$ are expanded up to the $1st$ order of $\epsilon_x$, $\epsilon_t$ and $\Bar{\lambda}_{2n}$. Next, from the remaining terms, all the terms proportional to $\Bar{\lambda}_8$ are expanded up to the $2nd$ order of $\epsilon_x$, $\epsilon_t$ and $\Bar{\lambda}_{2n}$. Afterwards, all the unused expressions proportional to $\Bar{\lambda}_6$ are expanded up to the $3rd$ order of $\epsilon_x$, $\epsilon_t$ and $\Bar{\lambda}_{2n}$. Finally, all leftover terms are expanded up to the $4th$ order of $\epsilon_x$, $\epsilon_t$ and $\Bar{\lambda}_{2n}$. For example,
\begin{align}
    \Bar{\lambda}_8 \, \frac{4 \, \mathcal{P}^{-\frac{1}{2}}}{\alpha(4) \, m_a^4} \left( \Psi_s |\Psi_s|^6 \right) \simeq& \, \Bar{\lambda}_8 \, \frac{4}{\alpha(4) \, m_a^4} \left( \psi_s |\psi_s|^6 \right) + \Bar{\lambda}_8 \, \frac{1}{\alpha(4) \, m_a^6}\nonumber \\
    & \times\left( 3 \, \psi_s^2 |\psi_s|^4 \bigtriangledown^2 \psi_s^* + 4 \, |\psi_s|^6 \bigtriangledown^2 \psi_s + \bigtriangledown^2 \left( \psi_s |\psi_s|^6 \right) \right) \,.
\end{align}

Once $\mathcal{P}$ has been expanded, the equation of motion takes the following form:
\begin{align}
    0 =& \, i \Dot{\psi}_s + \frac{1}{2 \, m_a} \, \bigtriangledown^2 \psi_s + \frac{1}{8  \, m_a^3} \, \bigtriangledown^4 \psi_s + \frac{1}{16 \, m_a^5} \, \bigtriangledown^6 \psi_s + \frac{5}{128 \, m_a^7} \, \bigtriangledown^8 \psi_s \nonumber\\
    & - \Bar{\lambda}_4 \, \frac{2}{\alpha(2) \, m_a^2} \left( \psi_s |\psi_s|^2 \right) - \Bar{\lambda}_4 \, \frac{1}{2 \, \alpha(2) \, m_a^4} \left( \psi_s^2 \bigtriangledown^2 \psi_s^* + 2 \, |\psi_s|^2 \bigtriangledown^2 \psi_s + \bigtriangledown^2 \left( \psi_s |\psi_s|^2 \right) \right) \nonumber \\
    &- \Bar{\lambda}_4 \, \frac{1}{16 \, \alpha(2) \, m_a^6} \Big[ 2 \, \psi_s^* (\bigtriangledown^2 \psi_s)^2 + 4 \, \psi_s |\bigtriangledown^2 \psi_s|^2 + 5 \, \psi_s^2 \bigtriangledown^4 \psi_s^* + 10 \, |\psi_s|^2 \bigtriangledown^4 \psi_s \nonumber \\
    &+ 5 \, \bigtriangledown^4 \left( \psi_s |\psi_s|^2 \right) + 2 \, \bigtriangledown^2 \left( \psi_s^2 \bigtriangledown^2 \psi_s^* + 2 \, |\psi_s|^2 \bigtriangledown^2 \psi_s \right) \Big] \nonumber
\end{align}
\begin{align}
    &- \Bar{\lambda}_4 \, \frac{1}{64 \, \alpha(2) \, m_a^8} \Big[ 2 \, \bigtriangledown^2 \psi_s |\bigtriangledown^2 \psi_s|^2 + 10 \, \psi_s^* \bigtriangledown^2 \psi_s \bigtriangledown^4 \psi_s + 10 \, \psi_s \bigtriangledown^2 \psi_s^* \bigtriangledown^4 \psi_s \nonumber \\
    &+ 10 \, \psi_s \bigtriangledown^2 \psi_s \bigtriangledown^4 \psi_s^* + 15 \, \psi_s^2 \bigtriangledown^6 \psi_s^*  + 30 \, |\psi_s|^2 \bigtriangledown^6 \psi_s + 15 \, \bigtriangledown^6 \left( \psi_s |\psi_s|^2 \right) \nonumber \\
    &+ 5 \, \bigtriangledown^4 \left( \psi_s^2 \bigtriangledown^2 \psi_s^* + 2 \, |\psi_s|^2 \bigtriangledown^2 \psi_s \right) + \bigtriangledown^2 \left( 2 \, \psi_s^* (\bigtriangledown^2 \psi_s)^2 + 4 \, \psi_s |\bigtriangledown^2 \psi_s|^2    + 5 \, \psi_s^2 \bigtriangledown^4 \psi_s^* \right. \nonumber \\
    &\left.+ 10 \, |\psi_s|^2 \bigtriangledown^4 \psi_s \right) \Big] - \Bar{\lambda}_6 \, \frac{3}{\alpha(3) \, m_a^3} \left( \psi_s |\psi_s|^4 \right) \nonumber \\ 
    &- \Bar{\lambda}_6 \, \frac{3}{4 \, \alpha(3) \, m_a^5} \left( 2 \, \psi_s^2 |\psi_s|^2 \bigtriangledown^2 \psi_s^* + 3 \, |\psi_s|^4 \bigtriangledown^2 \psi_s + \bigtriangledown^2 \left( \psi_s |\psi_s|^4 \right) \right) \nonumber \\
    &- \Bar{\lambda}_6 \, \frac{3}{16 \, \alpha(3) \, m_a^7} \, \bigtriangledown^2 \left( 2 \, \psi_s^2 |\psi_s|^2 \bigtriangledown^2 \psi_s^* + 3 \, |\psi_s|^4 \bigtriangledown^2 \psi_s \right)  \nonumber \\
     &- \Bar{\lambda}_6 \, \frac{3}{32 \, \alpha(3) \, m_a^7} \left( 2 \, \psi_s^3 (\bigtriangledown^2 \psi_s^*)^2 + 10 \, \psi_s^2 |\psi_s|^2 \bigtriangledown^4 \psi_s^* \right. \nonumber \\
    &+ \left. 12 \, \psi_s |\psi_s|^2 |\bigtriangledown^2 \psi_s|^2 + 6 \, \psi_s^* |\psi_s|^2 (\bigtriangledown^2 \psi_s)^2 + 15 \, |\psi_s|^4 \bigtriangledown^4 \psi_s + 5 \, \bigtriangledown^4 \left( \psi_s |\psi_s|^4 \right) \right) \nonumber\\ & + \Bar{\lambda}_4^2 \, \frac{51}{8 \, \alpha(3) \, m_a^5} \left( \psi_s |\psi_s|^4 \right) \nonumber \\
    &+ \Bar{\lambda}_4^2 \, \frac{51}{32 \, \alpha(3) \, m_a^7} \bigtriangledown^2 \left( \psi_s |\psi_s|^4 \right) + \Bar{\lambda}_4^2 \, \frac{51}{32 \, \alpha(3) \, m_a^7} \left( 2 \, \psi_s^2 |\psi_s|^2 \bigtriangledown^2 \psi_s^* + 3 \, |\psi_s|^4 \bigtriangledown^2 \psi_s \right) \nonumber \\
    &+ \Bar{\lambda}_4^2 \, \frac{3}{64 \, \alpha(3) \, m_a^7} \Big[ 36 \, \psi_s^2 \bigtriangledown^2 \left( \psi_s^* |\psi_s|^2 \right)  + 72 \, |\psi_s|^2 \bigtriangledown^2 \left( \psi_s |\psi_s|^2 \right) + {\psi_s^*}^2 \bigtriangledown^2 \left( \psi_s^3 \right) \Big] \nonumber \\
    &+ \Bar{\lambda}_4^2 \, \frac{3}{512 \, \alpha(3) \, m_a^9} \Big[ 170 \, \bigtriangledown^4 \left( \psi_s |\psi_s|^4 \right) + \bigtriangledown^2 \bigg( 72 \, \psi_s^2 \bigtriangledown^2 \left( \psi_s^* |\psi_s|^2 \right)  + 144 \, |\psi_s|^2 \bigtriangledown^2 \left( \psi_s |\psi_s|^2 \right)\nonumber \\
    &+ 2 \, {\psi_s^*}^2 \bigtriangledown^2 \left( \psi_s^3 \right) \bigg) + 240 \, \psi_s^2 \bigtriangledown^4 \left( \psi_s^* |\psi_s|^2 \right)  + 480 \, |\psi_s|^2 \bigtriangledown^4 \left( \psi_s |\psi_s|^2 \right) - {\psi_s^*}^2 \bigtriangledown^4 \left( \psi_s^3 \right) \nonumber \\
    &+ 68 \, \psi_s^3 (\bigtriangledown^2 \psi_s^*)^2 + 340 \, \psi_s^2 |\psi_s|^2 \bigtriangledown^4 \psi_s^* + 408 \, \psi_s |\psi_s|^2 |\bigtriangledown^2 \psi_s|^2 + 204 \, \psi_s^* |\psi_s|^2 (\bigtriangledown^2 \psi_s)^2 \nonumber \\
    &+ 510 \, |\psi_s|^4 \bigtriangledown^4 \psi_s + \bigtriangledown^2 \bigg( 136 \, \psi_s^2 |\psi_s|^2 \bigtriangledown^2 \psi_s^* + 204 \, |\psi_s|^4 \bigtriangledown^2 \psi_s  \bigg) \nonumber \\    
    &+ 144 \, \psi_s \, \bigtriangledown^2 \psi_s \bigtriangledown^2 ( \psi_s^* |\psi_s|^2) + \psi_s^2 \bigtriangledown^2 \left( 72 \, {\psi_s^*}^2 \bigtriangledown^2 \psi_s + 144 \, |\psi_s|^2 \bigtriangledown^2 \psi_s^* \right) \nonumber \\
    &+ 144 \, \psi_s \bigtriangledown^2 \psi_s^* \bigtriangledown^2 \left( \psi_s |\psi_s|^2 \right) + 144 \, \psi_s^* \bigtriangledown^2 \psi_s \bigtriangledown^2 \left( \psi_s |\psi_s|^2 \right) + |\psi_s|^2 \bigtriangledown^2 \left( 144 \, \psi_s^2 \bigtriangledown^2 \psi_s^* \right.\nonumber \\
    &\left.+ 288 \, |\psi_s|^2 \bigtriangledown^2 \psi_s \right) + 4 \, \psi_s^* \bigtriangledown^2 \psi_s^* \bigtriangledown^2 \psi_s^3 + 6 \, {\psi_s^*}^2 \bigtriangledown^2 (\psi_s^2 \bigtriangledown^2 \psi_s) \Big] - \Bar{\lambda}_8 \, \frac{4}{\alpha(4) \, m_a^4} \left( \psi_s |\psi_s|^6 \right)\nonumber \\
    &- \Bar{\lambda}_8 \, \frac{1}{\alpha(4) \, m_a^6} \left( 3 \, \psi_s^2 |\psi_s|^4 \bigtriangledown^2 \psi_s^* + 4 \, |\psi_s|^6 \bigtriangledown^2 \psi_s + \bigtriangledown^2 \left( \psi_s |\psi_s|^6 \right) \right) + \Bar{\lambda}_4 \Bar{\lambda}_6 \, \frac{44}{\alpha(4) \, m_a^6} (\psi_s |\psi_s|^6) \nonumber \\
    &+ \Bar{\lambda}_4 \Bar{\lambda}_6 \, \frac{1}{8 \, \alpha(4) \, m_a^8} \, |\psi_s|^2 \Big( 1035 \, \psi_s^2 |\psi_s|^2 \bigtriangledown^2 \psi_s^* + 1292 \, |\psi_s|^4 \bigtriangledown^2 \psi_s + 1035 \, \psi_s^3 \left( \bigtriangledown \psi_s^* \right)^2 \nonumber \\
    &+ 4140 \, \psi_s |\psi_s|^2 \left| \bigtriangledown \psi_s \right|^2 + 1806 \, \psi_s^* |\psi_s|^2 \left( \bigtriangledown \psi_s \right)^2 \Big) - \Bar{\lambda}_4^3 \, \frac{49}{\alpha(4) \, m_a^{8}} \left( \psi_s |\psi_s|^6 \right) \nonumber \\
    &- \Bar{\lambda}_4^3 \, \frac{1}{16 \, \alpha(4) \, m_a^{10}} \, \bigg(  588 \, \psi_s^2 |\psi_s|^4 \bigtriangledown^2 \psi_s^* + 784 \, |\psi_s|^6 \bigtriangledown^2 \psi_s + 196 \, \bigtriangledown^2 \left( \psi_s |\psi_s|^6 \right) \nonumber
\end{align}
\begin{align}
    &+ 153 \, \psi_s^2 \, \bigtriangledown^2 \left( \psi_s^* |\psi_s|^4 \right) + 306 \, |\psi_s|^2 \, \bigtriangledown^2 \left( \psi_s |\psi_s|^4 \right) + 3 \, {\psi_s^*}^2 \bigtriangledown^2 \left( \psi_s^3 |\psi_s|^2 \right) + \psi_s^4 \bigtriangledown^2 \left( {\psi_s^*}^3 \right) \nonumber\\&+ 306 \, \psi_s^2 |\psi_s|^2 \bigtriangledown^2 \left( \psi_s^* |\psi_s|^2  \right) + 459 \, |\psi_s|^4 \bigtriangledown^2 \left( \psi_s |\psi_s|^2  \right) +4 \, {\psi_s^*}^2 |\psi_s|^2 \bigtriangledown^2 \left( \psi_s^3 \right)  \bigg) \nonumber\\
    &- \Bar{\lambda}_{10} \, \frac{5}{\alpha(5) \, m_a^5} \left( \psi_s |\psi_s|^8 \right) + \Bar{\lambda}_4 \Bar{\lambda}_8 \, \frac{225}{2 \, \alpha(5) \, m_a^7} (\psi_s |\psi_s|^8) + \Bar{\lambda}_6^2 \, \frac{655}{6 \, \alpha(5) \, m_a^7} (\psi_s |\psi_s|^8) \nonumber \\
    &+ \Bar{\lambda}_6^2 \, \frac{5}{288 \, \alpha(5) \, m_a^9} \, |\psi_s|^4 \Big( 25716 \, \psi_s^2 |\psi_s|^2 \bigtriangledown^2 \psi_s^* + 30145 \, |\psi_s|^4 \bigtriangledown^2 \psi_s + 38574 \, \psi_s^3 \left( \bigtriangledown \psi_s^* \right)^2\nonumber \\
    & + 128580 \, \psi_s |\psi_s|^2 \left| \bigtriangledown \psi_s \right|^2 + 56290 \, \psi_s^* |\psi_s|^2 \left( \bigtriangledown \psi_s \right)^2 \Big) - \Bar{\lambda}_4^2 \Bar{\lambda}_6 \, \frac{21575}{24 \,  \alpha(5) \, m_a^{9}} ( \psi_s |\psi_s|^8 ) \nonumber \\
    & + \Bar{\lambda}_4^4 \, \frac{43125}{64 \, \alpha(5) \, m_a^{11}} ( \psi_s |\psi_s|^8 ) + \mathrm{ETT}_{\mathrm{eq. m.}} + \mathcal{O}(5) \label{eq_motion_4th_exp_P} \,,
\end{align}
where,
\begin{align}
    \mathrm{ETT}_{\mathrm{eq. m.}} =& - i \,  \Bar{\lambda}_4^2 \, \frac{81}{32 \, \alpha(3) \, m_a^6} ( \Dot{\psi}_s |\psi_s|^4 ) - i \,  \Bar{\lambda}_4^2 \, \frac{9}{128 \, \alpha(3) \, m_a^8} \Big[ 18 \, \Dot{\psi}_s \left( \psi_s |\psi_s|^2 \bigtriangledown^2 \psi_s^*\right. \nonumber \\
    &\left.+ \psi_s^* |\psi_s|^2 \bigtriangledown^2 \psi_s \right) + 9 \, |\psi_s|^4 \bigtriangledown^2 \Dot{\psi}_s - 16 \, \psi_s^2 \bigtriangledown^2 \left( 2 \, |\psi_s|^2 \Dot{\psi}_s^* + {\psi_s^*}^2 \Dot{\psi}_s \right)\nonumber \\
    &+ 32 \, |\psi_s|^2 \bigtriangledown^2 \left( 2 \, |\psi_s|^2 \Dot{\psi}_s + \psi_s^2 \Dot{\psi}_s^* \right)+ 3 \, {\psi_s^*}^2 \bigtriangledown^2 \left(  \psi_s^2 \Dot{\psi}_s \right) + 9 \, \bigtriangledown^2 \left( |\psi_s|^4 \Dot{\psi}_s \right) \Big] \nonumber \\
    & - \Bar{\lambda}_4^2 \, \frac{9}{128 \, \alpha(3) \, m_a^7} \Big( 32 \, \psi_s^2 |\psi_s|^2 \Ddot{\psi}_s^* + 33 \, |\psi_s|^4 \Ddot{\psi}_s + 16 \,  \psi_s^3 ( \Dot{\psi}_s^* )^2  \nonumber \\
    &+ 96 \,  \psi_s |\psi_s|^2 | \Dot{\psi}_s |^2 + 18 \, \psi_s^* |\psi_s|^2 ( \Dot{\psi}_s )^2 \Big) - i \, \Bar{\lambda}_4 \Bar{\lambda}_6 \, \frac{15}{\alpha(4) \, m_a^{7}} ( \Dot{\psi}_s |\psi_s|^6 )  \nonumber \\
    &+ i \, \Bar{\lambda}_4^3 \, \frac{42}{\alpha(4) \, m_a^{9}} ( \Dot{\psi}_s |\psi_s|^6 )- i \, \Bar{\lambda}_6^2 \, \frac{3875}{144 \, \alpha(5) \, m_a^{8}} ( \Dot{\psi}_s |\psi_s|^8 ) \label{extra-temporal-terms_eq_motion} \,,
\end{align}
are terms that contain temporal derivatives of $\psi_s$ that completely arise from the iterative process. Specifically, terms proportional to $\epsilon_t \psi_s$ and $\epsilon_t^2 \psi_s$ appear from the $2nd$ and $3rd$ iteration, respectively. These extra temporal terms must be somehow treated in order to not introduce new degrees of freedom. How exactly this is achieved is shown in the next subsection. Meanwhile, these terms will be treated as if no problem arises from their existence. Finally,
\begin{align}
    \mathcal{O}(5) =& \, \Bar{\lambda}_6 \Bar{\lambda}_8 \, \frac{762}{\alpha(6) \, m_a^8} (\psi_s |\psi_s|^{10}) - \Bar{\lambda}_4 \Bar{\lambda}_6^2 \, \frac{58293}{8 \, \alpha(6) \, m_a^{10}} ( \psi_s |\psi_s|^{10} ) \nonumber \\
    &+ \Bar{\lambda}_8^2 \, \frac{27895}{16 \, \alpha(7) \, m_a^{9}} \left(\psi_s |\psi_s|^{12} \right) - \Bar{\lambda}_6^3 \, \frac{810509}{32 \, \alpha(7) \, m_a^{11}} ( \psi_s |\psi_s|^{12} ) \,,
\end{align}
are terms of $6th$ or higher order, which have contributions beyond the amount of iterations and $\Bar{\lambda}_{2n}$ considered in this work. As a result, these expressions will not be considered from now on.

By applying the Euler-Lagrange equation it is possible to find the following hermitian Lagrangian that describes Eq.~\eqref{eq_motion_4th_exp_P}:
\begin{align}
    \mathcal{L} =& \, \frac{i}{2} \left( \Dot{\psi}_s \psi_s^* - \psi_s \Dot{\psi}_s^* \right)  - \frac{1}{2 \, m_a} \bigtriangledown \psi_s \cdot \bigtriangledown \psi_s^* + \frac{1}{8 \, m_a^3} \bigtriangledown^2 \psi_s \bigtriangledown^2 \psi_s^* \nonumber \\
    & - \frac{1}{16 \, m_a^5} \bigtriangledown \left( \bigtriangledown^2 \psi_s \right) \cdot \bigtriangledown \left( \bigtriangledown^2 \psi_s^* \right)+ \frac{5}{128 \, m_a^7} \bigtriangledown^4 \psi_s \bigtriangledown^4 \psi_s^*- \Bar{\lambda}_4 \, \frac{1}{\alpha(2) \, m_a^2} \, |\psi_s|^4\nonumber \\
    & - \Bar{\lambda}_4 \, \frac{1}{2 \, \alpha(2) \, m_a^4} \, |\psi_s|^2 \left( \psi_s^* \bigtriangledown^2 \psi_s + \psi_s \bigtriangledown^2 \psi_s^* \right) \nonumber \\
    &- \Bar{\lambda}_4 \, \frac{5}{16 \,  \alpha(2) \, m_a^6} \, |\psi_s|^2 \left( \psi_s^* \bigtriangledown^4 \psi_s + \psi_s \bigtriangledown^4 \psi_s^* \right) - \Bar{\lambda}_4 \, \frac{1}{16 \, \alpha(2) \, m_a^6} \Big( \psi_s^2 \left( \bigtriangledown^2 \psi_s^* \right)^2 \nonumber\\
    &+ 4 \, |\psi_s|^2 |\bigtriangledown^2 \psi_s|^2+ {\psi_s^*}^2 \left( \bigtriangledown^2 \psi_s \right)^2 \Big) - \Bar{\lambda}_4 \, \frac{1}{64 \,  \alpha(2) \, m_a^8}\Big[ 15 \, |\psi_s|^2 \left( \psi_s^* \bigtriangledown^6 \psi_s + \right. \nonumber \\
    &\left.\psi_s \bigtriangledown^6 \psi_s^* \right) + 5 \, ( \psi_s^2 \bigtriangledown^2 \psi_s^* \bigtriangledown^4 \psi_s^* + {\psi_s^*}^2 \bigtriangledown^2 \psi_s \bigtriangledown^4 \psi_s ) + 10 \, |\psi_s|^2 ( \bigtriangledown^2 \psi_s \bigtriangledown^4 \psi_s^*\nonumber \\
    &+ {\psi_s^*}^2 \bigtriangledown^2 \psi_s \bigtriangledown^4 \psi_s ) + 10 \, |\psi_s|^2 ( \bigtriangledown^2 \psi_s \bigtriangledown^4 \psi_s^* + \bigtriangledown^2 \psi_s^* \bigtriangledown^4 \psi_s ) + 2 \, |\bigtriangledown^2 \psi_s|^2 (\psi_s^* \bigtriangledown^2 \psi_s   \nonumber \\
    &+ \psi_s \bigtriangledown^2 \psi_s^* )  \Big]- \Bar{\lambda}_6 \, \frac{1}{\alpha(3) \, m_a^5} \, |\psi_s|^6 - \Bar{\lambda}_6 \, \frac{3}{4 \, \alpha(3) \, m_a^5} \, |\psi_s|^4 \left( \psi_s^* \bigtriangledown^2 \psi_s + \psi_s \bigtriangledown^2 \psi_s^* \right) \nonumber \\
    &- \Bar{\lambda}_6 \, \frac{15}{32 \, \alpha(3) \, m_a^7} \, |\psi_s|^4 \left( \psi_s^* \bigtriangledown^4 \psi_s + \psi_s \bigtriangledown^4 \psi_s^* \right) - \Bar{\lambda}_6 \, \frac{3}{16 \, \alpha(3) \, m_a^7} \, |\psi_s|^2 \Big( \psi_s^2 \left( \bigtriangledown^2 \psi_s^* \right)^2 \nonumber \\ &+ 3 |\psi_s|^2 |\bigtriangledown^2 \psi_s|^2 + {\psi_s^*}^2 \left( \bigtriangledown^2 \psi_s \right)^2 \Big) + \Bar{\lambda}_4^2 \, \frac{17}{8 \, \alpha(3) \, m_a^5} \, |\psi_s|^6\nonumber \\
    & + \Bar{\lambda}_4^2 \, \frac{3}{64 \, \alpha(3) \, m_a^7} \, |\psi_s|^2 \Big[ \left(6 \, \mathrm{C}_{4,4}^{\mathbf{[1]}} - 387\right) \, |\psi_s|^2 |\bigtriangledown \psi_s|^2 + \left(2 \, \mathrm{C}_{4,4}^{\mathbf{[1]}} - 140 \right) \left( {\psi_s^*}^2 \left( \bigtriangledown \psi_s \right)^2 \right. \nonumber \\
    &\left.+ \psi_s^2 \left( \bigtriangledown \psi_s^* \right)^2  \right)+ \mathrm{C}_{4,4}^{\mathbf{[1]}} \left( \psi_s^* |\psi_s|^2 \bigtriangledown^2 \psi_s + \psi_s |\psi_s|^2 \bigtriangledown^2 \psi_s^* \right) \Big] \nonumber \\
    &+ \Bar{\lambda}_4^2 \, \frac{1}{2048 \, \alpha(3) \, m_a^9} \bigg[ \left( 6306 + 4 \, \mathrm{C}_{4,4}^{\mathbf{[2]}} - 2 \, \mathrm{C}_{4,4}^{\mathbf{[3]}} \right) \, |\psi_s|^4 (\psi_s^* \bigtriangledown^4 \psi_s + \psi_s \bigtriangledown^4 \psi_s^*) \nonumber \\
    & + 12 \, \mathrm{C}_{4,4}^{\mathbf{[3]}} \, |\psi_s|^4 \left| \bigtriangledown^2 \psi_s \right|^2+ \left( 5436 + \mathrm{C}_{4,4}^{\mathbf{[5]}} + \mathrm{C}_{4,4}^{\mathbf{[6]}} - 4 \, \mathrm{C}_{4,4}^{\mathbf{[7]}} + 12 \, \mathrm{C}_{4,4}^{\mathbf{[8]}} \right) \, |\psi_s|^2 \left( \psi_s^2 (\bigtriangledown^2 \psi_s^*)^2 \right. \nonumber \\
    &\left.+ {\psi_s^*}^2 (\bigtriangledown^2 \psi_s)^2 \right) + 12 \, \mathrm{C}_{4,4}^{\mathbf{[4]}} \, |\psi_s|^4 \bigtriangledown^2 \left( |\bigtriangledown \psi_s|^2 \right)+ \left( 3936 + 8 \, \mathrm{C}_{4,4}^{\mathbf{[2]}} - 4 \, \mathrm{C}_{4,4}^{\mathbf{[3]}} + \mathrm{C}_{4,4}^{\mathbf{[5]}}\right. \nonumber \\
    &\left.+ \mathrm{C}_{4,4}^{\mathbf{[6]}} - 4 \, \mathrm{C}_{4,4}^{\mathbf{[7]}} + 12 \, \mathrm{C}_{4,4}^{\mathbf{[8]}} \right) \, |\psi_s|^2 ( \psi_s^2 \bigtriangledown \psi_s^* \cdot \bigtriangledown(\bigtriangledown^2 \psi_s^*)+ {\psi_s^*}^2 \bigtriangledown \psi_s \cdot \bigtriangledown(\bigtriangledown^2 \psi_s) ) \nonumber \\
    & + \left( -13392 + 12 \, \mathrm{C}_{4,4}^{\mathbf{[3]}} + 3 \, \mathrm{C}_{4,4}^{\mathbf{[6]}} \right) \, |\psi_s|^2 \left( \psi_s \bigtriangledown^2 \psi_s (\bigtriangledown \psi_s^*)^2 + \psi_s^* \bigtriangledown^2 \psi_s^* (\bigtriangledown \psi_s)^2 \right) \nonumber \\
    &+ \left( 2892 + \mathrm{C}_{4,4}^{\mathbf{[5]}} + \mathrm{C}_{4,4}^{\mathbf{[6]}} - 4 \, \mathrm{C}_{4,4}^{\mathbf{[7]}} + 12 \, \mathrm{C}_{4,4}^{\mathbf{[9]}} \right) \, |\psi_s|^2 \left( \psi_s^2 \bigtriangledown^2 \left( (\bigtriangledown \psi_s^*)^2 \right) \right.\nonumber \\
    &\left.+ {\psi_s^*}^2 \bigtriangledown^2 \left( (\bigtriangledown \psi_s)^2 \right) \right) + \left( -5976 + 12 \, \mathrm{C}_{4,4}^{\mathbf{[3]}} + 6 \, \mathrm{C}_{4,4}^{\mathbf{[5]}} + 3 \, \mathrm{C}_{4,4}^{\mathbf{[6]}} - 12 \, \mathrm{C}_{4,4}^{\mathbf{[7]}} \right.\nonumber \\
    &\left.+ 36 \, \mathrm{C}_{4,4}^{\mathbf{[8]}} \right) \, |\psi_s|^2 |\bigtriangledown \psi_s|^2 \left( \psi_s \bigtriangledown^2 \psi_s^* + \psi_s^* \bigtriangledown^2 \psi_s \right)+ \left( 5832 + 3 \, \mathrm{C}_{4,4}^{\mathbf{[5]}} + 6 \, \mathrm{C}_{4,4}^{\mathbf{[6]}} \right.\nonumber \\
    &\left.- 12 \, \mathrm{C}_{4,4}^{\mathbf{[7]}} + 36 \, \mathrm{C}_{4,4}^{\mathbf{[9]}} \right) \, |\psi_s|^2 \left( \psi_s \bigtriangledown \psi_s \cdot \bigtriangledown \left( (\bigtriangledown \psi_s^*)^2 \right) + \psi_s^* \bigtriangledown \psi_s^* \cdot \bigtriangledown \left( (\bigtriangledown \psi_s)^2 \right) \right)\nonumber \\
    &+ \left( -11520 + 24 \, \mathrm{C}_{4,4}^{\mathbf{[4]}} + 3 \, \mathrm{C}_{4,4}^{\mathbf{[5]}} \right) \, |\psi_s|^2 \left( \psi_s \bigtriangledown \psi_s^* \cdot \bigtriangledown \left( |\bigtriangledown \psi_s|^2 \right) \right. \nonumber
\end{align}
\begin{align}
    &\left.+ \psi_s^* \bigtriangledown \psi_s \cdot \bigtriangledown \left( |\bigtriangledown \psi_s|^2 \right) \right)+ 12 \, \mathrm{C}_{4,4}^{\mathbf{[2]}} |\psi_s|^4 \left( \bigtriangledown \psi_s \cdot \bigtriangledown ( \bigtriangledown^2 \psi_s^* ) + \bigtriangledown \psi_s^* \cdot \bigtriangledown ( \bigtriangledown^2 \psi_s ) \right) \nonumber \\
     &+ 12 \, \mathrm{C}_{4,4}^{\mathbf{[5]}} \, |\psi_s|^2 |\bigtriangledown \psi_s|^4 + 12 \, \mathrm{C}_{4,4}^{\mathbf{[6]}} \, |\psi_s|^2 (\bigtriangledown \psi_s)^2 (\bigtriangledown \psi_s^*)^2 \nonumber \\
    &+ 12 \, \mathrm{C}_{4,4}^{\mathbf{[7]}} \, |\bigtriangledown \psi_s|^2 \left( \psi_s^2  (\bigtriangledown \psi_s^*)^2  + {\psi_s^*}^2  (\bigtriangledown \psi_s)^2 \right) + 12 \, \mathrm{C}_{4,4}^{\mathbf{[9]}} \left( \psi_s^3 \bigtriangledown \psi_s^* \cdot \bigtriangledown \left( (\bigtriangledown \psi_s^*)^2 \right)\right.\nonumber\\
    &\left.+ {\psi_s^*}^3 \bigtriangledown \psi_s \cdot \bigtriangledown \left( (\bigtriangledown \psi_s)^2 \right) \right) + 12 \, \mathrm{C}_{4,4}^{\mathbf{[8]}} \left( \psi_s^3 \bigtriangledown^2 \psi_s^* (\bigtriangledown \psi_s^*)^2 + {\psi_s^*}^3 \bigtriangledown^2 \psi_s (\bigtriangledown \psi_s)^2 \right) \bigg]  \nonumber \\
    &- \Bar{\lambda}_8 \, \frac{1}{\alpha(4) \, m_a^4} \, |\psi_s|^8 + \Bar{\lambda}_4 \Bar{\lambda}_6 \, \frac{11}{\alpha(4) \, m_a^6} \, |\psi_s|^8- \Bar{\lambda}_8 \, \frac{1}{\alpha(4) \, m_a^6} \, |\psi_s|^6 \left( \psi_s^* \bigtriangledown^2 \psi_s + \psi_s \bigtriangledown^2 \psi_s^* \right)  \nonumber \\
    &+ \Bar{\lambda}_4 \Bar{\lambda}_6 \, \frac{1}{16 \, \alpha(4) \, m_a^8} \, |\psi_s|^4 \Big[ \left(8 \, \mathrm{C}_{4,6} - 2584\right) \, |\psi_s|^2 |\bigtriangledown \psi_s|^2\nonumber \\
    &+ \mathrm{C}_{4,6} \, |\psi_s|^2 (\psi_s^* \bigtriangledown^2 \psi_s + \psi_s \bigtriangledown^2 \psi_s^*) + \left(3 \, \mathrm{C}_{4,6} - 1035\right) \left( {\psi_s^*}^2 \left( \bigtriangledown \psi_s \right)^2 + \psi_s^2 \left( \bigtriangledown \psi_s^* \right)^2 \right) \Big]  \nonumber \\
    &- \Bar{\lambda}_4^3 \, \frac{49}{4 \, \alpha(4) \, m_a^{8}} \, |\psi_s|^8  - \Bar{\lambda}_4^3 \, \frac{1}{8 \, \alpha(4) \, m_a^{10}} |\psi_s|^4 \Big[ \left(8 \, \mathrm{C}_{4,4,4} - 2020 \right) |\psi_s|^2 |\bigtriangledown \psi_s|^2 \nonumber \\
    &+ \mathrm{C}_{4,4,4} |\psi_s|^2 (\psi_s^* \bigtriangledown^2 \psi_s + \psi_s \bigtriangledown^2 \psi_s^*) \nonumber \\
    &+ \left(3 \, \mathrm{C}_{4,4,4} - 831 \right) \left( {\psi_s^*}^2 \left( \bigtriangledown \psi_s \right)^2 + \psi_s^2 \left( \bigtriangledown \psi_s^* \right)^2 \right) \Big] - \Bar{\lambda}_{10} \, \frac{1}{\alpha(5) \, m_a^5} \, |\psi_s|^{10} + \Bar{\lambda}_4 \Bar{\lambda}_8 \, \frac{45}{2 \, \alpha(5) \, m_a^7} \, |\psi_s|^{10} \nonumber \\
    &+ \Bar{\lambda}_6^2 \, \frac{131}{6 \, \alpha(5) \, m_a^7} \, |\psi_s|^{10} + \Bar{\lambda}_6^2 \, \frac{5}{288 \, \alpha(5) \, m_a^9} \, |\psi_s|^6 \Big[ (4 \, \mathrm{C}_{6,6} - 12858) ( {\psi_s^*}^2 \left( \bigtriangledown \psi_s \right)^2 + \psi_s^2 \left( \bigtriangledown \psi_s^* \right)^2 ) \nonumber \\
    &+ \mathrm{C}_{6,6} \, |\psi_s|^2 (\psi_s^* \bigtriangledown^2 \psi_s + \psi_s \bigtriangledown^2 \psi_s^*) + (10 \, \mathrm{C}_{6,6} - 30145) \, |\psi_s|^2 |\bigtriangledown \psi_s|^2  \Big]  \nonumber \\
    &- \Bar{\lambda}_4^2 \Bar{\lambda}_6 \, \frac{4315}{24 \, \alpha(5) \, m_a^9} \, |\psi_s|^{10}+ \Bar{\lambda}_4^4 \, \frac{8625}{64 \, \alpha(5) \, m_a^{11}} \, |\psi_s|^{10} + \mathrm{ETT}_{\mathrm{\mathcal{L}}} \label{lagrangian_before_ETT} \,,
\end{align}
where,
\begin{align}
    \mathrm{ETT}_{\mathrm{\mathcal{L}}} =& - i \, \Bar{\lambda}_4^2 \, \frac{27}{64 \, \alpha(3) \, m_a^{6}} |\psi_s|^4 \left( \Dot{\psi}_s \psi_s^* - \psi_s \Dot{\psi}_s^* \right)\nonumber \\
    &- i \, \Bar{\lambda}_4^2 \, \frac{9}{256 \, \alpha(3) \, m_a^{8}} \bigg[ 2 \, \mathrm{C}_{4,4,\textbf{\RNum{1}}}^{\mathbf{[1]}} \, |\psi_s|^2 |\bigtriangledown \psi_s|^2 (\Dot{\psi}_s \psi_s^* - \psi_s \Dot{\psi}_s^*) \nonumber \\
    &+ \left( -32 + \mathrm{C}_{4,4,\textbf{\RNum{1}}}^{\mathbf{[2]}} - 2 \, \mathrm{C}_{4,4,\textbf{\RNum{1}}}^{\mathbf{[3]}} \right ) \, |\psi_s|^4 (\psi_s^* \bigtriangledown^2 \Dot{\psi}_s - \psi_s \bigtriangledown^2 \Dot{\psi}_s^*) \nonumber \\
    &+ 2 \, \mathrm{C}_{4,4,\textbf{\RNum{1}}}^{\mathbf{[2]}} \, |\psi_s|^2 ({\psi_s^*}^2 \Dot{\psi}_s \bigtriangledown^2 \psi_s - \psi_s^2 \Dot{\psi}_s^* \bigtriangledown^2 \psi_s^*) \nonumber \\
    &+ \left( - 128 + \mathrm{C}_{4,4,\textbf{\RNum{1}}}^{\mathbf{[1]}} - 6 \, \mathrm{C}_{4,4,\textbf{\RNum{1}}}^{\mathbf{[3]}} \right ) \, |\psi_s|^4 (\bigtriangledown \psi_s^* \cdot \bigtriangledown \Dot{\psi}_s - \bigtriangledown \psi_s \cdot \bigtriangledown \Dot{\psi}_s^*) \nonumber \\
    &+ 2 \, \mathrm{C}_{4,4,\textbf{\RNum{1}}}^{\mathbf{[3]}} \left( {\psi_s^*}^3 \Dot{\psi}_s (\bigtriangledown \psi_s)^2 - \psi_s^3 \Dot{\psi}_s^* (\bigtriangledown \psi_s^*)^2 \right) \nonumber \\
    &+ \left( 37 + \mathrm{C}_{4,4,\textbf{\RNum{1}}}^{\mathbf{[1]}} - 3 \, \mathrm{C}_{4,4,\textbf{\RNum{1}}}^{\mathbf{[2]}} \right ) \, |\psi_s|^4 (\Dot{\psi}_s \bigtriangledown^2 \psi_s^* - \Dot{\psi}_s^* \bigtriangledown^2 \psi_s)\nonumber \\
    & + \left( -90 + 2 \, \mathrm{C}_{4,4,\textbf{\RNum{1}}}^{\mathbf{[1]}} - 6 \, \mathrm{C}_{4,4,\textbf{\RNum{1}}}^{\mathbf{[3]}} \right ) \, |\psi_s|^2 (\psi_s \Dot{\psi}_s (\bigtriangledown \psi_s^*)^2 \nonumber \\
    &- \psi_s^* \Dot{\psi}_s^* (\bigtriangledown \psi_s)^2) \bigg] - \Bar{\lambda}_4^2 \, \frac{9}{128 \, \alpha(3) \, m_a^7} \Big[ \mathrm{C}_{4,4,\textbf{\RNum{2}}} \, |\psi_s|^4 (\Ddot{\psi}_s \psi_s^* + \psi_s \Ddot{\psi}_s^*) + \left( 6 \, \mathrm{C}_{4,4,\textbf{\RNum{2}}} - 33 \right) |\psi_s|^4 |\Dot{\psi}_s|^2 \nonumber
\end{align}
\begin{align}
    &+ \left( 2 \, \mathrm{C}_{4,4,\textbf{\RNum{2}}} - 16 \right) |\psi_s|^2 \left((\Dot{\psi}_s)^2 {\psi_s^*}^2 + \psi_s^2 (\Dot{\psi}_s^*)^2 \right) \Big] - i \, \Bar{\lambda}_4 \Bar{\lambda}_6 \, \frac{15}{8 \, \alpha(4) \, m_a^{7}} |\psi_s|^6 \left( \Dot{\psi}_s \psi_s^* - \psi_s \Dot{\psi}_s^* \right) \nonumber \\
    &+ i \, \Bar{\lambda}_4^3 \, \frac{21}{4 \, \alpha(4) \, m_a^{9}} |\psi_s|^6 \left( \Dot{\psi}_s \psi_s^* - \psi_s \Dot{\psi}_s^* \right) - i \, \Bar{\lambda}_6^2 \, \frac{775}{288 \, \alpha(5) \, m_a^{8}} |\psi_s|^8 \left( \Dot{\psi}_s \psi_s^* - \psi_s \Dot{\psi}_s^* \right) \label{extra-temporal-terms_lagrangian} \,,
\end{align}
is the part of the lagrangian corresponding to Eq.~\eqref{extra-temporal-terms_eq_motion}. Additionally, $\mathrm{C}_{4,4,\textbf{\RNum{1}}}^{\mathbf{[i]}}$, $\mathrm{C}_{4,4,6}$, $\mathrm{C}_{4,4,4}$, $\mathrm{C}_{6,6}$, $\mathrm{C}_{4,4,\textbf{\RNum{1}}}^{\mathbf{[j]}}$, and $\mathrm{C}_{4,4,\textbf{\RNum{2}}}$, where $i=1,2,..,9$ and $j=1,2,3$, are all constants.

%%%%%%%%%%%%%%%%%%%%%%%%%%%%%%%%%%%%%%%%%%%%%%%%%%%%%%%%%%%%%%%%%%%%%%%%%%%%%%%%

\subsection{Treatment of the extra temporal terms}

In order to write the additional temporal terms as expressions depending only on $\psi_s$, $\epsilon_x$, and $\Bar{\lambda}_{2n}$, the equation of motion from a previous iteration is replaced in Eq.~\eqref{extra-temporal-terms_lagrangian}. However, this must be done carefully in order to not introduce incomplete terms. In this work the following expressions are needed
\begin{align}
    i \Dot{\psi}_s =& - \frac{1}{2 \, m_a} \, \bigtriangledown^2 \psi_s + \Bar{\lambda}_4 \, \frac{2}{\alpha(2) \, m_a^2} \left( \psi_s |\psi_s|^2 \right) \label{eq_motion-0th-ett} \,, \\
    i \Dot{\psi}_s =& - \frac{1}{2 \, m_a} \, \bigtriangledown^2 \psi_s -  \frac{1}{8  \, m_a^3} \, \bigtriangledown^4 \psi_s + \Bar{\lambda}_4 \, \frac{2}{\alpha(2) \, m_a^2} \left( \psi_s |\psi_s|^2 \right) \nonumber \\ 
    &+ \Bar{\lambda}_4 \, \frac{1}{2 \, \alpha(2) \, m_a^4} \left( \psi_s^2 \bigtriangledown^2 \psi_s^* + 2 \, |\psi_s|^2 \bigtriangledown^2 \psi_s \right) + \Bar{\lambda}_4 \, \frac{1}{2 \, \alpha(2) \, m_a^4} \left( \bigtriangledown^2 \left( \psi_s |\psi_s|^2 \right) \right) \nonumber \\ 
    &+ \Bar{\lambda}_6 \, \frac{3}{\alpha(3) \, m_a^3} \left( \psi_s |\psi_s|^4 \right) - \Bar{\lambda}_4^2 \, \frac{51}{8 \, \alpha(3) \, m_a^5} \left( \psi_s |\psi_s|^4 \right) \label{eq_motion-1st-ett} \,.
\end{align}
where Eqs.~\eqref{eq_motion-0th-ett} and~\eqref{eq_motion-1st-ett} are the equations of motion when the iterative process, as shown through this section, ended at the $0th$ and $1st$ iteration, respectively. Additionally, it is necessary to compute $\Ddot{\psi}_s$. This is done by deriving with respect time Eq.~\eqref{eq_motion-0th-ett} and then implementing Eq.~\eqref{eq_motion-0th-ett} itself to replace the resulting $\Dot{\psi}_s$. Doing as such results in
\begin{align}
    \Ddot{\psi}_s =& - \frac{1}{4 \, m_a^2} \bigtriangledown^4 \psi_s - \Bar{\lambda}_4^2 \, \frac{9}{2 \, \alpha(3) \, m_a^4} \left( \psi_s |\psi_s|^4 \right) \nonumber \\
    &+ \Bar{\lambda}_4 \, \frac{4}{\alpha(2) \, m_a^3} \left( |\psi_s|^2 \bigtriangledown^2 \psi_s+ \psi_s |\bigtriangledown \psi_s|^2  \right) \nonumber \\
    &+ \Bar{\lambda}_4 \, \frac{2}{\alpha(2) \, m_a^3} \left(\psi_s^* (\bigtriangledown^2 \psi_s)^2 \right) \label{ddot_psi-0th-ett}
\end{align}
If Eq.~\eqref{eq_motion-1st-ett} were replaced in the last term of Eq.~\eqref{extra-temporal-terms_lagrangian}, then a term proportional to $\Bar{\lambda}_4^2 \Bar{\lambda}_6^2$ would arise. However, this new term is beyond $5th$ order. Consequently, even though this contribution would not be wrong, it would be fruitless to added it to the final lagrangian as it is not complete. Taking this into account, by replacing Eqs.~\eqref{eq_motion-0th-ett}, \eqref{eq_motion-1st-ett}, and \eqref{ddot_psi-0th-ett} in Eq.~\eqref{extra-temporal-terms_lagrangian}, one arrives at
\begin{align}
    \mathrm{ETT}_{\mathrm{\mathcal{L}}} =& \, \Bar{\lambda}_4^2 \, \frac{27}{128 \, \alpha(3) \, m_a^{7}} |\psi_s|^2 \left( \psi_s^* |\psi_s|^2 \bigtriangledown^2 \psi_s + \psi_s |\psi_s|^2 \bigtriangledown^2 \psi_s^* \right)  \nonumber \\
    &+ \Bar{\lambda}_4^2 \, \frac{9}{512 \, \alpha(3) \, m_a^9} \bigg[ \left( -29 + \mathrm{C}_{4,4,\textbf{\RNum{2}}} + \mathrm{C}_{4,4,\textbf{\RNum{1}}}^{\mathbf{[2]}} - 2 \, \mathrm{C}_{4,4,\textbf{\RNum{1}}}^{\mathbf{[3]}} \right) \, |\psi_s|^4 (\psi_s^* \bigtriangledown^4 \psi_s + \psi_s \bigtriangledown^4 \psi_s^*) \nonumber \\
    &+ \left( 107 - 6 \, \mathrm{C}_{4,4,\textbf{\RNum{2}}} + 2 \, \mathrm{C}_{4,4,\textbf{\RNum{1}}}^{\mathbf{[1]}} - 6 \, \mathrm{C}_{4,4,\textbf{\RNum{1}}}^{\mathbf{[2]}}  \right) \, |\psi_s|^4 \left| \bigtriangledown^2 \psi_s \right|^2 \nonumber \\
    &+ \left( - 16 + 2 \, \mathrm{C}_{4,4,\textbf{\RNum{2}}} + 2 \, \mathrm{C}_{4,4,\textbf{\RNum{1}}}^{\mathbf{[2]}} \right) \, |\psi_s|^2 \left( \psi_s^2 (\bigtriangledown^2 \psi_s^*)^2 + {\psi_s^*}^2 (\bigtriangledown^2 \psi_s)^2 \right) \nonumber \\
    &+ \left( -90 + 2 \, \mathrm{C}_{4,4,\textbf{\RNum{1}}}^{\mathbf{[1]}} - 6 \, \mathrm{C}_{4,4,\textbf{\RNum{1}}}^{\mathbf{[3]}} \right ) \, |\psi_s|^2 \left( \psi_s \bigtriangledown^2 \psi_s (\bigtriangledown \psi_s^*)^2 + \psi_s^* \bigtriangledown^2 \psi_s^* (\bigtriangledown \psi_s)^2 \right) \nonumber \\
    &+ 2 \, \mathrm{C}_{4,4,\textbf{\RNum{1}}}^{\mathbf{[1]}} \ |\psi_s|^2 |\bigtriangledown \psi_s|^2 \left( \psi_s \bigtriangledown^2 \psi_s^* + \psi_s^* \bigtriangledown^2 \psi_s \right) \nonumber \\
    &+ \left( - 128 + \mathrm{C}_{4,4,\textbf{\RNum{1}}}^{\mathbf{[1]}} - 6 \, \mathrm{C}_{4,4,\textbf{\RNum{1}}}^{\mathbf{[3]}} \right ) \, |\psi_s|^4 \left( \bigtriangledown \psi_s \cdot \bigtriangledown ( \bigtriangledown^2 \psi_s^* ) + \bigtriangledown \psi_s^* \cdot \bigtriangledown ( \bigtriangledown^2 \psi_s ) \right) \nonumber \\
    &+ 2 \, \mathrm{C}_{4,4,\textbf{\RNum{1}}}^{\mathbf{[3]}}\, \left( \psi_s^3 \bigtriangledown^2 \psi_s^* (\bigtriangledown \psi_s^*)^2 + {\psi_s^*}^3 \bigtriangledown^2 \psi_s (\bigtriangledown \psi_s)^2 \right) \bigg] \nonumber \\
    &+ \Bar{\lambda}_4 \Bar{\lambda}_6 \, \frac{15}{16 \, \alpha(4) \, m_a^{8}} \, |\psi_s|^4 \left( \psi_s^* |\psi_s|^2 \bigtriangledown^2 \psi_s + \psi_s |\psi_s|^2 \bigtriangledown^2 \psi_s^* \right) - \Bar{\lambda}_4^3 \, \frac{27}{8 \, \alpha(4) \, m_a^8} \, |\psi_s|^8 \nonumber \\
    & - \Bar{\lambda}_4^3 \, \frac{3}{64 \, \alpha(4) \, m_a^{10}} \Big[ \left( -2232 + 24 \, \mathrm{C}_{4,4,\textbf{\RNum{2}}} + 24 \, \mathrm{C}_{4,4,\textbf{\RNum{1}}}^{\mathbf{[1]}}+ 24 \, \mathrm{C}_{4,4,\textbf{\RNum{1}}}^{\mathbf{[2]}} - 120 \, \mathrm{C}_{4,4,\textbf{\RNum{1}}}^{\mathbf{[3]}} \right) \, |\psi_s|^2 |\bigtriangledown \psi_s|^2 \nonumber \\
    &+ \left( -64 + 6 \, \mathrm{C}_{4,4,\textbf{\RNum{2}}} + 3 \, \mathrm{C}_{4,4,\textbf{\RNum{1}}}^{\mathbf{[1]}}+ 6 \, \mathrm{C}_{4,4,\textbf{\RNum{1}}}^{\mathbf{[2]}} - 18 \, \mathrm{C}_{4,4,\textbf{\RNum{1}}}^{\mathbf{[3]}} \right) \, |\psi_s|^2 \left( \psi_s^* \bigtriangledown^2 \psi_s + \psi_s  \bigtriangledown^2 \psi_s^* \right) \nonumber \\
    &+ \left( -828 + 6 \, \mathrm{C}_{4,4,\textbf{\RNum{2}}} + 9 \, \mathrm{C}_{4,4,\textbf{\RNum{1}}}^{\mathbf{[1]}}+ 6 \, \mathrm{C}_{4,4,\textbf{\RNum{1}}}^{\mathbf{[2]}} - 24 \, \mathrm{C}_{4,4,\textbf{\RNum{1}}}^{\mathbf{[3]}} \right) \, \left( \psi_s^2 (\bigtriangledown \psi_s^*)^2 + {\psi_s^*}^2 (\bigtriangledown \psi_s)^2 \right) \Big] \nonumber \\
    &+ \Bar{\lambda}_6^2 \, \frac{775}{576 \, \alpha(5) \, m_a^{9}} |\psi_s|^6 \left( \psi_s^* |\psi_s|^2 \bigtriangledown^2 \psi_s + \psi_s |\psi_s|^2 \bigtriangledown^2 \psi_s^* \right)  - \Bar{\lambda}_4^2 \Bar{\lambda}_6 \, \frac{75}{2 \, \alpha(5) \, m_a^9} \, |\psi_s|^{10} \nonumber \\
    &+ \Bar{\lambda}_4^4 \, \frac{12225}{128 \, \alpha(5) \, m_a^{11}} \, |\psi_s|^{10}  \,. \label{extra-temporal-terms_lagrangian-finalform}
\end{align}
Replacing Eq.~\eqref{extra-temporal-terms_lagrangian-finalform} in Eq.~\eqref{lagrangian_before_ETT} results in
\begin{align}
\mathcal{L} =& \, \frac{i}{2} \left( \Dot{\psi}_s \psi_s^* - \psi_s \Dot{\psi}_s^* \right) - \mathcal{H}_{\text{eff}} \,,
\end{align}
where,
\begin{equation}
\mathcal{H}_{\text{eff}} = \mathcal{T}_\text{eff} + V_{\text{eff}} + W_\text{eff} \,,
\end{equation}
and
\begin{align}
    \mathcal{T}_\text{eff} =& \, \frac{1}{2 \, m_a} \bigtriangledown \psi_s \cdot \bigtriangledown \psi_s^* - \frac{1}{8 \, m_a^3} \bigtriangledown^2 \psi_s \bigtriangledown^2 \psi_s^* + \frac{1}{16 \, m_a^5} \bigtriangledown \left( \bigtriangledown^2 \psi_s \right) \cdot \bigtriangledown \left( \bigtriangledown^2 \psi_s^* \right)\nonumber \\
    &- \frac{5}{128 \, m_a^7} \bigtriangledown^4 \psi_s \bigtriangledown^4 \psi_s^* \,,
\end{align}
\begin{align}
    V_{\text{eff}} =& \, \Bar{\lambda}_4 \, \frac{1}{\alpha(2) \, m_a^2} \, |\psi_s|^4 + \Bar{\lambda}_6 \, \frac{1}{\alpha(3) \, m_a^5} \, |\psi_s|^6 - \Bar{\lambda}_4^2 \, \frac{17}{8 \, \alpha(3) \, m_a^5} \, |\psi_s|^6 + \Bar{\lambda}_8 \, \frac{1}{\alpha(4) \, m_a^4} \, |\psi_s|^8  \nonumber \\
    &- \Bar{\lambda}_4 \Bar{\lambda}_6 \, \frac{11}{\alpha(4) \, m_a^6} \, |\psi_s|^8+ \Bar{\lambda}_4^3 \, \frac{125}{8 \, \alpha(4) \, m_a^{8}} \, |\psi_s|^8  + \Bar{\lambda}_{10} \, \frac{1}{\alpha(5) \, m_a^5} \, |\psi_s|^{10}  \nonumber \\
    &- \Bar{\lambda}_4 \Bar{\lambda}_8 \, \frac{45}{2 \, \alpha(5) \, m_a^7} \, |\psi_s|^{10} - \Bar{\lambda}_6^2 \, \frac{131}{6 \, \alpha(5) \, m_a^7} \, |\psi_s|^{10}\nonumber \\
    &+ \Bar{\lambda}_4^2 \Bar{\lambda}_6 \, \frac{5215}{24 \, \alpha(5) \, m_a^9} \, |\psi_s|^{10} - \Bar{\lambda}_4^4 \, \frac{29475}{128 \, \alpha(5) \, m_a^{11}} \, |\psi_s|^{10} \,,
\end{align}
\begin{align}
    W_\text{eff} =& \, \Bar{\lambda}_4 \, \frac{1}{2 \, \alpha(2) \, m_a^4} \, |\psi_s|^2 \left( \psi_s^* \bigtriangledown^2 \psi_s + \psi_s \bigtriangledown^2 \psi_s^* \right) + \Bar{\lambda}_4 \, \frac{5}{16 \,  \alpha(2) \, m_a^6} \, |\psi_s|^2 \left( \psi_s^* \bigtriangledown^4 \psi_s \right.\nonumber \\
    &\left.+ \psi_s \bigtriangledown^4 \psi_s^* \right) + \Bar{\lambda}_4 \, \frac{1}{16 \, \alpha(2) \, m_a^6} \Big( \psi_s^2 \left( \bigtriangledown^2 \psi_s^* \right)^2 + 4 \, |\psi_s|^2 |\bigtriangledown^2 \psi_s|^2 + {\psi_s^*}^2 \left( \bigtriangledown^2 \psi_s \right)^2 \Big) \nonumber \\
    &+ \Bar{\lambda}_4 \, \frac{1}{64 \,  \alpha(2) \, m_a^8} \Big[ 15 \, |\psi_s|^2 \left( \psi_s^* \bigtriangledown^6 \psi_s + \psi_s \bigtriangledown^6 \psi_s^* \right) + 5 \, ( \psi_s^2 \bigtriangledown^2 \psi_s^* \bigtriangledown^4 \psi_s^*  \nonumber \\
    &+ {\psi_s^*}^2 \bigtriangledown^2 \psi_s \bigtriangledown^4 \psi_s )( \psi_s^2 \bigtriangledown^2 \psi_s^* \bigtriangledown^4 \psi_s^* + {\psi_s^*}^2 \bigtriangledown^2 \psi_s \bigtriangledown^4 \psi_s )+ 10 \, |\psi_s|^2 ( \bigtriangledown^2 \psi_s \bigtriangledown^4 \psi_s^* \nonumber \\ 
    & + \bigtriangledown^2 \psi_s^* \bigtriangledown^4 \psi_s ) + 2 \, |\bigtriangledown^2 \psi_s|^2 (\psi_s^* \bigtriangledown^2 \psi_s  + \psi_s \bigtriangledown^2 \psi_s^* )  \Big] \nonumber \\
    &+ \Bar{\lambda}_6 \, \frac{3}{4 \, \alpha(3) \, m_a^5} \, |\psi_s|^4 \left( \psi_s^* \bigtriangledown^2 \psi_s + \psi_s \bigtriangledown^2 \psi_s^* \right) \nonumber \\
    &+ \Bar{\lambda}_6 \, \frac{15}{32 \, \alpha(3) \, m_a^7} \, |\psi_s|^4 \left( \psi_s^* \bigtriangledown^4 \psi_s + \psi_s \bigtriangledown^4 \psi_s^* \right) \nonumber \\
    &+ \Bar{\lambda}_6 \, \frac{3}{16 \, \alpha(3) \, m_a^7} \, |\psi_s|^2 \Big( \psi_s^2 \left( \bigtriangledown^2 \psi_s^* \right)^2 + 3 |\psi_s|^2 |\bigtriangledown^2 \psi_s|^2 + {\psi_s^*}^2 \left( \bigtriangledown^2 \psi_s \right)^2 \Big) \nonumber \\
    &- \Bar{\lambda}_4^2 \, \frac{3}{128 \, \alpha(3) \, m_a^7} \, |\psi_s|^2 \Big[ \left(12 \, \mathrm{C}_{4,4}^{\mathbf{[1]}} - 774 \right) \, |\psi_s|^2 |\bigtriangledown \psi_s|^2 \nonumber \\
    &+ \left(4 \, \mathrm{C}_{4,4}^{\mathbf{[1]}} - 280 \right) \left( {\psi_s^*}^2 \left( \bigtriangledown \psi_s \right)^2 + \psi_s^2 \left( \bigtriangledown \psi_s^* \right)^2  \right) \nonumber \\
    &+ \left( 2 \, \mathrm{C}_{4,4}^{\mathbf{[1]}} + 9 \right) \left( \psi_s^* |\psi_s|^2 \bigtriangledown^2 \psi_s + \psi_s |\psi_s|^2 \bigtriangledown^2 \psi_s^* \right) \Big] \nonumber \\
    &- \Bar{\lambda}_4^2 \, \frac{1}{2048 \, \alpha(3) \, m_a^9} \bigg[ \left( 5262 + 36 \, \mathrm{C}_{4,4,\mathbf{\RNum{2}}} + 36 \, \mathrm{C}_{4,4,\mathbf{\RNum{1}}}^{\mathbf{[2]}} - 72 \, \mathrm{C}_{4,4,\mathbf{\RNum{1}}}^{\mathbf{[3]}} + 4 \, \mathrm{C}_{4,4}^{\mathbf{[2]}} - 2 \, \mathrm{C}_{4,4}^{\mathbf{[3]}} \right) \, |\psi_s|^4 \nonumber \\
    &\times(\psi_s^* \bigtriangledown^4 \psi_s + \psi_s \bigtriangledown^4 \psi_s^*) \nonumber \\
    &+ \left( 3852 - 216 \, \mathrm{C}_{4,4,\mathbf{\RNum{2}}} + 72 \, \mathrm{C}_{4,4,\mathbf{\RNum{1}}}^{\mathbf{[1]}} - 216 \, \mathrm{C}_{4,4,\mathbf{\RNum{1}}}^{\mathbf{[2]}} + 12 \, \mathrm{C}_{4,4}^{\mathbf{[3]}} \right)\, |\psi_s|^4 \left| \bigtriangledown^2 \psi_s \right|^2 \nonumber \\
    &+ \left( 4860 + 72 \, \, \mathrm{C}_{4,4,\mathbf{\RNum{2}}} + 72 \, \mathrm{C}_{4,4,\mathbf{\RNum{1}}}^{\mathbf{[2]}} + \mathrm{C}_{4,4}^{\mathbf{[5]}} + \mathrm{C}_{4,4}^{\mathbf{[6]}} - 4 \, \mathrm{C}_{4,4}^{\mathbf{[7]}} + 12 \, \mathrm{C}_{4,4}^{\mathbf{[8]}} \right) \nonumber \\
    &\times |\psi_s|^2 \left( \psi_s^2 (\bigtriangledown^2 \psi_s^*)^2 + {\psi_s^*}^2 (\bigtriangledown^2 \psi_s)^2 \right) \nonumber
\end{align}
\begin{align}
    &+ \left( 3936 + 8 \, \mathrm{C}_{4,4}^{\mathbf{[2]}} - 4 \, \mathrm{C}_{4,4}^{\mathbf{[3]}} + \mathrm{C}_{4,4}^{\mathbf{[5]}} + \mathrm{C}_{4,4}^{\mathbf{[6]}} - 4 \, \mathrm{C}_{4,4}^{\mathbf{[7]}} + 12 \, \mathrm{C}_{4,4}^{\mathbf{[8]}} \right) \, |\psi_s|^2 ( \psi_s^2 \bigtriangledown \psi_s^* \cdot \bigtriangledown(\bigtriangledown^2 \psi_s^*) \nonumber \\
    &+ {\psi_s^*}^2 \bigtriangledown \psi_s \cdot \bigtriangledown(\bigtriangledown^2 \psi_s) ) + 12 \, \mathrm{C}_{4,4}^{\mathbf{[4]}} \, |\psi_s|^4 \bigtriangledown^2 \left( |\bigtriangledown \psi_s|^2 \right) + 12 \, \mathrm{C}_{4,4}^{\mathbf{[5]}} \, |\psi_s|^2 |\bigtriangledown \psi_s|^4  \nonumber \\
    &+ 12 \, \mathrm{C}_{4,4}^{\mathbf{[6]}} \, |\psi_s|^2 (\bigtriangledown \psi_s)^2 (\bigtriangledown \psi_s^*)^2+ \left( -16632 + 72 \, \mathrm{C}_{4,4,\mathbf{\RNum{1}}}^{\mathbf{[1]}} - 216 \, \mathrm{C}_{4,4,\mathbf{\RNum{1}}}^{\mathbf{[3]}} + 12 \, \mathrm{C}_{4,4}^{\mathbf{[3]}} + 3 \, \mathrm{C}_{4,4}^{\mathbf{[6]}} \right) \nonumber \\
    &\times\, |\psi_s|^2 \left( \psi_s \bigtriangledown^2 \psi_s (\bigtriangledown \psi_s^*)^2 + \psi_s^* \bigtriangledown^2 \psi_s^* (\bigtriangledown \psi_s)^2 \right) \nonumber \\
    &+ \left( 2892 + \mathrm{C}_{4,4}^{\mathbf{[5]}} + \mathrm{C}_{4,4}^{\mathbf{[6]}} - 4 \, \mathrm{C}_{4,4}^{\mathbf{[7]}} + 12 \, \mathrm{C}_{4,4}^{\mathbf{[9]}} \right) \, |\psi_s|^2 \left( \psi_s^2 \bigtriangledown^2 \left( (\bigtriangledown \psi_s^*)^2 \right) + {\psi_s^*}^2 \bigtriangledown^2 \left( (\bigtriangledown \psi_s)^2 \right) \right) \nonumber \\
    &+ \left( -5976 + 72 \, \mathrm{C}_{4,4,\mathbf{\RNum{1}}}^{\mathbf{[1]}} + 12 \, \mathrm{C}_{4,4}^{\mathbf{[3]}} + 6 \, \mathrm{C}_{4,4}^{\mathbf{[5]}} + 3 \, \mathrm{C}_{4,4}^{\mathbf{[6]}} - 12 \, \mathrm{C}_{4,4}^{\mathbf{[7]}} + 36 \, \mathrm{C}_{4,4}^{\mathbf{[8]}} \right) \nonumber \\
    &\times\, |\psi_s|^2 |\bigtriangledown \psi_s|^2 \left( \psi_s \bigtriangledown^2 \psi_s^* + \psi_s^* \bigtriangledown^2 \psi_s \right) \nonumber \\
    &+ \left( 5832 + 3 \, \mathrm{C}_{4,4}^{\mathbf{[5]}} + 6 \, \mathrm{C}_{4,4}^{\mathbf{[6]}} - 12 \, \mathrm{C}_{4,4}^{\mathbf{[7]}} + 36 \, \mathrm{C}_{4,4}^{\mathbf{[9]}} \right) \, |\psi_s|^2 \left( \psi_s \bigtriangledown \psi_s \cdot \bigtriangledown \left( (\bigtriangledown \psi_s^*)^2 \right)\right. \nonumber \\
    &\left.+ \psi_s^* \bigtriangledown \psi_s^* \cdot \bigtriangledown \left( (\bigtriangledown \psi_s)^2 \right) \right)+ \left( -11520 + 24 \, \mathrm{C}_{4,4}^{\mathbf{[4]}} + 3 \, \mathrm{C}_{4,4}^{\mathbf{[5]}} \right) \, |\psi_s|^2 \left( \psi_s \bigtriangledown \psi_s^* \cdot \bigtriangledown \left( |\bigtriangledown \psi_s|^2 \right) \right. \nonumber \\
    &\left.+ \psi_s^* \bigtriangledown \psi_s \cdot \bigtriangledown \left( |\bigtriangledown \psi_s|^2 \right) \right)+ \left( -4608 + 36 \, \mathrm{C}_{4,4,\mathbf{\RNum{1}}}^{\mathbf{[1]}} - 216 \, \mathrm{C}_{4,4,\mathbf{\RNum{1}}}^{\mathbf{[3]}} + 12 \, \mathrm{C}_{4,4}^{\mathbf{[2]}} \right) |\psi_s|^4 \nonumber \\
    &\times\left( \bigtriangledown \psi_s \cdot \bigtriangledown ( \bigtriangledown^2 \psi_s^* ) + \bigtriangledown \psi_s^* \cdot \bigtriangledown ( \bigtriangledown^2 \psi_s ) \right) \nonumber \\
    &+ 12 \, \mathrm{C}_{4,4}^{\mathbf{[7]}} \, |\bigtriangledown \psi_s|^2 \left( \psi_s^2  (\bigtriangledown \psi_s^*)^2  + {\psi_s^*}^2  (\bigtriangledown \psi_s)^2 \right) + 12 \, \mathrm{C}_{4,4}^{\mathbf{[9]}} \left( \psi_s^3 \bigtriangledown \psi_s^* \cdot \bigtriangledown \left( (\bigtriangledown \psi_s^*)^2 \right) \right. \nonumber \\
    &\left.+ {\psi_s^*}^3 \bigtriangledown \psi_s \cdot \bigtriangledown \left( (\bigtriangledown \psi_s)^2 \right) \right)+ \left( 72 \, \mathrm{C}_{4,4,\mathbf{\RNum{1}}}^{\mathbf{[3]}} +12 \, \mathrm{C}_{4,4}^{\mathbf{[8]}} \right)\left( \psi_s^3 \bigtriangledown^2 \psi_s^* (\bigtriangledown \psi_s^*)^2 \right. \nonumber \\
    &\left.+ {\psi_s^*}^3 \bigtriangledown^2 \psi_s (\bigtriangledown \psi_s)^2 \right) \bigg]+ \Bar{\lambda}_8 \, \frac{1}{\alpha(4) \, m_a^6} \, |\psi_s|^6 \left( \psi_s^* \bigtriangledown^2 \psi_s + \psi_s \bigtriangledown^2 \psi_s^* \right)\nonumber \\
    &- \Bar{\lambda}_4 \Bar{\lambda}_6 \, \frac{1}{16 \, \alpha(4) \, m_a^8} \, |\psi_s|^4 \Big[ \left(8 \, \mathrm{C}_{4,6} - 2584\right) \, |\psi_s|^2 |\bigtriangledown \psi_s|^2 \nonumber \\
    &+ \mathrm{C}_{4,6} \, |\psi_s|^2 (\psi_s^* \bigtriangledown^2 \psi_s + \psi_s \bigtriangledown^2 \psi_s^*) + \left(3 \, \mathrm{C}_{4,6} - 1035\right) \left( {\psi_s^*}^2 \left( \bigtriangledown \psi_s \right)^2 + \psi_s^2 \left( \bigtriangledown \psi_s^* \right)^2 \right) \Big]  \nonumber \\
    &+ \Bar{\lambda}_4^3 \, \frac{1}{64 \, \alpha(4) \, m_a^{10}} |\psi_s|^4 \Big[ \left(-22856 + 64 \, \mathrm{C}_{4,4,4} + 72 \, \mathrm{C}_{4,4,\textbf{\RNum{2}}} + 72 \, \mathrm{C}_{4,4,\textbf{\RNum{1}}}^{\mathbf{[1]}}+ 72 \, \mathrm{C}_{4,4,\textbf{\RNum{1}}}^{\mathbf{[2]}} \right.\nonumber \\
    &\left.- 360 \, \mathrm{C}_{4,4,\textbf{\RNum{1}}}^{\mathbf{[3]}} \right) |\psi_s|^2 |\bigtriangledown \psi_s|^2 + \left( - 192 \, + 8 \, \mathrm{C}_{4,4,4} + 18 \, \mathrm{C}_{4,4,\textbf{\RNum{2}}} + 9 \, \mathrm{C}_{4,4,\textbf{\RNum{1}}}^{\mathbf{[1]}}+ 18 \, \mathrm{C}_{4,4,\textbf{\RNum{1}}}^{\mathbf{[2]}} \right.\nonumber \\
    &\left.- 54 \, \mathrm{C}_{4,4,\textbf{\RNum{1}}}^{\mathbf{[3]}} \right) |\psi_s|^2 (\psi_s^* \bigtriangledown^2 \psi_s + \psi_s \bigtriangledown^2 \psi_s^*) \nonumber \\
    &+ \left(- 9132 + 24 \, \mathrm{C}_{4,4,4} + 18 \, \mathrm{C}_{4,4,\textbf{\RNum{2}}} + 27 \, \mathrm{C}_{4,4,\textbf{\RNum{1}}}^{\mathbf{[1]}}+ 18 \, \mathrm{C}_{4,4,\textbf{\RNum{1}}}^{\mathbf{[2]}} - 72 \, \mathrm{C}_{4,4,\textbf{\RNum{1}}}^{\mathbf{[3]}} \right) \left( {\psi_s^*}^2 \left( \bigtriangledown \psi_s \right)^2 \right.  \nonumber \\
    &\left. + \psi_s^2 \left( \bigtriangledown \psi_s^* \right)^2 \right) \Big]- \Bar{\lambda}_6^2 \, \frac{5}{576 \, \alpha(5) \, m_a^9} \, |\psi_s|^6 \Big[ (8 \, \mathrm{C}_{6,6} - 25716) ( {\psi_s^*}^2 \left( \bigtriangledown \psi_s \right)^2 + \psi_s^2 \left( \bigtriangledown \psi_s^* \right)^2 ) \nonumber \\
    &+ \left( 2 \, \mathrm{C}_{6,6} + 155 \right) \, |\psi_s|^2 (\psi_s^* \bigtriangledown^2 \psi_s + \psi_s \bigtriangledown^2 \psi_s^*) + (20 \, \mathrm{C}_{6,6} - 60290) \, |\psi_s|^2 |\bigtriangledown \psi_s|^2  \Big] \,.
\end{align}

%%%%%%%%%%%%%%%%%%%%%%%%%%%%%%%%%%%%%%%%%%%%%%%%%%%%%%%%%%%%%%%%%%%%%%%%%%%%%%%%
%%%%%%%%%%%%%%%%%%%%%%%%%%%%%%%%%%%%%%%%%%%%%%%%%%%%%%%%%%%%%%%%%%%%%%%%%%%%%%%%

\section{Summary}
\label{sec:summary}

Axions are one of the most strongly motivated candidates for making up the dark matter of the universe. They were originally introduced as the pseudo-Goldstone bosons associated with the spontaneously broken Peccei-Quinn symmetry that address the strong $CP$-problem of QCD. Therefore, they could link, from their weaknesses, two of the most successful theories of particle physics: Standard Model and General Relativity.

Axions can be produced in the early universe with an abundance compatible with the observed dark matter density by appealing to some physical mechanisms that produce them at extremely low-energy conditions. Besides, it has been argued that non-relativistic axions which weakly interact among themselves can produce Bose-Einstein condensates that exhibit short-range correlations with either scarce occupancy or low coherence. Having at hand a derived effective field theory for low-energy axions which seems to be unique, since the predicted Bose-Eintein condensates demands higher order corrections, we have computed them up to $(\psi^\ast\psi)^5$, considering also the effects of fast-oscillating fluctuations that influence the dominant slowly-varying non-relativistic field.

Once the unique non-relativistic EFT for low-energy axions has been reviewed and meticulously derived up to fifth order corrections. We have also shown the systematic, improvable and iterative way of computing even higher order terms, considering the back-reaction of fast-oscillating fields. Now it remains to consider the possible future steps to follow in this line of research, which shall be the exploration of axion's coupling with gravity and the development of computational algorithms that allows us to treat possible many-body axion bound-state systems. Both are under consideration already and we expect to communicate results soon. 

%%%%%%%%%%%%%%%%%%%%%%%%%%%%%%%%%%%%%%%%%%%%%%%%%%%%%%%%%%%%%%%%%%%%%%%%%%%%%%%%
%%%%%%%%%%%%%%%%%%%%%%%%%%%%%%%%%%%%%%%%%%%%%%%%%%%%%%%%%%%%%%%%%%%%%%%%%%%%%%%%

\begin{acknowledgments}
This work has been partially funded by 
% Bryan Cordero-Patino
%
% Alvaro Dueñas-Vidal
the Escuela Polit\'ecnica Nacional under projects PII-DFIS-2022-01 and PIM 19-01; 
% Jorge Segovia
the Ministerio Espa\~nol de Ciencia e Innovaci\'on under grant No. PID2019-107844GB-C22; the Junta de Andaluc\'ia under contract Nos. Operativo FEDER Andaluc\'ia 2014-2020 UHU-1264517, P18-FR-5057 and also PAIDI FQM-370.
\end{acknowledgments}

%%%%%%%%%%%%%%%%%%%%%%%%%%%%%%%%%%%%%%%%%%%%%%%%%%%%%%%%%%%%%%%%%%%%%%%%%%%%%%%%
%%%%%%%%%%%%%%%%%%%%%%%%%%%%%%%%%%%%%%%%%%%%%%%%%%%%%%%%%%%%%%%%%%%%%%%%%%%%%%%%

\appendix

\section{Mode functions in Eqs.~\eqref{GFDR_exp2}}
\label{App-A}

\begin{align}
    \widetilde{G}_\nu = \, \mathcal{P}^{-\frac{1}{2}} \sum_{\nu_1,\nu_2} \bigg[ &\Psi_{\nu_1} \Psi_{\nu_2} \Psi_{2 + \nu - \nu_1 - \nu_2}+ 3 \, \Psi_{\nu_1} \Psi_{\nu_2} \Psi_{-\nu + \nu_1 + \nu_2}^*  \nonumber\\
    &+ 3 \, \Psi_{\nu_1}^* \Psi_{\nu_2}^* \Psi_{-2 + \nu + \nu_1 + \nu_2} + \Psi_{\nu_1}^* \Psi_{\nu_2}^* \Psi_{4-\nu-\nu_1 -\nu_2}^* \bigg] \,, \label{G_exp3}
\end{align}

\begin{align}
    \widetilde{F}_\nu = \, \mathcal{P}^{-\frac{1}{2}} \sum_{\nu_1, ..., \nu_4} \bigg[ &\Psi_{\nu_1} \Psi_{\nu_2} \Psi_{\nu_3} \Psi_{\nu_4} \Psi_{4+\nu-\nu_1-\nu_2-\nu_3-\nu_4} \nonumber\\ 
    &+ 5 \, \Psi_{\nu_1} \Psi_{\nu_2} \Psi_{\nu_3} \Psi_{\nu_4} \Psi_{-2-\nu+\nu_1+\nu_2+\nu_3+\nu_4}^*\nonumber\\
    &+ 10 \, \Psi_{\nu_1} \Psi_{\nu_2} \Psi_{\nu_3} \Psi_{\nu_4}^* \Psi_{-\nu+\nu_1+\nu_2+\nu_3-\nu_4}^* \nonumber\\ 
    &+ 10 \, \Psi_{\nu_1}^* \Psi_{\nu_2}^* \Psi_{\nu_3}^* \Psi_{\nu_4} \Psi_{-2+\nu+\nu_1+\nu_2+\nu_3-\nu_4} \nonumber \\
    &+ 5 \, \Psi_{\nu_1}^* \Psi_{\nu_2}^* \Psi_{\nu_3}^* \Psi_{\nu_4}^* \Psi_{-4+\nu+\nu_1+\nu_2+\nu_3+\nu_4} \nonumber \\ 
    &+ \Psi_{\nu_1}^* \Psi_{\nu_2}^* \Psi_{\nu_3}^* \Psi_{\nu_4}^* \Psi_{6-\nu-\nu_1-\nu_2-\nu_3-\nu_4}^* \bigg]  \,, \label{F_exp3}
\end{align}

\begin{align}
    \widetilde{D}_\nu = \, \mathcal{P}^{-\frac{1}{2}} \sum_{\nu_1, ..., \nu_6} \bigg[ &\Psi_{\nu_1} \Psi_{\nu_2} \Psi_{\nu_3} \Psi_{\nu_4} \Psi_{\nu_5} \Psi_{\nu_6} \Psi_{6+\nu- \sum_{i=1}^{6}\nu_i} \nonumber \\
    &+ 7 \, \Psi_{\nu_1} \Psi_{\nu_2} \Psi_{\nu_3} \Psi_{\nu_4} \Psi_{\nu_5} \Psi_{\nu_6} \Psi_{-4-\nu+\sum_{i=1}^{6}\nu_i}^* \nonumber \\
    &+ 21 \, \Psi_{\nu_1} \Psi_{\nu_2} \Psi_{\nu_3} \Psi_{\nu_4} \Psi_{\nu_5} \Psi_{\nu_6}^* \Psi_{-2-\nu+\sum_{i=1}^{5}\nu_i-\nu_6}^* \nonumber \\
    &+ 35 \, \Psi_{\nu_1} \Psi_{\nu_2} \Psi_{\nu_3} \Psi_{\nu_4} \Psi_{\nu_5}^* \Psi_{\nu_6}^* \Psi_{-\nu+\sum_{i=1}^{4}\nu_i-\nu_5-\nu_6}^* \nonumber \\
    &+ 35 \, \Psi_{\nu_1}^* \Psi_{\nu_2}^* \Psi_{\nu_3}^* \Psi_{\nu_4}^* \Psi_{\nu_5} \Psi_{\nu_6} \Psi_{-2+\nu+\sum_{i=1}^{4}\nu_i-\nu_5-\nu_6} \nonumber\\ 
    &+ 21 \, \Psi_{\nu_1}^* \Psi_{\nu_2}^* \Psi_{\nu_3}^* \Psi_{\nu_4}^* \Psi_{\nu_5}^* \Psi_{\nu_6} \Psi_{-4+\nu+\sum_{i=1}^{5}\nu_i-\nu_6} \nonumber \\
    &+ 7  \, \Psi_{\nu_1}^* \Psi_{\nu_2}^* \Psi_{\nu_3}^* \Psi_{\nu_4}^* \Psi_{\nu_5}^* \Psi_{\nu_6}^* \Psi_{-6+\nu+\sum_{i=1}^{6}\nu_i}\nonumber \\
    &+ \Psi_{\nu_1}^* \Psi_{\nu_2}^* \Psi_{\nu_3}^* \Psi_{\nu_4}^* \Psi_{\nu_5}^* \Psi_{\nu_6}^* \Psi_{8-\nu-\sum_{i=1}^{6}\nu_i}^* \bigg]  \,, \label{D_exp3}
\end{align}

\begin{align}
    \widetilde{R}_\nu =& \, \mathcal{P}^{-\frac{1}{2}} \sum_{\nu_1, ..., \nu_8} \bigg[ \Psi_{\nu_1} \Psi_{\nu_2} \Psi_{\nu_3} \Psi_{\nu_4} \Psi_{\nu_5} \Psi_{\nu_6} \Psi_{\nu_7} \Psi_{\nu_8} \Psi_{8+\nu- \sum_{i=1}^{8}\nu_i} \nonumber \\
    &+ 9 \, \Psi_{\nu_1} \Psi_{\nu_2} \Psi_{\nu_3} \Psi_{\nu_4} \Psi_{\nu_5} \Psi_{\nu_6} \Psi_{\nu_7} \Psi_{\nu_8} \Psi_{-6-\nu+\sum_{i=1}^{8}\nu_i}^* \nonumber \\
    &+ 36 \, \Psi_{\nu_1} \Psi_{\nu_2} \Psi_{\nu_3} \Psi_{\nu_4} \Psi_{\nu_5} \Psi_{\nu_6} \Psi_{\nu_7} \Psi_{\nu_8}^* \Psi_{-4-\nu+\sum_{i=1}^{7}\nu_i-\nu_8}^* \nonumber \\
    &+ 84 \, \Psi_{\nu_1} \Psi_{\nu_2} \Psi_{\nu_3} \Psi_{\nu_4} \Psi_{\nu_5} \Psi_{\nu_6} \Psi_{\nu_7}^* \Psi_{\nu_8}^* \Psi_{-2-\nu+\sum_{i=1}^{6}\nu_i-\nu_7-\nu_8}^* \nonumber \\
    &+ 126 \, \Psi_{\nu_1} \Psi_{\nu_2} \Psi_{\nu_3} \Psi_{\nu_4} \Psi_{\nu_5} \Psi_{\nu_6}^* \Psi_{\nu_7}^* \Psi_{\nu_8}^* \Psi_{-\nu+\sum_{i=1}^{5}\nu_i-\sum_{j=6}^{8}\nu_j}^* \nonumber \\
    &+ 126 \, \Psi_{\nu_1}^* \Psi_{\nu_2}^* \Psi_{\nu_3}^* \Psi_{\nu_4}^* \Psi_{\nu_5}^* \Psi_{\nu_6} \Psi_{\nu_7} \Psi_{\nu_8} \Psi_{-2+\nu+\sum_{i=1}^{5}\nu_i-\sum_{j=6}^{8}\nu_j} \nonumber \\
    &+ 84 \, \Psi_{\nu_1}^* \Psi_{\nu_2}^* \Psi_{\nu_3}^* \Psi_{\nu_4}^* \Psi_{\nu_5}^* \Psi_{\nu_6}^* \Psi_{\nu_7} \Psi_{\nu_8} \Psi_{-4+\nu+\sum_{i=1}^{6}\nu_i-\nu_7-\nu_8} \nonumber \\
    &+ 36 \, \Psi_{\nu_1}^* \Psi_{\nu_2}^* \Psi_{\nu_3}^* \Psi_{\nu_4}^* \Psi_{\nu_5}^* \Psi_{\nu_6}^* \Psi_{\nu_7}^* \Psi_{\nu_8} \Psi_{-6+\nu+\sum_{i=1}^{7}\nu_i-\nu_8} \nonumber \\
    &+ 9  \, \Psi_{\nu_1}^* \Psi_{\nu_2}^* \Psi_{\nu_3}^* \Psi_{\nu_4}^* \Psi_{\nu_5}^* \Psi_{\nu_6}^* \Psi_{\nu_7}^* \Psi_{\nu_8}^* \Psi_{-8+\nu+\sum_{i=1}^{8}\nu_i} \nonumber \\
    &+ \Psi_{\nu_1}^* \Psi_{\nu_2}^* \Psi_{\nu_3}^* \Psi_{\nu_4}^* \Psi_{\nu_5}^* \Psi_{\nu_6}^* \Psi_{\nu_7}^* \Psi_{\nu_8}^* \Psi_{10-\nu-\sum_{i=1}^{8}\nu_i}^* \bigg]  \,. \label{R_exp3}
\end{align}

%%%%%%%%%%%%%%%%%%%%%%%%%%%%%%%%%%%%%%%%%%%%%%%%%%%%%%%%%%%%%%%%%%%%%%%%%%%%%%%%

\section{1\textit{st} order iteration expressions}
\label{App-B}
    
\begin{align}
    \mathrm{G} {\scriptstyle (4)}_{0,(1)}^{(1)} = \, \mathcal{P}^{-\frac{1}{2}} \bigg[ &3 \, \Psi_s^2 \Gamma_{2} \mathcal{P}^{-1} \left( \Psi_s^* |\Psi_s|^2 \right)  + 6 \, |\Psi_s|^2 \Gamma_{2} \mathcal{P}^{-1} \left( \Psi_s |\Psi_s|^2 \right) \nonumber\\ 
    &+ {\Psi_s^*}^2 (\Gamma_{4} + \Gamma_{-2}) \mathcal{P}^{-1} \left( \Psi_s^3 \right) \bigg]  \,,
\end{align}
\begin{align}
    \mathrm{G} {\scriptstyle (6)}_{0,(1)}^{(1)} = \, \mathcal{P}^{-\frac{1}{2}} \bigg[ & 2 \, \Psi_s^2 \Gamma_{2} \mathcal{P}^{-1} \left( \Psi_s^* |\Psi_s|^4 \right) + 4 \, |\Psi_s|^2 \Gamma_{2} \mathcal{P}^{-1} \left( \Psi_s |\Psi_s|^4 \right) \nonumber \\
    &+ {\Psi_s^*}^2 (\Gamma_{4} + \Gamma_{-2}) \mathcal{P}^{-1} \left( \Psi_s^3 |\Psi_s|^2 \right) \bigg]  \,,
\end{align}
\begin{align}
    \mathrm{G} {\scriptstyle (8)}_{0,(1)}^{(1)} = \, \mathcal{P}^{-\frac{1}{2}} \bigg[ &5 \, \Psi_s^2 \Gamma_{2} \mathcal{P}^{-1} \left( \Psi_s^* |\Psi_s|^6  \right) + 10 \, |\Psi_s|^2 \Gamma_{2} \mathcal{P}^{-1} \left( \Psi_s |\Psi_s|^6  \right)\nonumber \\ 
    &+ 3 \, {\Psi_s^*}^2 (\Gamma_{4} + \Gamma_{-2}) \mathcal{P}^{-1} \left( \Psi_s^3 |\Psi_s|^4 \right) \bigg]  \,,
\end{align}
\begin{align}
    \mathrm{F} {\scriptstyle (4)}_{0,(1)}^{(1)} = \, \mathcal{P}^{-\frac{1}{2}} \bigg[ &\Psi_s^4 (\Gamma_{4} + \Gamma_{-2}) \mathcal{P}^{-1} \left( {\Psi_s^*}^3 \right) + 12 \, \Psi_s^2 |\Psi_s|^2 \Gamma_{2} 
    \mathcal{P}^{-1} \left( \Psi_s^* |\Psi_s|^2 \right) \nonumber \\
    &+ 18 \, |\Psi_s|^4 \Gamma_{2} \mathcal{P}^{-1} \left( \Psi_s |\Psi_s|^2 \right)  + 4 \, {\Psi_s^*}^2 |\Psi_s|^2 (\Gamma_{4} + \Gamma_{-2}) \mathcal{P}^{-1} \left( \Psi_s^3 \right) \bigg]  \,,
\end{align}
\begin{align}
    \mathrm{F} {\scriptstyle (6)}_{0,(1)}^{(1)} = \, \mathcal{P}^{-\frac{1}{2}} \bigg[&5 \, \Psi_s^4 (\Gamma_{4} + \Gamma_{-2}) \mathcal{P}^{-1} \left( {\Psi_s^*}^3 |\Psi_s|^2 \right)\nonumber\\
    &+ 40 \, \Psi_s^2 |\Psi_s|^2 \Gamma_{2} \mathcal{P}^{-1} \left( \Psi_s^* |\Psi_s|^4 \right) \nonumber \\
    &+ 60 \, |\Psi_s|^4 \Gamma_{2} \mathcal{P}^{-1} \left( \Psi_s |\Psi_s|^4 \right) \nonumber \\
    &+ 20 \, {\Psi_s^*}^2 |\Psi_s|^2 (\Gamma_{4} + \Gamma_{-2}) \mathcal{P}^{-1} \left( \Psi_s^3 |\Psi_s|^2 \right) \nonumber \\
    &+ {\Psi_s^*}^4 (\Gamma_{6} + \Gamma_{-4}) \mathcal{P}^{-1} \left( \Psi_s^5 \right) \bigg]  \,,
\end{align}
\begin{align}
    \mathrm{F} {\scriptstyle (8)}_{0,(1)}^{(1)} = \, \mathcal{P}^{-\frac{1}{2}} \bigg[ &3 \, \Psi_s^4 (\Gamma_{4} + \Gamma_{-2}) \mathcal{P}^{-1} \left( {\Psi_s^*}^3 |\Psi_s|^4   \right) \nonumber \\
    &+ 20 \, \Psi_s^2 |\Psi_s|^2 \Gamma_{2} \mathcal{P}^{-1} \left( \Psi_s^* |\Psi_s|^6  \right)\nonumber \\ 
    &+ 30 \, |\Psi_s|^4 \Gamma_{2} \mathcal{P}^{-1} \left( \Psi_s |\Psi_s|^6  \right) \nonumber \\
    &+ 12 \, {\Psi_s^*}^2 |\Psi_s|^2 (\Gamma_{4} + \Gamma_{-2}) \mathcal{P}^{-1} \left( \Psi_s^3 |\Psi_s|^4 \right) \nonumber \\
    &+ {\Psi_s^*}^4 (\Gamma_{6} + \Gamma_{-4}) \mathcal{P}^{-1} \left( \Psi_s^5 |\Psi_s|^2 \right) \bigg]  \,,
\end{align}
\begin{align}
    \mathrm{D} {\scriptstyle (4)}_{0,(1)}^{(1)} = \, \mathcal{P}^{-\frac{1}{2}} \bigg[& 2 \, \Psi_s^4 |\Psi_s|^2 (\Gamma_{4} + \Gamma_{-2}) \mathcal{P}^{-1} \left( {\Psi_s^*}^3 \right)\nonumber \\
    &+ 15 \, \Psi_s^2 |\Psi_s|^4 \Gamma_{2} \mathcal{P}^{-1} \left( \Psi_s^* |\Psi_s|^2 \right) \nonumber \\
    &+ 20 \, |\Psi_s|^6 \Gamma_{2} \mathcal{P}^{-1} \left( \Psi_s |\Psi_s|^2 \right) \nonumber \\
    &+ 5 \, {\Psi_s^*}^2 |\Psi_s|^4 (\Gamma_{4} + \Gamma_{-2}) \mathcal{P}^{-1} \left( \Psi_s^3 \right) \bigg]  \,,
\end{align}
\begin{align}
    \mathrm{D} {\scriptstyle (6)}_{0,(1)}^{(1)} = \, \mathcal{P}^{-\frac{1}{2}} \bigg[& \Psi_s^6 (\Gamma_{6} + \Gamma_{-4}) \mathcal{P}^{-1} \left( {\Psi_s^*}^5 \right)\nonumber \\
    &+ 30 \, \Psi_s^4 |\Psi_s|^2 (\Gamma_{4} + \Gamma_{-2}) \mathcal{P}^{-1} \left( {\Psi_s^*}^3 |\Psi_s|^2 \right) \nonumber \\
    &+ 150 \, \Psi_s^2 |\Psi_s|^4 \Gamma_{2} \mathcal{P}^{-1} \left( \Psi_s^* |\Psi_s|^4 \right) \nonumber \\
    &+ 200 \, |\Psi_s|^6 \Gamma_{2} \mathcal{P}^{-1} \left( \Psi_s |\Psi_s|^4 \right) \nonumber \\
    &+ 75 \, {\Psi_s^*}^2 |\Psi_s|^4 (\Gamma_{4} + \Gamma_{-2}) \mathcal{P}^{-1} \left( \Psi_s^3 |\Psi_s|^2 \right)\nonumber \\
    &+ 6 \,{\Psi_s^*}^4 |\Psi_s|^2 (\Gamma_{6} + \Gamma_{-4}) \mathcal{P}^{-1} \left( \Psi_s^5 \right) \bigg]  \,,
\end{align}
\begin{align}
    \mathrm{D} {\scriptstyle (8)}_{0,(1)}^{(1)} =\, \mathcal{P}^{-\frac{1}{2}} \bigg[& 7 \, \Psi_s^6 (\Gamma_{6} + \Gamma_{-4}) \mathcal{P}^{-1} \left( {\Psi_s^*}^5 |\Psi_s|^2 \right)\nonumber \\
    &+ 126 \, \Psi_s^4 |\Psi_s|^2 (\Gamma_{4} + \Gamma_{-2}) \mathcal{P}^{-1} \left( {\Psi_s^*}^3 |\Psi_s|^4   \right) \nonumber \\
    &+ 525 \, \Psi_s^2 |\Psi_s|^4 \Gamma_{2} \mathcal{P}^{-1} \left( \Psi_s^* |\Psi_s|^6  \right)\nonumber\\
    &+ 700 \, |\Psi_s|^6 \Gamma_{2} \mathcal{P}^{-1} \left( \Psi_s |\Psi_s|^6  \right) \nonumber \\
    &+ 315 \, {\Psi_s^*}^2 |\Psi_s|^4 (\Gamma_{4} + \Gamma_{-2}) \mathcal{P}^{-1} \left( \Psi_s^3 |\Psi_s|^4 \right) \nonumber \\
    &+ 42 \,{\Psi_s^*}^4 |\Psi_s|^2 (\Gamma_{6} + \Gamma_{-4}) \mathcal{P}^{-1} \left( \Psi_s^5 |\Psi_s|^2 \right) \nonumber \\
    &+ {\Psi_s^*}^6 (\Gamma_{8} + \Gamma_{-6}) \mathcal{P}^{-1} \left( \Psi_s^7 \right) \bigg]  \,.
\end{align}

%%%%%%%%%%%%%%%%%%%%%%%%%%%%%%%%%%%%%%%%%%%%%%%%%%%%%%%%%%%%%%%%%%%%%%%%%%%%%%%%

\section{2\textit{nd} order iteration expressions}
\label{App-C}

\begin{align}
    \mathrm{G} {\scriptstyle (4,4)}_{0,(1)}^{(2)} =& \,  \mathcal{P}^{-\frac{1}{2}} \bigg( 2 \, \Psi_s \bigg[  (\Gamma_{4} + \Gamma_{-2}) \mathcal{P}^{-1} \left( \Psi_s^3 \right) \times (\Gamma_{4} + \Gamma_{-2}) \mathcal{P}^{-1} \left( {\Psi_s^*}^3 \right) \nonumber \\ 
    &+ 9 \, \Gamma_{2} \mathcal{P}^{-1} \left( \Psi_s |\Psi_s|^2 \right)  \times \Gamma_{2} \mathcal{P}^{-1} \left( \Psi_s^* |\Psi_s|^2 \right) \bigg] \nonumber \\
    &+ 3 \, \Psi_s^* \bigg[ 3 \left( \Gamma_{2} \mathcal{P}^{-1} \left( \Psi_s |\Psi_s|^2 \right) \right)^2 + 2 \, \Gamma_{2} \mathcal{P}^{-1} \left( \Psi_s^* |\Psi_s|^2 \right)\nonumber \\
    &\times (\Gamma_{4} + \Gamma_{-2}) \mathcal{P}^{-1} \left( \Psi_s^3 \right) \bigg] \bigg) \,,
\end{align}
\begin{align}
    \mathrm{G} {\scriptstyle (4,6)}_{0,(1)}^{(2)} =& \, \mathcal{P}^{-\frac{1}{2}} \bigg( 5 \, \Psi_s \bigg[ (\Gamma_{4} + \Gamma_{-2}) \mathcal{P}^{-1} \left( \Psi_s^3 \right) \times (\Gamma_{4} + \Gamma_{-2}) \mathcal{P}^{-1} \left( {\Psi_s^*}^3 |\Psi_s|^2 \right) \nonumber \\ 
    &+ (\Gamma_{4} + \Gamma_{-2}) \mathcal{P}^{-1} \left( \Psi_s^3 |\Psi_s|^2 \right) \times (\Gamma_{4} + \Gamma_{-2}) \mathcal{P}^{-1} \left( {\Psi_s^*}^3 \right) \nonumber \\
    &+ 6 \, \Gamma_{2} \mathcal{P}^{-1} \left( \Psi_s |\Psi_s|^2 \right) \times \Gamma_{2} \mathcal{P}^{-1} \left( \Psi_s^* |\Psi_s|^4 \right) \nonumber \\ 
    &+ 6 \, \Gamma_{2} \mathcal{P}^{-1} \left( \Psi_s |\Psi_s|^4 \right) \times \Gamma_{2} \mathcal{P}^{-1} \left( \Psi_s^* |\Psi_s|^2 \right) \bigg]  \nonumber \\
    &+ \Psi_s^* \bigg[ (\Gamma_{4} + \Gamma_{-2}) \mathcal{P}^{-1} \left( {\Psi_s^*}^3 \right) \times (\Gamma_{6} + \Gamma_{-4}) \mathcal{P}^{-1} \left( \Psi_s^5 \right) \nonumber \\ 
    &+ 30 \, \Gamma_{2} \mathcal{P}^{-1} \left( \Psi_s |\Psi_s|^2 \right) \times \Gamma_{2} \mathcal{P}^{-1} \left( \Psi_s |\Psi_s|^4 \right) \nonumber \\
    &+ 15 \, \Gamma_{2} \mathcal{P}^{-1} \left( \Psi_s^* |\Psi_s|^2 \right) \times (\Gamma_{4} + \Gamma_{-2}) \mathcal{P}^{-1} \left( \Psi_s^3 |\Psi_s|^2 \right) \nonumber \\ 
    &+ 10 \, \Gamma_{2} \mathcal{P}^{-1} \left( \Psi_s^* |\Psi_s|^4 \right) \times (\Gamma_{4} + \Gamma_{-2}) \mathcal{P}^{-1} \left( \Psi_s^3 \right) \bigg] \bigg) \,,
\end{align}
\begin{align}
    \mathrm{G} {\scriptstyle (6,6)}_{0,(1)}^{(2)} =& \, \mathcal{P}^{-\frac{1}{2}} \bigg( \Psi_s \bigg[ (\Gamma_{6} + \Gamma_{-4}) \mathcal{P}^{-1} \left( \Psi_s^5 \right) \times (\Gamma_{6} + \Gamma_{-4}) \mathcal{P}^{-1} \left( {\Psi_s^*}^5 \right) \nonumber \\
    &+ 25 \, (\Gamma_{4} + \Gamma_{-2}) \mathcal{P}^{-1} \left( \Psi_s^3 |\Psi_s|^2 \right) \times (\Gamma_{4} + \Gamma_{-2}) \mathcal{P}^{-1} \left( {\Psi_s^*}^3 |\Psi_s|^2 \right) \nonumber \\ 
    &+ 100 \, \Gamma_{2} \mathcal{P}^{-1} \left( \Psi_s |\Psi_s|^4 \right) \times \Gamma_{2} \mathcal{P}^{-1} \left( \Psi_s^* |\Psi_s|^4 \right) \bigg] \nonumber \\ 
    &+ 5 \, \Psi_s^* \bigg[ (\Gamma_{4} + \Gamma_{-2}) \mathcal{P}^{-1} \left( {\Psi_s^*}^3 |\Psi_s|^2 \right) \times (\Gamma_{6} + \Gamma_{-4}) \mathcal{P}^{-1} \left( \Psi_s^5 \right)  \nonumber \\
    &+ 10 \left( \Gamma_{2} \mathcal{P}^{-1} \left( \Psi_s |\Psi_s|^4 \right) \right)^2\nonumber \\
    &+ 10 \, \Gamma_{2} \mathcal{P}^{-1} \left( \Psi_s^* |\Psi_s|^4 \right) \times (\Gamma_{4} + \Gamma_{-2}) \mathcal{P}^{-1} \left( \Psi_s^3 |\Psi_s|^2 \right) \bigg] \bigg) \,,
\end{align}

\begin{align}
    \mathrm{F} {\scriptstyle (4,4)}_{0,(1)}^{(2)} =& \, \mathcal{P}^{-\frac{1}{2}} \, \bigg( \Psi_s^3 \bigg[ 3 \left( \Gamma_{2} \mathcal{P}^{-1} \left( \Psi_s^* |\Psi_s|^2 \right) \right)^2 \nonumber \\
    &+ 2 \, \Gamma_{2} \mathcal{P}^{-1} \left( \Psi_s |\Psi_s|^2 \right) \times (\Gamma_{4} + \Gamma_{-2}) \mathcal{P}^{-1} \left( {\Psi_s^*}^3 \right) \bigg] \nonumber \\ 
    &+ 2 \, \Psi_s |\Psi_s|^2 \bigg[ (\Gamma_{4} + \Gamma_{-2}) \mathcal{P}^{-1} \left( \Psi_s^3 \right) \times (\Gamma_{4} + \Gamma_{-2}) \mathcal{P}^{-1} \left( {\Psi_s^*}^3 \right) \nonumber \\ 
    &+ 9 \, \Gamma_{2} \mathcal{P}^{-1} \left( \Psi_s |\Psi_s|^2 \right)  \times \Gamma_{2} \mathcal{P}^{-1} \left( \Psi_s^* |\Psi_s|^2 \right)  \bigg] \nonumber \\ 
    &+ 3 \, \Psi_s^* |\Psi_s|^2 \bigg[ 3 \left( \Gamma_{2} \mathcal{P}^{-1} \left( \Psi_s |\Psi_s|^2 \right) \right)^2 \nonumber \\
    &+ 2 \, \Gamma_{2} \mathcal{P}^{-1} \left( \Psi_s^* |\Psi_s|^2 \right) \times (\Gamma_{4} + \Gamma_{-2}) \mathcal{P}^{-1} \left( \Psi_s^3 \right) \bigg] \nonumber \\ 
    &+ 2 \, {\Psi_s^*}^3 \bigg[ \Gamma_{2} \mathcal{P}^{-1} \left( \Psi_s |\Psi_s|^2 \right) \times (\Gamma_{4} + \Gamma_{-2}) \mathcal{P}^{-1} \left( \Psi_s^3 \right) \bigg] \bigg) \,,
\end{align}
\begin{align}
    \mathrm{F} {\scriptstyle (4,6)}_{0,(1)}^{(2)} =& \, \mathcal{P}^{-\frac{1}{2}} \, \bigg( \Psi_s^3 \bigg[ (\Gamma_{4} + \Gamma_{-2}) \mathcal{P}^{-1} \left( \Psi_s^3 \right) \times (\Gamma_{6} + \Gamma_{-4}) \mathcal{P}^{-1} \left( {\Psi_s^*}^5 \right) \nonumber \\ 
    &+ 30 \, \Gamma_{2} \mathcal{P}^{-1} \left( \Psi_s^* |\Psi_s|^2 \right) \times \Gamma_{2} \mathcal{P}^{-1} \left( \Psi_s^* |\Psi_s|^4 \right) \nonumber \\ 
    &+ 15 \, \Gamma_{2} \mathcal{P}^{-1} \left( \Psi_s |\Psi_s|^2 \right) \times (\Gamma_{4} + \Gamma_{-2}) \mathcal{P}^{-1} \left( {\Psi_s^*}^3 |\Psi_s|^2 \right) \nonumber \\ 
    &+ 10 \, \Gamma_{2} \mathcal{P}^{-1} \left( \Psi_s |\Psi_s|^4 \right) \times (\Gamma_{4} + \Gamma_{-2}) \mathcal{P}^{-1} \left( {\Psi_s^*}^3 \right) \bigg] \nonumber \\ 
    &+ 15 \, \Psi_s |\Psi_s|^2 \bigg[ (\Gamma_{4} + \Gamma_{-2}) \mathcal{P}^{-1} \left( \Psi_s^3 \right) \times (\Gamma_{4} + \Gamma_{-2}) \mathcal{P}^{-1} \left( {\Psi_s^*}^3 |\Psi_s|^2 \right) \nonumber \\ 
    &+ (\Gamma_{4} + \Gamma_{-2}) \mathcal{P}^{-1} \left( \Psi_s^3 |\Psi_s|^2 \right) \times (\Gamma_{4} + \Gamma_{-2}) \mathcal{P}^{-1} \left( {\Psi_s^*}^3 \right) \nonumber \\
    &+ 6 \, \Gamma_{2} \mathcal{P}^{-1} \left( \Psi_s |\Psi_s|^2 \right) \times \Gamma_{2} \mathcal{P}^{-1} \left( \Psi_s^* |\Psi_s|^4 \right) \nonumber \\ 
    &+ 6 \, \Gamma_{2} \mathcal{P}^{-1} \left( \Psi_s |\Psi_s|^4 \right) \times \Gamma_{2} \mathcal{P}^{-1} \left( \Psi_s^* |\Psi_s|^2 \right) \bigg] \nonumber \\ 
    &+ 3 \, \Psi_s^* |\Psi_s|^2 \bigg[ (\Gamma_{4} + \Gamma_{-2}) \mathcal{P}^{-1} \left( {\Psi_s^*}^3 \right) \times (\Gamma_{6} + \Gamma_{-4}) \mathcal{P}^{-1} \left( \Psi_s^5 \right) \nonumber \\ 
    &+ 30 \, \Gamma_{2} \mathcal{P}^{-1} \left( \Psi_s |\Psi_s|^2 \right) \times \Gamma_{2} \mathcal{P}^{-1} \left( \Psi_s |\Psi_s|^4 \right) \nonumber \\ 
    &+ 15 \, \Gamma_{2} \mathcal{P}^{-1} \left( \Psi_s^* |\Psi_s|^2 \right) \times (\Gamma_{4} + \Gamma_{-2}) \mathcal{P}^{-1} \left( \Psi_s^3 |\Psi_s|^2 \right) \nonumber \\ 
    &+ 10 \, \Gamma_{2} \mathcal{P}^{-1} \left( \Psi_s^* |\Psi_s|^4 \right) \times (\Gamma_{4} + \Gamma_{-2}) \mathcal{P}^{-1} \left( \Psi_s^3 \right) \bigg] \nonumber \\ 
    &+ {\Psi_s^*}^3 \bigg[  15 \, \Gamma_{2} \mathcal{P}^{-1} \left( \Psi_s |\Psi_s|^2 \right) \times (\Gamma_{4} + \Gamma_{-2}) \mathcal{P}^{-1} \left( \Psi_s^3 |\Psi_s|^2 \right) \nonumber \\ 
    &+ 10 \, \Gamma_{2} \mathcal{P}^{-1} \left( \Psi_s |\Psi_s|^4 \right) \times (\Gamma_{4} + \Gamma_{-2}) \mathcal{P}^{-1} \left( \Psi_s^3 \right) \nonumber \\ 
    &+ 3 \, \Gamma_{2} \mathcal{P}^{-1} \left( \Psi_s^* |\Psi_s|^2 \right) \times (\Gamma_{6} + \Gamma_{-4}) \mathcal{P}^{-1} \left( \Psi_s^5 \right) \bigg] \bigg) \,,
\end{align}
\begin{align}
    \mathrm{F} {\scriptstyle (6,6)}_{0,(1)}^{(2)} =& \, \mathcal{P}^{-\frac{1}{2}} \, \bigg( 5 \, \Psi_s^3 \bigg[ (\Gamma_{4} + \Gamma_{-2}) \mathcal{P}^{-1} \left( \Psi_s^3 |\Psi_s|^2 \right) \times (\Gamma_{6} + \Gamma_{-4}) \mathcal{P}^{-1} \left( {\Psi_s^*}^5 \right) \nonumber \\
    &+ 10 \left( \Gamma_{2} \mathcal{P}^{-1} \left( \Psi_s^* |\Psi_s|^4 \right) \right)^2 \nonumber \\
    &+ 10 \, \Gamma_{2} \mathcal{P}^{-1} \left( \Psi_s |\Psi_s|^4 \right) \times (\Gamma_{4} + \Gamma_{-2}) \mathcal{P}^{-1} \left( {\Psi_s^*}^3 |\Psi_s|^2 \right) \bigg] \nonumber \\ 
    &+ 3\, \Psi_s |\Psi_s|^2 \bigg[ (\Gamma_{6} + \Gamma_{-4}) \mathcal{P}^{-1} \left( \Psi_s^5 \right) \times (\Gamma_{6} + \Gamma_{-4}) \mathcal{P}^{-1} \left( {\Psi_s^*}^5 \right) \nonumber \\
    &+ 25 \, (\Gamma_{4} + \Gamma_{-2}) \mathcal{P}^{-1} \left( \Psi_s^3 |\Psi_s|^2 \right) \times (\Gamma_{4} + \Gamma_{-2}) \mathcal{P}^{-1} \left( {\Psi_s^*}^3 |\Psi_s|^2 \right) \nonumber \\ 
    &+ 100 \, \Gamma_{2} \mathcal{P}^{-1} \left( \Psi_s |\Psi_s|^4 \right) \times \Gamma_{2} \mathcal{P}^{-1} \left( \Psi_s^* |\Psi_s|^4 \right) \bigg] \nonumber \\ 
    &+ 15 \, \Psi_s^* |\Psi_s|^2 \bigg[ (\Gamma_{4} + \Gamma_{-2}) \mathcal{P}^{-1} \left( {\Psi_s^*}^3 |\Psi_s|^2 \right) \times (\Gamma_{6} + \Gamma_{-4}) \mathcal{P}^{-1} \left( \Psi_s^5 \right) \nonumber \\
    &+ 10 \left( \Gamma_{2} \mathcal{P}^{-1} \left( \Psi_s |\Psi_s|^4 \right) \right)^2\nonumber \\
    &+ 10 \, \Gamma_{2} \mathcal{P}^{-1} \left( \Psi_s^* |\Psi_s|^4 \right) \times (\Gamma_{4} + \Gamma_{-2}) \mathcal{P}^{-1} \left( \Psi_s^3 |\Psi_s|^2 \right) \bigg] \nonumber \\ 
    &+ 10 \, {\Psi_s^*}^3 \bigg[ 5 \, \Gamma_{2} \mathcal{P}^{-1} \left( \Psi_s |\Psi_s|^4 \right) \times (\Gamma_{4} + \Gamma_{-2}) \mathcal{P}^{-1} \left( \Psi_s^3 |\Psi_s|^2 \right) \nonumber \\ 
    &+ \Gamma_{2} \mathcal{P}^{-1} \left( \Psi_s^* |\Psi_s|^4 \right) \times (\Gamma_{6} + \Gamma_{-4}) \mathcal{P}^{-1} \left( \Psi_s^5 \right) \bigg] \bigg) \,,
\end{align}

\begin{align}
    \mathrm{G} {\scriptstyle (4, \textbf{\RNum{1}})}_{0,(2)}^{(2)} =& \, \mathcal{P}^{-\frac{1}{2}} \, \bigg( - \Psi_s^2 (\Gamma_{2})^2 \mathcal{P}^{-1}\bigg[ 2 \, |\Psi_s|^2 \Dot{\Psi}_s^* + {\Psi_s^*}^2 \Dot{\Psi}_s \bigg] \nonumber \\
    &+ 2 \, |\Psi_s|^2 (\Gamma_{2})^2 \mathcal{P}^{-1} \bigg[ 2 \, |\Psi_s|^2 \Dot{\Psi}_s + \Psi_s^2 \Dot{\Psi}_s^* \bigg] \nonumber \\ 
    &+ {\Psi_s^*}^2 \left((\Gamma_{4})^2 - (\Gamma_{-2})^2 \right) \mathcal{P}^{-1} \bigg[  \Psi_s^2 \Dot{\Psi}_s \bigg] \bigg) \,,
\end{align}
\begin{align}
    \mathrm{G} {\scriptstyle (6, \textbf{\RNum{1}})}_{0,(2)}^{(2)} =& \, \mathcal{P}^{-\frac{1}{2}} \, \bigg( - 2 \, \Psi_s^2 (\Gamma_{2})^2  \mathcal{P}^{-1} \bigg[ 3 \, |\Psi_s|^4 \Dot{\Psi}_s^* + 2 \, {\Psi_s^*}^2 |\Psi_s|^2 \Dot{\Psi}_s \bigg] \nonumber \\ 
    &+ 4 \, |\Psi_s|^2 (\Gamma_{2})^2 \mathcal{P}^{-1} \bigg[ 3 \, |\Psi_s|^4 \Dot{\Psi}_s + 2 \, \Psi_s^2 |\Psi_s|^2 \Dot{\Psi}_s^* \bigg] \nonumber \\ 
    &+ {\Psi_s^*}^2 \left((\Gamma_{4})^2 - (\Gamma_{-2})^2 \right) \mathcal{P}^{-1} \bigg[ 4 \, \Psi_s^2 |\Psi_s|^2 \Dot{\Psi}_s + \Psi_s^4 \Dot{\Psi}_s^* \bigg] \bigg) \,,
\end{align}
\begin{align}
    \mathrm{G} {\scriptstyle (4,4)}_{0,(2)}^{(2)} =& \, \mathcal{P}^{-\frac{1}{2}} \, \bigg( \Psi_s^2 \, \Gamma_{2} \mathcal{P}^{-1} \bigg[ \Psi_s^2 (\Gamma_{4} + \Gamma_{-2}) \mathcal{P}^{-1} \left( {\Psi_s^*}^3 \right) + 6 \, |\Psi_s|^2 \Gamma_{2} \mathcal{P}^{-1} \left( \Psi_s^* |\Psi_s|^2  \right) \nonumber \\ 
    &+ 3 \, {\Psi_s^*}^2 \Gamma_{2} \mathcal{P}^{-1} \left( \Psi_s |\Psi_s|^2  \right) \bigg] + 2 \, |\Psi_s|^2 \, \Gamma_{2} \mathcal{P}^{-1} \bigg[ {\Psi_s^*}^2 (\Gamma_{4} + \Gamma_{-2}) \mathcal{P}^{-1} \left( \Psi_s^3 \right) \nonumber \\
    &+ 6 \, |\Psi_s|^2 \Gamma_{2} \mathcal{P}^{-1} \left( \Psi_s |\Psi_s|^2  \right) + 3 \, \Psi_s^2 \Gamma_{2} \mathcal{P}^{-1} \left( \Psi_s^* |\Psi_s|^2  \right) \bigg] \nonumber \\ 
    &+ {\Psi_s^*}^2 (\Gamma_{4} + \Gamma_{-2}) \mathcal{P}^{-1} \bigg[ 2 \, |\Psi_s|^2 (\Gamma_{4} + \Gamma_{-2}) \mathcal{P}^{-1} \left( \Psi_s^3 \right)\nonumber \\
    &+ 3 \, \Psi_s^2 \Gamma_{2} \mathcal{P}^{-1} \left( \Psi_s |\Psi_s|^2  \right) \bigg] \bigg) \,,
\end{align}
\begin{align}
    \mathrm{G} {\scriptstyle (4,6)}_{0,(2)}^{(2)} =& \, \mathcal{P}^{-\frac{1}{2}} \, \bigg( 5 \, \Psi_s^2 \, \Gamma_{2} \mathcal{P}^{-1} \bigg[ 3 \, \Psi_s^2 (\Gamma_{4} + \Gamma_{-2}) \mathcal{P}^{-1} \left( {\Psi_s^*}^3 |\Psi_s|^2   \right) \nonumber \\
    &+ 12 \, |\Psi_s|^2 \Gamma_{2} \mathcal{P}^{-1} \left( \Psi_s^* |\Psi_s|^4  \right)+ 6 \, {\Psi_s^*}^2 \Gamma_{2} \mathcal{P}^{-1} \left( \Psi_s |\Psi_s|^4  \right)  \nonumber \\ 
    & + 4 \, \Psi_s^2 |\Psi_s|^2 (\Gamma_{4} + \Gamma_{-2}) \mathcal{P}^{-1} \left( {\Psi_s^*}^3   \right) + 18 \, |\Psi_s|^4 \Gamma_{2} \mathcal{P}^{-1} \left( \Psi_s^* |\Psi_s|^2  \right) \nonumber \\
    &+ 12 \, {\Psi_s^*}^2 |\Psi_s|^2 \Gamma_{2} \mathcal{P}^{-1} \left( \Psi_s |\Psi_s|^2  \right) + {\Psi_s^*}^4 (\Gamma_{4} + \Gamma_{-2}) \mathcal{P}^{-1} \left( \Psi_s^3   \right) \bigg] \nonumber \\
    &+ 10 |\Psi_s|^2 \, \Gamma_{2} \mathcal{P}^{-1} \bigg[ 3 \, {\Psi_s^*}^2 (\Gamma_{4} + \Gamma_{-2}) \mathcal{P}^{-1} \left( \Psi_s^3 |\Psi_s|^2   \right) + 12 \, |\Psi_s|^2 \Gamma_{2} \mathcal{P}^{-1} \left( \Psi_s |\Psi_s|^4  \right) \nonumber \\
    &+ 6 \, \Psi_s^2 \Gamma_{2} \mathcal{P}^{-1} \left( \Psi_s^* |\Psi_s|^4  \right)  + 4 \, {\Psi_s^*}^2 |\Psi_s|^2 (\Gamma_{4} + \Gamma_{-2}) \mathcal{P}^{-1} \left( \Psi_s^3   \right)  \nonumber \\
    &+ 18 \, |\Psi_s|^4 \Gamma_{2} \mathcal{P}^{-1} \left( \Psi_s |\Psi_s|^2  \right)+ 12 \, \Psi_s^2 |\Psi_s|^2 \Gamma_{2} \mathcal{P}^{-1} \left( \Psi_s^* |\Psi_s|^2  \right) \nonumber \\
    &+ \Psi_s^4 (\Gamma_{4} + \Gamma_{-2}) \mathcal{P}^{-1} \left( {\Psi_s^*}^3   \right) \bigg]+ 3 \, {\Psi_s^*}^2 (\Gamma_{4} + \Gamma_{-2}) \mathcal{P}^{-1} \bigg[ {\Psi_s^*}^2 (\Gamma_{6} + \Gamma_{-4}) \mathcal{P}^{-1} \left( \Psi_s^5 \right) \nonumber \\
    & + 10 \, |\Psi_s|^2 (\Gamma_{4} + \Gamma_{-2}) \mathcal{P}^{-1} \left( \Psi_s^3 |\Psi_s|^2   \right)+ 10 \, \Psi_s^2 \Gamma_{2} \mathcal{P}^{-1} \left( \Psi_s |\Psi_s|^4  \right) \nonumber \\
    & + 10 \, |\Psi_s|^4 (\Gamma_{4} + \Gamma_{-2}) \mathcal{P}^{-1} \left( \Psi_s^3  \right) + 20 \, \Psi_s^2 |\Psi_s|^2 \Gamma_{2} \mathcal{P}^{-1} \left( \Psi_s |\Psi_s|^2  \right) \nonumber \\
    &+ 5 \, \Psi_s^4 \Gamma_{2} \mathcal{P}^{-1} \left( \Psi_s^* |\Psi_s|^2  \right) \bigg] \bigg) \,,
\end{align}
\begin{align}
    \mathrm{G} {\scriptstyle (6,6)}_{0,(2)}^{(2)} =& \, \mathcal{P}^{-\frac{1}{2}} \, \bigg( \Psi_s^2 \, \Gamma_{2} \mathcal{P}^{-1} \bigg[ \Psi_s^4 (\Gamma_{6} + \Gamma_{-4}) \mathcal{P}^{-1} \left( {\Psi_s^*}^5  \right) \nonumber \\
    &+ 20 \, \Psi_s^2 |\Psi_s|^2 (\Gamma_{4} + \Gamma_{-2}) \mathcal{P}^{-1} \left( {\Psi_s^*}^3 |\Psi_s|^2  \right) \nonumber \\
    &+ 60 \, |\Psi_s|^4 \Gamma_{2} \mathcal{P}^{-1} \left( \Psi_s^* |\Psi_s|^4  \right) + 40 \, {\Psi_s^*}^2 |\Psi_s|^2 \Gamma_{2} \mathcal{P}^{-1} \left( \Psi_s |\Psi_s|^4  \right) \nonumber \\
    &+ 5 \, {\Psi_s^*}^4 (\Gamma_{4} + \Gamma_{-2}) \mathcal{P}^{-1} \left( \Psi_s^3 |\Psi_s|^2   \right) \bigg] + 2 \, |\Psi_s|^2 \, \Gamma_{2} \mathcal{P}^{-1} \bigg[ {\Psi_s^*}^4 (\Gamma_{6} + \Gamma_{-4}) \mathcal{P}^{-1} \left( \Psi_s^5  \right)  \nonumber \\ 
    &+ 20 \, {\Psi_s^*}^2 |\Psi_s|^2 (\Gamma_{4} + \Gamma_{-2}) \mathcal{P}^{-1} \left( \Psi_s^3 |\Psi_s|^2  \right) + 60 \, |\Psi_s|^4 \Gamma_{2} \mathcal{P}^{-1} \left( \Psi_s |\Psi_s|^4  \right) \nonumber \\ 
    &+ 40 \, \Psi_s^2 |\Psi_s|^2 \Gamma_{2} \mathcal{P}^{-1} \left( \Psi_s^* |\Psi_s|^4  \right) + 5 \, \Psi_s^4 (\Gamma_{4} + \Gamma_{-2}) \mathcal{P}^{-1} \left( {\Psi_s^*}^3 |\Psi_s|^2   \right) \bigg] \nonumber \\
    &+ 2 \, {\Psi_s^*}^2 (\Gamma_{4} + \Gamma_{-2}) \mathcal{P}^{-1} \bigg[ 2 \, {\Psi_s^*}^2 |\Psi_s|^2 (\Gamma_{6} + \Gamma_{-4}) \mathcal{P}^{-1} \left( \Psi_s^5 \right) \nonumber \\
    &+ 15 \, |\Psi_s|^4 (\Gamma_{4} + \Gamma_{-2}) \mathcal{P}^{-1} \left( \Psi_s^3 |\Psi_s|^2   \right) + 20 \, \Psi_s^2 |\Psi_s|^2 \Gamma_{2} \mathcal{P}^{-1} \left( \Psi_s |\Psi_s|^4  \right) \nonumber \\
    &+ 5 \, \Psi_s^4 \Gamma_{2} \mathcal{P}^{-1} \left( \Psi_s^* |\Psi_s|^4  \right) \bigg] \bigg) \,,
\end{align}

\begin{align}
    \mathrm{F} {\scriptstyle (4, \textbf{\RNum{1}})}_{0,(2)}^{(2)} =& \, \mathcal{P}^{-\frac{1}{2}} \, \bigg( - \Psi_s^4 \left((\Gamma_{4})^2 - (\Gamma_{-2})^2 \right) \mathcal{P}^{-1} \bigg[ {\Psi_s^*}^2 \Dot{\Psi}_s^* \bigg]\nonumber \\
    &- 4 \, \Psi_s^2 |\Psi_s|^2 (\Gamma_{2})^2 \mathcal{P}^{-1} \bigg[ 2 \, |\Psi_s|^2 \Dot{\Psi}_s^* + {\Psi_s^*}^2 \Dot{\Psi}_s \bigg] \nonumber \\
    &+ 4 \, {\Psi_s^*}^2 |\Psi_s|^2 \left((\Gamma_{4})^2 - (\Gamma_{-2})^2 \right) \mathcal{P}^{-1} \bigg[ \Psi_s^2 \Dot{\Psi}_s \bigg] \nonumber\\ 
    &+ 6 \, |\Psi_s|^4 (\Gamma_{2})^2 \mathcal{P}^{-1} \bigg[ 2 \, |\Psi_s|^2 \Dot{\Psi}_s + \Psi_s^2 \Dot{\Psi}_s^* \bigg] \bigg) \,,
\end{align}
\begin{align}
    \mathrm{F} {\scriptstyle (6, \textbf{\RNum{1}})}_{0,(2)}^{(2)} =& \, \mathcal{P}^{-\frac{1}{2}} \, \bigg( - \Psi_s^4 \left((\Gamma_{4})^2 - (\Gamma_{-2})^2 \right) \mathcal{P}^{-1} \bigg[ 4 \, {\Psi_s^*}^2 |\Psi_s|^2 \Dot{\Psi}_s^* + {\Psi_s^*}^4 \Dot{\Psi}_s \bigg] \nonumber \\
    &- 8 \, \Psi_s^2 |\Psi_s|^2 (\Gamma_{2})^2 \mathcal{P}^{-1} \bigg[ 3 \, |\Psi_s|^4 \Dot{\Psi}_s^* + 2 \, {\Psi_s^*}^2 |\Psi_s|^2 \Dot{\Psi}_s \bigg] \nonumber \\
    &+ 4 \, {\Psi_s^*}^2 |\Psi_s|^2 \left((\Gamma_{4})^2 - (\Gamma_{-2})^2 \right) \mathcal{P}^{-1} \bigg[ 4 \, \Psi_s^2 |\Psi_s|^2 \Dot{\Psi}_s + \Psi_s^4 \Dot{\Psi}_s^* \bigg] \nonumber \\
    &+ 12 \, |\Psi_s|^4 (\Gamma_{2})^2 \mathcal{P}^{-1} \bigg[ 3 \, |\Psi_s|^4 \Dot{\Psi}_s + 2 \, \Psi_s^2 |\Psi_s|^2 \Dot{\Psi}_s^* \bigg] \nonumber \\
    &+ {\Psi_s^*}^4 \left((\Gamma_{6})^2 - (\Gamma_{-4})^2 \right) \mathcal{P}^{-1} \bigg[ \Psi_s^4 \Dot{\Psi}_s \bigg] \bigg) \,,
\end{align}
\begin{align}
    \mathrm{F} {\scriptstyle (4,4)}_{0,(2)}^{(2)} =& \, \mathcal{P}^{-\frac{1}{2}} \, \bigg( \Psi_s^4 (\Gamma_{4} + \Gamma_{-2}) \mathcal{P}^{-1} \bigg[ 2 \, |\Psi_s|^2 (\Gamma_{4} + \Gamma_{-2}) \mathcal{P}^{-1} \left( {\Psi_s^*}^3 \right)\nonumber \\
    &+ 3 \, {\Psi_s^*}^2 \Gamma_{2} \mathcal{P}^{-1} \left( \Psi_s^* |\Psi_s|^2  \right) \bigg]  \nonumber \\
    &+ 4 \, \Psi_s^2 |\Psi_s|^2 \, \Gamma_{2} \mathcal{P}^{-1} \bigg[ \Psi_s^2 (\Gamma_{4} + \Gamma_{-2}) \mathcal{P}^{-1} \left( {\Psi_s^*}^3 \right) + 6 \, |\Psi_s|^2 \Gamma_{2} \mathcal{P}^{-1} \left( \Psi_s^* |\Psi_s|^2  \right) \nonumber \\ 
    &+ 3 \, {\Psi_s^*}^2 \Gamma_{2} \mathcal{P}^{-1} \left( \Psi_s |\Psi_s|^2  \right) \bigg]\nonumber \\
    &+ 4 \, {\Psi_s^*}^2 |\Psi_s|^2 (\Gamma_{4} + \Gamma_{-2}) \mathcal{P}^{-1} \bigg[ 2 \, |\Psi_s|^2 (\Gamma_{4} + \Gamma_{-2}) \mathcal{P}^{-1} \left( \Psi_s^3 \right) \nonumber \\
    &+ 3 \, \Psi_s^2 \Gamma_{2} \mathcal{P}^{-1} \left( \Psi_s |\Psi_s|^2  \right) \bigg] + 6 \, |\Psi_s|^4 \, \Gamma_{2} \mathcal{P}^{-1} \bigg[ {\Psi_s^*}^2 (\Gamma_{4} + \Gamma_{-2}) \mathcal{P}^{-1} \left( \Psi_s^3 \right) \nonumber \\
    &+ 6 \, |\Psi_s|^2 \Gamma_{2} \mathcal{P}^{-1} \left( \Psi_s |\Psi_s|^2  \right) + 3 \, \Psi_s^2 \Gamma_{2} \mathcal{P}^{-1} \left( \Psi_s^* |\Psi_s|^2  \right) \bigg] \nonumber \\ 
    &+ {\Psi_s^*}^4 (\Gamma_{6} + \Gamma_{-4}) \mathcal{P}^{-1} \bigg[ \Psi_s^2 (\Gamma_{4} + \Gamma_{-2}) \mathcal{P}^{-1} \left( \Psi_s^3  \right) \bigg] \bigg) \,,
\end{align}
\begin{align}
    \mathrm{F} {\scriptstyle (4,6)}_{0,(2)}^{(2)} =& \, \mathcal{P}^{-\frac{1}{2}} \, \bigg( 3 \, \Psi_s^4 (\Gamma_{4} + \Gamma_{-2}) \mathcal{P}^{-1} \bigg[ \Psi_s^2 (\Gamma_{6} + \Gamma_{-4}) \mathcal{P}^{-1} \left( {\Psi_s^*}^5 \right) \nonumber \\
    &+ 10 \, |\Psi_s|^2 (\Gamma_{4} + \Gamma_{-2}) \mathcal{P}^{-1} \left( {\Psi_s^*}^3 |\Psi_s|^2   \right) + 10 \, {\Psi_s^*}^2 \Gamma_{2} \mathcal{P}^{-1} \left( \Psi_s^* |\Psi_s|^4  \right) \nonumber \\
    &+ 10 \, |\Psi_s|^4 (\Gamma_{4} + \Gamma_{-2}) \mathcal{P}^{-1} \left( {\Psi_s^*}^3  \right) + 20 \, {\Psi_s^*}^2 |\Psi_s|^2 \Gamma_{2} \mathcal{P}^{-1} \left( \Psi_s^* |\Psi_s|^2  \right) \nonumber \\
    &+ 5 \, {\Psi_s^*}^4 \Gamma_{2} \mathcal{P}^{-1} \left( \Psi_s |\Psi_s|^2  \right) \bigg] + 20 \, \Psi_s^2 |\Psi_s|^2 \, \Gamma_{2} \mathcal{P}^{-1} \bigg[ 3 \, \Psi_s^2 (\Gamma_{4} + \Gamma_{-2}) \mathcal{P}^{-1} \left( {\Psi_s^*}^3 |\Psi_s|^2   \right) \nonumber \\
    &+ 12 \, |\Psi_s|^2 \Gamma_{2} \mathcal{P}^{-1} \left( \Psi_s^* |\Psi_s|^4  \right) + 6 \, {\Psi_s^*}^2 \Gamma_{2} \mathcal{P}^{-1} \left( \Psi_s |\Psi_s|^4  \right)  \nonumber \\
    &+ 4 \, \Psi_s^2 |\Psi_s|^2 (\Gamma_{4} + \Gamma_{-2}) \mathcal{P}^{-1} \left( {\Psi_s^*}^3   \right)+ 18 \, |\Psi_s|^4 \Gamma_{2} \mathcal{P}^{-1} \left( \Psi_s^* |\Psi_s|^2  \right) \nonumber \\ 
    & + 12 \, {\Psi_s^*}^2 |\Psi_s|^2 \Gamma_{2} \mathcal{P}^{-1} \left( \Psi_s |\Psi_s|^2  \right) + {\Psi_s^*}^4 (\Gamma_{4} + \Gamma_{-2}) \mathcal{P}^{-1} \left( \Psi_s^3 \right) \bigg] \nonumber \\
    &+ 12 \, {\Psi_s^*}^2 |\Psi_s|^2 (\Gamma_{4} + \Gamma_{-2}) \mathcal{P}^{-1} \bigg[ {\Psi_s^*}^2 (\Gamma_{6} + \Gamma_{-4}) \mathcal{P}^{-1} \left( \Psi_s^5 \right) \nonumber\\
    &+ 10 \, |\Psi_s|^2 (\Gamma_{4} + \Gamma_{-2}) \mathcal{P}^{-1} \left( \Psi_s^3 |\Psi_s|^2   \right) + 10 \, \Psi_s^2 \Gamma_{2} \mathcal{P}^{-1} \left( \Psi_s |\Psi_s|^4  \right) \nonumber\\
    &+ 10 \, |\Psi_s|^4 (\Gamma_{4} + \Gamma_{-2}) \mathcal{P}^{-1} \left( \Psi_s^3  \right) + 20 \, \Psi_s^2 |\Psi_s|^2 \Gamma_{2} \mathcal{P}^{-1} \left( \Psi_s |\Psi_s|^2  \right) \nonumber \\
     & + 5 \, \Psi_s^4 \Gamma_{2} \mathcal{P}^{-1} \left( \Psi_s^* |\Psi_s|^2  \right) \bigg] \nonumber\\
      &+ 30 \, |\Psi_s|^4 \, \Gamma_{2} \mathcal{P}^{-1} \bigg[ 3 \, {\Psi_s^*}^2 (\Gamma_{4} + \Gamma_{-2}) \mathcal{P}^{-1} \left( \Psi_s^3 |\Psi_s|^2   \right) + 12 \, |\Psi_s|^2 \Gamma_{2} \mathcal{P}^{-1} \left( \Psi_s |\Psi_s|^4  \right) \nonumber
\end{align}
\begin{align}
    &+ 6 \, \Psi_s^2 \Gamma_{2} \mathcal{P}^{-1} \left( \Psi_s^* |\Psi_s|^4  \right)  + 4 \, {\Psi_s^*}^2 |\Psi_s|^2 (\Gamma_{4} + \Gamma_{-2}) \mathcal{P}^{-1} \left( \Psi_s^3   \right)  \nonumber \\ 
    &+ 18 \, |\Psi_s|^4 \Gamma_{2} \mathcal{P}^{-1} \left( \Psi_s |\Psi_s|^2  \right)+ 12 \, \Psi_s^2 |\Psi_s|^2 \Gamma_{2} \mathcal{P}^{-1} \left( \Psi_s^* |\Psi_s|^2  \right) + \Psi_s^4 (\Gamma_{4} + \Gamma_{-2}) \mathcal{P}^{-1} \left( {\Psi_s^*}^3   \right) \bigg] \nonumber \\
    &+ {\Psi_s^*}^4 (\Gamma_{6} + \Gamma_{-4}) \mathcal{P}^{-1} \bigg[ 6 \, |\Psi_s|^2 (\Gamma_{6} + \Gamma_{-4}) \mathcal{P}^{-1} \left( \Psi_s^5 \right) + 15 \, \Psi_s^2 (\Gamma_{4} + \Gamma_{-2}) \mathcal{P}^{-1} \left( \Psi_s^3 |\Psi_s|^2   \right) \nonumber \\
    &+ 20 \, \Psi_s^2 |\Psi_s|^2 (\Gamma_{4} + \Gamma_{-2}) \mathcal{P}^{-1} \left( \Psi_s^3  \right) + 15 \, \Psi_s^4 \Gamma_{2} \mathcal{P}^{-1} \left( \Psi_s |\Psi_s|^2  \right) \bigg] \bigg)  \,,
\end{align}
\begin{align}
    \mathrm{F} {\scriptstyle (6,6)}_{0,(2)}^{(2)} =& \, \mathcal{P}^{-\frac{1}{2}} \, \bigg( \Psi_s^4 (\Gamma_{4} + \Gamma_{-2}) \mathcal{P}^{-1} \bigg[ 2 \, \Psi_s^2 |\Psi_s|^2 (\Gamma_{6} + \Gamma_{-4}) \mathcal{P}^{-1} \left( {\Psi_s^*}^5 \right) \nonumber \\
    &+ 15 \, |\Psi_s|^4 (\Gamma_{4} + \Gamma_{-2}) \mathcal{P}^{-1} \left( {\Psi_s^*}^3 |\Psi_s|^2   \right) + 20 \, {\Psi_s^*}^2 |\Psi_s|^2 \Gamma_{2} \mathcal{P}^{-1} \left( \Psi_s^* |\Psi_s|^4  \right) \nonumber \\
    &+ 5 \, {\Psi_s^*}^4 \Gamma_{2} \mathcal{P}^{-1} \left( \Psi_s |\Psi_s|^4  \right) \bigg] + 2 \, \Psi_s^2 |\Psi_s|^2 \, \Gamma_{2} \mathcal{P}^{-1} \bigg[ \Psi_s^4 (\Gamma_{6} + \Gamma_{-4}) \mathcal{P}^{-1} \left( {\Psi_s^*}^5  \right)  \nonumber \\ 
    &+ 20 \, \Psi_s^2 |\Psi_s|^2 (\Gamma_{4} + \Gamma_{-2}) \mathcal{P}^{-1} \left( {\Psi_s^*}^3 |\Psi_s|^2  \right) + 60 \, |\Psi_s|^4 \Gamma_{2} \mathcal{P}^{-1} \left( \Psi_s^* |\Psi_s|^4  \right) \nonumber \\ 
    &+ 40 \, {\Psi_s^*}^2 |\Psi_s|^2 \Gamma_{2} \mathcal{P}^{-1} \left( \Psi_s |\Psi_s|^4  \right) + 5 \, {\Psi_s^*}^4 (\Gamma_{4} + \Gamma_{-2}) \mathcal{P}^{-1} \left( \Psi_s^3 |\Psi_s|^2   \right) \bigg] \nonumber \\
    &+ 4 \, {\Psi_s^*}^2 |\Psi_s|^2 (\Gamma_{4} + \Gamma_{-2}) \mathcal{P}^{-1} \bigg[ 2 \, {\Psi_s^*}^2 |\Psi_s|^2 (\Gamma_{6} + \Gamma_{-4}) \mathcal{P}^{-1} \left( \Psi_s^5 \right) \nonumber \\
    &+ 15 \, |\Psi_s|^4 (\Gamma_{4} + \Gamma_{-2}) \mathcal{P}^{-1} \left( \Psi_s^3 |\Psi_s|^2   \right) + 20 \, \Psi_s^2 |\Psi_s|^2 \Gamma_{2} \mathcal{P}^{-1} \left( \Psi_s |\Psi_s|^4  \right) \nonumber \\
    &+ 5 \, \Psi_s^4 \Gamma_{2} \mathcal{P}^{-1} \left( \Psi_s^* |\Psi_s|^4  \right) \bigg] + 3 \, |\Psi_s|^4 \, \Gamma_{2} \mathcal{P}^{-1} \bigg[ {\Psi_s^*}^4 (\Gamma_{6} + \Gamma_{-4}) \mathcal{P}^{-1} \left( \Psi_s^5  \right)  \nonumber \\ 
    &+ 20 \, {\Psi_s^*}^2 |\Psi_s|^2 (\Gamma_{4} + \Gamma_{-2}) \mathcal{P}^{-1} \left( \Psi_s^3 |\Psi_s|^2  \right) + 60 \, |\Psi_s|^4 \Gamma_{2} \mathcal{P}^{-1} \left( \Psi_s |\Psi_s|^4  \right) \nonumber \\ 
    &+ 40 \, \Psi_s^2 |\Psi_s|^2 \Gamma_{2} \mathcal{P}^{-1} \left( \Psi_s^* |\Psi_s|^4  \right) + 5 \, \Psi_s^4 (\Gamma_{4} + \Gamma_{-2}) \mathcal{P}^{-1} \left( {\Psi_s^*}^3 |\Psi_s|^2   \right) \bigg] \nonumber \\
    &+ {\Psi_s^*}^4 (\Gamma_{6} + \Gamma_{-4}) \mathcal{P}^{-1} \bigg[ 3 \, |\Psi_s|^4 (\Gamma_{6} + \Gamma_{-4}) \mathcal{P}^{-1} \left( \Psi_s^5 \right) \nonumber \\ 
    &+ 10 \, \Psi_s^2 |\Psi_s|^2 (\Gamma_{4} + \Gamma_{-2}) \mathcal{P}^{-1} \left( \Psi_s^3 |\Psi_s|^2   \right) + 5 \, \Psi_s^4 \Gamma_{2} \mathcal{P}^{-1} \left( \Psi_s |\Psi_s|^4  \right) \bigg] \bigg)  \,.
\end{align}

%%%%%%%%%%%%%%%%%%%%%%%%%%%%%%%%%%%%%%%%%%%%%%%%%%%%%%%%%%%%%%%%%%%%%%%%%%%%%%%%

\section{3\textit{rd} order iteration expressions}
\label{App-D}

\begin{align}
    \mathrm{G} {\scriptstyle (4,4,4)}_{0,(1)}^{(3)} =& \, \mathcal{P}^{-\frac{1}{2}} \bigg[ 3 \left( \Gamma_{2} \mathcal{P}^{-1} \left( \Psi_s^* |\Psi_s|^2 \right) \right)^2 \times (\Gamma_{4} + \Gamma_{-2}) \mathcal{P}^{-1} \left( \Psi_s^3 \right) \nonumber \\
    &+ 2 \left| (\Gamma_{4} + \Gamma_{-2}) \mathcal{P}^{-1} \left( \Psi_s^3 \right) \right|^2 \times \Gamma_{2} \mathcal{P}^{-1} \left( \Psi_s |\Psi_s|^2 \right) \nonumber \\
    &+ 9 \left| \Gamma_{2} \mathcal{P}^{-1} \left( \Psi_s |\Psi_s|^2 \right) \right|^2 \times \Gamma_{2} \mathcal{P}^{-1} \left( \Psi_s |\Psi_s|^2 \right) \bigg]  \,,
\end{align}

\begin{align}
    \mathrm{G} {\scriptstyle (4,4, \textbf{\RNum{1}})}_{0,(1)(2)}^{(3)} =& \, \mathcal{P}^{-\frac{1}{2}} \bigg[ \Psi_s \bigg( 3 \,(\Gamma_{2})^2 \mathcal{P}^{-1} \left( 2 \, |\Psi_s|^2 \Dot{\Psi}_s + \Psi_s^2 \Dot{\Psi}_s^* \right) \times \Gamma_{2} \mathcal{P}^{-1} \left( \Psi_s^* |\Psi_s|^2 \right) \nonumber \\
    &- 3 \,(\Gamma_{2})^2 \mathcal{P}^{-1} \left( 2 \, |\Psi_s|^2 \Dot{\Psi}_s^* + {\Psi_s^*}^2 \Dot{\Psi}_s \right) \times \Gamma_{2} \mathcal{P}^{-1} \left( \Psi_s |\Psi_s|^2 \right) \nonumber \\
    &+ ((\Gamma_{4})^2 - (\Gamma_{-2})^2) \mathcal{P}^{-1} \left( \Psi_s^2 \Dot{\Psi}_s \right) \times (\Gamma_{4} + \Gamma_{-2}) \mathcal{P}^{-1} \left( {\Psi_s^*}^3 \right) \nonumber \\
    &- ((\Gamma_{4})^2 - (\Gamma_{-2})^2) \mathcal{P}^{-1} \left( {\Psi_s^*}^2 \Dot{\Psi}_s^* \right) \times (\Gamma_{4} + \Gamma_{-2}) \mathcal{P}^{-1} \left( \Psi_s^3 \right)  \bigg) \nonumber \\
    &+ \Psi_s^* \bigg( 3 \, ((\Gamma_{4})^2 - (\Gamma_{-2})^2) \mathcal{P}^{-1} \left( \Psi_s^2 \Dot{\Psi}_s \right) \times \Gamma_{2} \mathcal{P}^{-1} \left( \Psi_s^* |\Psi_s|^2 \right) \nonumber \\
    &- (\Gamma_{2})^2 \mathcal{P}^{-1} \left( 2 \, |\Psi_s|^2 \Dot{\Psi}_s^* + {\Psi_s^*}^2 \Dot{\Psi}_s \right) \times (\Gamma_{4} + \Gamma_{-2}) \mathcal{P}^{-1} \left( \Psi_s^3 \right) \nonumber \\
    &+ 3 \,(\Gamma_{2})^2 \mathcal{P}^{-1} \left( 2 \, |\Psi_s|^2 \Dot{\Psi}_s + \Psi_s^2 \Dot{\Psi}_s^* \right) \times \Gamma_{2} \mathcal{P}^{-1} \left( \Psi_s |\Psi_s|^2 \right) \bigg) \bigg]  \,,
\end{align}
\begin{align}
    \mathrm{G} {\scriptstyle (4,4,4)}_{0,(1)(2)}^{(3)} =& \, \mathcal{P}^{-\frac{1}{2}} \bigg[ \Psi_s \bigg( 3 \, \Gamma_{2} \mathcal{P}^{-1} \bigg( {\Psi_s^*}^2 (\Gamma_{4} + \Gamma_{-2}) \mathcal{P}^{-1} \left( \Psi_s^3 \right) + 6 \, |\Psi_s|^2 \Gamma_{2} \mathcal{P}^{-1} \left( \Psi_s |\Psi_s|^2  \right) \nonumber \\ 
    &+ 3 \, \Psi_s^2 \Gamma_{2} \mathcal{P}^{-1} \left( \Psi_s^* |\Psi_s|^2  \right) \bigg) \times \Gamma_{2} \mathcal{P}^{-1} \left( \Psi_s^* |\Psi_s|^2 \right) \nonumber \\
    &+ 3 \, \Gamma_{2} \mathcal{P}^{-1} \bigg( \Psi_s^2 (\Gamma_{4} + \Gamma_{-2}) \mathcal{P}^{-1} \left( {\Psi_s^*}^3 \right)+ 6 \, |\Psi_s|^2 \Gamma_{2} \mathcal{P}^{-1} \left( \Psi_s^* |\Psi_s|^2  \right) \nonumber \\ 
    & + 3 \, {\Psi_s^*}^2 \Gamma_{2} \mathcal{P}^{-1} \left( \Psi_s |\Psi_s|^2  \right) \bigg) \times \Gamma_{2} \mathcal{P}^{-1} \left( \Psi_s |\Psi_s|^2 \right) \nonumber \\ 
    &+ (\Gamma_{4} + \Gamma_{-2}) \mathcal{P}^{-1} \bigg( 2 \, |\Psi_s|^2 (\Gamma_{4} + \Gamma_{-2}) \mathcal{P}^{-1} \left( \Psi_s^3 \right) + 3 \, \Psi_s^2 \Gamma_{2} \mathcal{P}^{-1} \left( \Psi_s |\Psi_s|^2  \right) \bigg) \nonumber \\
    &\times (\Gamma_{4} + \Gamma_{-2}) \mathcal{P}^{-1} \left( {\Psi_s^*}^3 \right) + (\Gamma_{4} + \Gamma_{-2}) \mathcal{P}^{-1} \bigg( 2 \, |\Psi_s|^2 (\Gamma_{4} + \Gamma_{-2}) \mathcal{P}^{-1} \left( {\Psi_s^*}^3 \right) \nonumber \\ 
    &+ 3 \, {\Psi_s^*}^2 \Gamma_{2} \mathcal{P}^{-1} \left( \Psi_s^* |\Psi_s|^2  \right) \bigg) \times (\Gamma_{4} + \Gamma_{-2}) \mathcal{P}^{-1} \left( \Psi_s^3 \right) \bigg) \nonumber \\ 
    &+ \Psi_s^* \bigg( 3 \,(\Gamma_{4} + \Gamma_{-2}) \mathcal{P}^{-1} \bigg( 2 \, |\Psi_s|^2 (\Gamma_{4} + \Gamma_{-2}) \mathcal{P}^{-1} \left( \Psi_s^3 \right)  \nonumber \\
    &+ 3 \, \Psi_s^2 \Gamma_{2} \mathcal{P}^{-1} \left( \Psi_s |\Psi_s|^2  \right) \bigg)\times \Gamma_{2} \mathcal{P}^{-1} \left( \Psi_s^* |\Psi_s|^2 \right) + \Gamma_{2} \mathcal{P}^{-1} \bigg( \Psi_s^2 (\Gamma_{4}\nonumber \\
    &+ \Gamma_{-2}) \mathcal{P}^{-1} \left( {\Psi_s^*}^3 \right) + 6 \, |\Psi_s|^2 \Gamma_{2} \mathcal{P}^{-1} \left( \Psi_s^* |\Psi_s|^2  \right) \nonumber \\ 
    &+ 3 \, {\Psi_s^*}^2 \Gamma_{2} \mathcal{P}^{-1} \left( \Psi_s |\Psi_s|^2  \right) \bigg) \times (\Gamma_{4} + \Gamma_{-2}) \left( \Psi_s^3 \right) \nonumber\\
    &+ 3 \, \Gamma_{2} \mathcal{P}^{-1} \bigg( {\Psi_s^*}^2 (\Gamma_{4} + \Gamma_{-2}) \mathcal{P}^{-1} \left( \Psi_s^3 \right)+ 6 \, |\Psi_s|^2 \Gamma_{2} \mathcal{P}^{-1} \left( \Psi_s |\Psi_s|^2  \right) \nonumber \\ 
    & + 3 \, \Psi_s^2 \Gamma_{2} \mathcal{P}^{-1} \left( \Psi_s^* |\Psi_s|^2  \right) \bigg) \times \Gamma_{2} \mathcal{P}^{-1} \left( \Psi_s |\Psi_s|^2 \right) \bigg) \bigg]  \,,
\end{align}

\begin{align}
    \mathrm{G} {\scriptstyle (4, \textbf{\RNum{2}})}_{0,(3)}^{(3)} =& \, \mathcal{P}^{-\frac{1}{2}} \bigg[ \Psi_s^2 (\Gamma_{2})^3 \mathcal{P}^{-1} \left( 2 \, |\Psi_s|^2 \Ddot{\Psi}_s^* + {\Psi_s^*}^2 \Ddot{\Psi}_s \right)  \nonumber \\
    &+ 2  \, |\Psi_s|^2 (\Gamma_{2})^3 \mathcal{P}^{-1} \left( 2 \, |\Psi_s|^2 \Ddot{\Psi}_s + \Psi_s^2 \Ddot{\Psi}_s^* \right)\nonumber \\
    &+ {\Psi_s^*}^2 ((\Gamma_{4})^3 +(\Gamma_{-2})^3) \mathcal{P}^{-1} \left( \Psi_s^2 \Ddot{\Psi}_s \right) \bigg]  \,,
\end{align}
\begin{align}
    \mathrm{G} {\scriptstyle (4, \textbf{\RNum{1}}, \textbf{\RNum{1}})}_{0,(3)}^{(3)} =& \, \mathcal{P}^{-\frac{1}{2}} \bigg[ \Psi_s^2 (\Gamma_{2})^3 \mathcal{P}^{-1} \left( \Psi_s \left( \Dot{\Psi}_s^* \right)^2 + 2 \, \Psi_s^* \left| \Dot{\Psi}_s \right|^2 \right) \nonumber \\
    &+ 2  \, |\Psi_s|^2 (\Gamma_{2})^3 \mathcal{P}^{-1} \left( \Psi_s^* \left( \Dot{\Psi}_s \right)^2 + 2 \, \Psi_s \left| \Dot{\Psi}_s \right|^2 \right)\nonumber \\
    &+ {\Psi_s^*}^2 ((\Gamma_{4})^3 +(\Gamma_{-2})^3) \mathcal{P}^{-1} \left( \Psi_s \left( \Dot{\Psi}_s \right)^2 \right) \bigg]  \,,
\end{align}
\begin{align}
    \mathrm{G} {\scriptstyle (4,4, \textbf{\RNum{1}})}_{0,(3)}^{(3)} =& \, \mathcal{P}^{-\frac{1}{2}} \bigg[ - \Psi_s^2 \bigg[ (\Gamma_{2})^2 \mathcal{P}^{-1} \bigg( 2 \, \Psi_s \Dot{\Psi}_s (\Gamma_{4} + \Gamma_{-2}) \mathcal{P}^{-1} \left( {\Psi_s^*}^3 \right)\nonumber \\
    &+ 3 \, \Psi_s^2 (\Gamma_{4} + \Gamma_{-2}) \mathcal{P}^{-1} \left( {\Psi_s^*}^2 \Dot{\Psi}_s^* \right)+ 6 \, \Psi_s \Dot{\Psi}_s^* \Gamma_{2} \mathcal{P}^{-1} \left( \Psi_s^* |\Psi_s|^2  \right) \nonumber \\
    & + 6 \, \Psi_s^* \Dot{\Psi}_s \Gamma_{2} \mathcal{P}^{-1} \left( \Psi_s^* |\Psi_s|^2  \right) + 12 \, |\Psi_s|^2 \Gamma_{2} \mathcal{P}^{-1} \left( \Dot{\Psi}_s^* |\Psi_s|^2  \right) \nonumber \\
    &+ 6 \, |\Psi_s|^2 \Gamma_{2} \mathcal{P}^{-1} \left( {\Psi_s^*}^2 \Dot{\Psi}_s  \right) + 6 \, \Psi_s^* \Dot{\Psi}_s^* \Gamma_{2} \mathcal{P}^{-1} \left( \Psi_s |\Psi_s|^2  \right)  \nonumber \\
    &+ 6 \, {\Psi_s^*}^2 \Gamma_{2} \mathcal{P}^{-1} \left( |\Psi_s|^2 \Dot{\Psi}_s \right)+ 3 \, {\Psi_s^*}^2 \Gamma_{2} \mathcal{P}^{-1} \left( \Psi_s^2 \Dot{\Psi}_s^*  \right) \bigg)\nonumber \\
    &+ 3 \, \Gamma_2 \mathcal{P}^{-1} \bigg( \Psi_s^2 ((\Gamma_{4})^2 - (\Gamma_{-2})^2) \mathcal{P}^{-1} \left( {\Psi_s^*}^2 \Dot{\Psi}_s^* \right) \nonumber \\ 
    &+ 2  \, |\Psi_s|^2 (\Gamma_{2})^2 \mathcal{P}^{-1} \left( 2 \, |\Psi_s|^2 \Dot{\Psi}_s^* + {\Psi_s^*}^2 \Dot{\Psi}_s \right) \nonumber \\
    &- {\Psi_s^*}^2 (\Gamma_{2})^2 \mathcal{P}^{-1} \left( 2 \, |\Psi_s|^2 \Dot{\Psi}_s + \Psi_s^2 \Dot{\Psi}_s^* \right) \bigg) \bigg] \nonumber \\ 
    &+ 2  \, |\Psi_s|^2 \bigg[ (\Gamma_{2})^2 \mathcal{P}^{-1} \bigg( 2 \, \Psi_s^* \Dot{\Psi}_s^* (\Gamma_{4} + \Gamma_{-2}) \mathcal{P}^{-1} \left( \Psi_s^3 \right)\nonumber \\
    &+ 3 \, {\Psi_s^*}^2 (\Gamma_{4} + \Gamma_{-2}) \mathcal{P}^{-1} \left( \Psi_s^2 \Dot{\Psi}_s \right)+ 6 \, \Psi_s^* \Dot{\Psi}_s \Gamma_{2} \mathcal{P}^{-1} \left( \Psi_s |\Psi_s|^2  \right) \nonumber \\
    & + 6 \, \Psi_s \Dot{\Psi}_s^* \Gamma_{2} \mathcal{P}^{-1} \left( \Psi_s |\Psi_s|^2  \right) + 12 \, |\Psi_s|^2 \Gamma_{2} \mathcal{P}^{-1} \left( \Dot{\Psi}_s |\Psi_s|^2  \right) \nonumber \\
    &+ 6 \, |\Psi_s|^2 \Gamma_{2} \mathcal{P}^{-1} \left( \Psi_s^2 \Dot{\Psi}_s^*  \right) + 6 \, \Psi_s \Dot{\Psi}_s \Gamma_{2} \mathcal{P}^{-1} \left( \Psi_s^* |\Psi_s|^2  \right)  \nonumber \\
    &+ 6 \, \Psi_s^2 \Gamma_{2} \mathcal{P}^{-1} \left( |\Psi_s|^2 \Dot{\Psi}_s^* \right)+ 3 \, \Psi_s^2 \Gamma_{2} \mathcal{P}^{-1} \left( {\Psi_s^*}^2 \Dot{\Psi}_s  \right) \bigg) \nonumber \\
    &+ 3 \, \Gamma_2 \mathcal{P}^{-1} \bigg( {\Psi_s^*}^2 ((\Gamma_{4})^2 - (\Gamma_{-2})^2) \mathcal{P}^{-1} \left( \Psi_s^2 \Dot{\Psi}_s \right) \nonumber 
\end{align}
\begin{align}
    &+ 2  \, |\Psi_s|^2 (\Gamma_{2})^2 \mathcal{P}^{-1} \left( 2 \, |\Psi_s|^2 \Dot{\Psi}_s + \Psi_s^2 \Dot{\Psi}_s^* \right) - \Psi_s^2 (\Gamma_{2})^2 \mathcal{P}^{-1} \left( 2 \, |\Psi_s|^2 \Dot{\Psi}_s^*  + {\Psi_s^*}^2 \Dot{\Psi}_s \right) \bigg) \bigg] \nonumber \\ 
    &+ {\Psi_s^*}^2 \bigg[ ((\Gamma_{4})^2 - (\Gamma_{-2})^2) \mathcal{P}^{-1} \bigg( 2 \, \Psi_s \Dot{\Psi}_s^* (\Gamma_{4} + \Gamma_{-2}) \mathcal{P}^{-1} \left( \Psi_s^3 \right) \nonumber \\ 
    &+ 2 \, \Psi_s^* \Dot{\Psi}_s (\Gamma_{4} + \Gamma_{-2}) \mathcal{P}^{-1} \left( \Psi_s^3 \right) + 6 \, |\Psi_s|^2 (\Gamma_{4} + \Gamma_{-2}) \mathcal{P}^{-1} \left( \Psi_s^2 \Dot{\Psi}_s \right)\nonumber \\
    &+ 6 \, \Psi_s \Dot{\Psi}_s \Gamma_{2} \mathcal{P}^{-1} \left( \Psi_s |\Psi_s|^2  \right) + 6 \, \Psi_s^2 \Gamma_{2} \mathcal{P}^{-1} \left( |\Psi_s|^2 \Dot{\Psi}_s \right) + 3 \, \Psi_s^2 \Gamma_{2} \mathcal{P}^{-1} \left( \Psi_s^2 \Dot{\Psi}_s^*  \right) \bigg) \nonumber \\
    &+ 3 (\Gamma_{4} + \Gamma_{-2}) \mathcal{P}^{-1} \bigg( \Psi_s^2 (\Gamma_{2})^2 \mathcal{P}^{-1} \left( 2 \, |\Psi_s|^2 \Dot{\Psi}_s + \Psi_s^2 \Dot{\Psi}_s^* \right) \nonumber \\
    &+ 2  \, |\Psi_s|^2 ((\Gamma_{4})^2 - (\Gamma_{-2})^2) \mathcal{P}^{-1} \left( \Psi_s^2 \Dot{\Psi}_s \right) \bigg) \bigg] \bigg] \,,
\end{align}
\begin{align}
    \mathrm{G} {\scriptstyle (4,4,4)}_{0,(3)}^{(3)} =& \, \mathcal{P}^{-\frac{1}{2}} \bigg[ \Psi_s^2 \Gamma_2 \mathcal{P}^{-1} \bigg[ 3 \, \Psi_s \Big( 3 \left( \Gamma_{2} \mathcal{P}^{-1} \left( \Psi_s^* |\Psi_s|^2 \right) \right)^2 \nonumber \\
    &+ 2 \, \Gamma_{2} \mathcal{P}^{-1} \left( \Psi_s |\Psi_s|^2 \right) \times (\Gamma_{4} + \Gamma_{-2}) \mathcal{P}^{-1} \left( {\Psi_s^*}^3 \right) \Big) \nonumber \\
    &+ 2 \, \Psi_s^* \Big( 9 \left| \Gamma_{2} \mathcal{P}^{-1} \left( \Psi_s |\Psi_s|^2 \right)  \right|^2+ \left| (\Gamma_{4} + \Gamma_{-2}) \mathcal{P}^{-1} \left( \Psi_s^3 \right) \right|^2 \Big) \nonumber \\
    & + 3 \, \Psi_s^2 (\Gamma_{4} + \Gamma_{-2}) \mathcal{P}^{-1} \bigg( 2 \, |\Psi_s|^2 (\Gamma_{4} + \Gamma_{-2}) \mathcal{P}^{-1} \left( {\Psi_s^*}^3 \right) \nonumber \\
    &+ 3 \, {\Psi_s^*}^2 \Gamma_{2} \mathcal{P}^{-1} \left( \Psi_s^* |\Psi_s|^2  \right) \bigg) + 6  \, |\Psi_s|^2 \Gamma_{2} \mathcal{P}^{-1} \bigg( \Psi_s^2 (\Gamma_{4} + \Gamma_{-2}) \mathcal{P}^{-1} \left( {\Psi_s^*}^3 \right) \nonumber \\
    &+ 6 \, |\Psi_s|^2 \Gamma_{2} \mathcal{P}^{-1} \left( \Psi_s^* |\Psi_s|^2  \right) + 3 \, {\Psi_s^*}^2 \Gamma_{2} \mathcal{P}^{-1} \left( \Psi_s |\Psi_s|^2  \right) \bigg) \nonumber \\ 
    &+ 3 \, {\Psi_s^*}^2 \Gamma_{2} \mathcal{P}^{-1} \bigg( {\Psi_s^*}^2 (\Gamma_{4} + \Gamma_{-2}) \mathcal{P}^{-1} \left( \Psi_s^3 \right) + 6 \, |\Psi_s|^2 \Gamma_{2} \mathcal{P}^{-1} \left( \Psi_s |\Psi_s|^2  \right) \nonumber \\ 
    &+ 3 \, \Psi_s^2 \Gamma_{2} \mathcal{P}^{-1} \left( \Psi_s^* |\Psi_s|^2  \right) \bigg) \bigg] + 2  \, |\Psi_s|^2  \Gamma_2 \mathcal{P}^{-1} \bigg[ 3 \, \Psi_s^* \Big( 3 \left( \Gamma_{2} \mathcal{P}^{-1} \left( \Psi_s |\Psi_s|^2 \right) \right)^2 \nonumber \\ 
    &+ 2 \, \Gamma_{2} \mathcal{P}^{-1} \left( \Psi_s^* |\Psi_s|^2 \right) \times (\Gamma_{4} + \Gamma_{-2}) \mathcal{P}^{-1} \left( \Psi_s^3 \right) \Big)  \nonumber
\end{align}
\begin{align}    
    &+ 2 \, \Psi_s \Big( 9 \left| \Gamma_{2} \mathcal{P}^{-1} \left( \Psi_s |\Psi_s|^2 \right)  \right|^2+ \left| (\Gamma_{4} + \Gamma_{-2}) \mathcal{P}^{-1} \left( \Psi_s^3 \right) \right|^2 \Big) \nonumber \\
    &+ 3 \, {\Psi_s^*}^2 (\Gamma_{4} + \Gamma_{-2}) \mathcal{P}^{-1} \bigg( 2 \, |\Psi_s|^2 (\Gamma_{4} + \Gamma_{-2}) \mathcal{P}^{-1} \left( \Psi_s^3 \right) \nonumber \\
    &+ 3 \, \Psi_s^2 \Gamma_{2} \mathcal{P}^{-1} \left( \Psi_s |\Psi_s|^2  \right) \bigg) + 6  \, |\Psi_s|^2 \Gamma_{2} \mathcal{P}^{-1} \bigg( {\Psi_s^*}^2 (\Gamma_{4} + \Gamma_{-2}) \mathcal{P}^{-1} \left( \Psi_s^3 \right) \nonumber \\
    &+ 6 \, |\Psi_s|^2 \Gamma_{2} \mathcal{P}^{-1} \left( \Psi_s |\Psi_s|^2  \right) + 3 \, \Psi_s^2 \Gamma_{2} \mathcal{P}^{-1} \left( \Psi_s^* |\Psi_s|^2  \right) \bigg) \nonumber \\
    &+ 3 \, \Psi_s^2 \Gamma_{2} \mathcal{P}^{-1} \bigg( \Psi_s^2 (\Gamma_{4} + \Gamma_{-2}) \mathcal{P}^{-1} \left( {\Psi_s^*}^3 \right) + 6 \, |\Psi_s|^2 \Gamma_{2} \mathcal{P}^{-1} \left( \Psi_s^* |\Psi_s|^2  \right) \nonumber \\ 
    &+ 3 \, {\Psi_s^*}^2 \Gamma_{2} \mathcal{P}^{-1} \left( \Psi_s |\Psi_s|^2  \right) \bigg) \bigg] \nonumber \\ 
    &+ 3 \, {\Psi_s^*}^2 (\Gamma_{4} + \Gamma_{-2}) \mathcal{P}^{-1} \bigg[ \Psi_s \Big( 2 \, \Gamma_{2} \mathcal{P}^{-1} \left( \Psi_s^* |\Psi_s|^2 \right) \times (\Gamma_{4} + \Gamma_{-2}) \mathcal{P}^{-1} \left( \Psi_s^3 \right) \nonumber \\
    &+ 3 \left( \Gamma_{2} \mathcal{P}^{-1} \left( \Psi_s |\Psi_s|^2 \right) \right)^2 \Big) + 2 \, \Psi_s^* \Gamma_{2} \mathcal{P}^{-1} \left( \Psi_s |\Psi_s|^2 \right) \times (\Gamma_{4} + \Gamma_{-2}) \mathcal{P}^{-1} \left( \Psi_s^3 \right) \nonumber \\
    &+ \Psi_s^2 \Gamma_{2} \mathcal{P}^{-1} \bigg( {\Psi_s^*}^2 (\Gamma_{4} + \Gamma_{-2}) \mathcal{P}^{-1} \left( \Psi_s^3 \right) + 6 \, |\Psi_s|^2 \Gamma_{2} \mathcal{P}^{-1} \left( \Psi_s |\Psi_s|^2  \right) \nonumber \\ 
    &+ 3 \, \Psi_s^2 \Gamma_{2} \mathcal{P}^{-1} \left( \Psi_s^* |\Psi_s|^2  \right) \bigg) + 2  \, |\Psi_s|^2 (\Gamma_{4} + \Gamma_{-2}) \mathcal{P}^{-1} \bigg( 2 \, |\Psi_s|^2 (\Gamma_{4} + \Gamma_{-2}) \mathcal{P}^{-1} \left( \Psi_s^3 \right) \nonumber \\ 
    &+ 3 \, \Psi_s^2 \Gamma_{2} \mathcal{P}^{-1} \left( \Psi_s |\Psi_s|^2  \right) \bigg) \bigg] \bigg] \,.
\end{align}

%%%%%%%%%%%%%%%%%%%%%%%%%%%%%%%%%%%%%%%%%%%%%%%%%%%%%%%%%%%%%%%%%%%%%%%%%%%%%%%%
%%%%%%%%%%%%%%%%%%%%%%%%%%%%%%%%%%%%%%%%%%%%%%%%%%%%%%%%%%%%%%%%%%%%%%%%%%%%%%%%

\bibliography{PrintDarkEFTHO}

\begin{thebibliography}{29}%
\makeatletter
\providecommand \@ifxundefined [1]{%
 \@ifx{#1\undefined}
}%
\providecommand \@ifnum [1]{%
 \ifnum #1\expandafter \@firstoftwo
 \else \expandafter \@secondoftwo
 \fi
}%
\providecommand \@ifx [1]{%
 \ifx #1\expandafter \@firstoftwo
 \else \expandafter \@secondoftwo
 \fi
}%
\providecommand \natexlab [1]{#1}%
\providecommand \enquote  [1]{``#1''}%
\providecommand \bibnamefont  [1]{#1}%
\providecommand \bibfnamefont [1]{#1}%
\providecommand \citenamefont [1]{#1}%
\providecommand \href@noop [0]{\@secondoftwo}%
\providecommand \href [0]{\begingroup \@sanitize@url \@href}%
\providecommand \@href[1]{\@@startlink{#1}\@@href}%
\providecommand \@@href[1]{\endgroup#1\@@endlink}%
\providecommand \@sanitize@url [0]{\catcode `\\12\catcode `\$12\catcode
  `\&12\catcode `\#12\catcode `\^12\catcode `\_12\catcode `\%12\relax}%
\providecommand \@@startlink[1]{}%
\providecommand \@@endlink[0]{}%
\providecommand \url  [0]{\begingroup\@sanitize@url \@url }%
\providecommand \@url [1]{\endgroup\@href {#1}{\urlprefix }}%
\providecommand \urlprefix  [0]{URL }%
\providecommand \Eprint [0]{\href }%
\providecommand \doibase [0]{http://dx.doi.org/}%
\providecommand \selectlanguage [0]{\@gobble}%
\providecommand \bibinfo  [0]{\@secondoftwo}%
\providecommand \bibfield  [0]{\@secondoftwo}%
\providecommand \translation [1]{[#1]}%
\providecommand \BibitemOpen [0]{}%
\providecommand \bibitemStop [0]{}%
\providecommand \bibitemNoStop [0]{.\EOS\space}%
\providecommand \EOS [0]{\spacefactor3000\relax}%
\providecommand \BibitemShut  [1]{\csname bibitem#1\endcsname}%
\let\auto@bib@innerbib\@empty
%</preamble>
\bibitem [{\citenamefont {Workman}\ \emph {et~al.}(2022)\citenamefont {Workman}
  \emph {et~al.}}]{ParticleDataGroup:2022pth}%
  \BibitemOpen
  \bibfield  {author} {\bibinfo {author} {\bibfnamefont {R.~L.}\ \bibnamefont
  {Workman}} \emph {et~al.} (\bibinfo {collaboration} {Particle Data Group}),\
  }\href {\doibase 10.1093/ptep/ptac097} {\bibfield  {journal} {\bibinfo
  {journal} {PTEP}\ }\textbf {\bibinfo {volume} {2022}},\ \bibinfo {pages}
  {083C01} (\bibinfo {year} {2022})}\BibitemShut {NoStop}%
\bibitem [{\citenamefont {Kim}\ and\ \citenamefont
  {Carosi}(2010)}]{Kim:2008hd}%
  \BibitemOpen
  \bibfield  {author} {\bibinfo {author} {\bibfnamefont {J.~E.}\ \bibnamefont
  {Kim}}\ and\ \bibinfo {author} {\bibfnamefont {G.}~\bibnamefont {Carosi}},\
  }\href {\doibase 10.1103/RevModPhys.82.557} {\bibfield  {journal} {\bibinfo
  {journal} {Rev. Mod. Phys.}\ }\textbf {\bibinfo {volume} {82}},\ \bibinfo
  {pages} {557} (\bibinfo {year} {2010})},\ \bibinfo {note} {[Erratum:
  Rev.Mod.Phys. 91, 049902 (2019)]},\ \Eprint {http://arxiv.org/abs/0807.3125}
  {arXiv:0807.3125 [hep-ph]} \BibitemShut {NoStop}%
\bibitem [{\citenamefont {Peccei}\ and\ \citenamefont
  {Quinn}(1977{\natexlab{a}})}]{Peccei:1977hh}%
  \BibitemOpen
  \bibfield  {author} {\bibinfo {author} {\bibfnamefont {R.~D.}\ \bibnamefont
  {Peccei}}\ and\ \bibinfo {author} {\bibfnamefont {H.~R.}\ \bibnamefont
  {Quinn}},\ }\href {\doibase 10.1103/PhysRevLett.38.1440} {\bibfield
  {journal} {\bibinfo  {journal} {Phys. Rev. Lett.}\ }\textbf {\bibinfo
  {volume} {38}},\ \bibinfo {pages} {1440} (\bibinfo {year}
  {1977}{\natexlab{a}})}\BibitemShut {NoStop}%
\bibitem [{\citenamefont {Weinberg}(1978)}]{Weinberg:1977ma}%
  \BibitemOpen
  \bibfield  {author} {\bibinfo {author} {\bibfnamefont {S.}~\bibnamefont
  {Weinberg}},\ }\href {\doibase 10.1103/PhysRevLett.40.223} {\bibfield
  {journal} {\bibinfo  {journal} {Phys. Rev. Lett.}\ }\textbf {\bibinfo
  {volume} {40}},\ \bibinfo {pages} {223} (\bibinfo {year} {1978})}\BibitemShut
  {NoStop}%
\bibitem [{\citenamefont {Wilczek}(1978)}]{Wilczek:1977pj}%
  \BibitemOpen
  \bibfield  {author} {\bibinfo {author} {\bibfnamefont {F.}~\bibnamefont
  {Wilczek}},\ }\href {\doibase 10.1103/PhysRevLett.40.279} {\bibfield
  {journal} {\bibinfo  {journal} {Phys. Rev. Lett.}\ }\textbf {\bibinfo
  {volume} {40}},\ \bibinfo {pages} {279} (\bibinfo {year} {1978})}\BibitemShut
  {NoStop}%
\bibitem [{\citenamefont {Davis}(1986)}]{Davis:1986xc}%
  \BibitemOpen
  \bibfield  {author} {\bibinfo {author} {\bibfnamefont {R.~L.}\ \bibnamefont
  {Davis}},\ }\href {\doibase 10.1016/0370-2693(86)90300-X} {\bibfield
  {journal} {\bibinfo  {journal} {Phys. Lett. B}\ }\textbf {\bibinfo {volume}
  {180}},\ \bibinfo {pages} {225} (\bibinfo {year} {1986})}\BibitemShut
  {NoStop}%
\bibitem [{\citenamefont {Harari}\ and\ \citenamefont
  {Sikivie}(1987)}]{Harari:1987ht}%
  \BibitemOpen
  \bibfield  {author} {\bibinfo {author} {\bibfnamefont {D.}~\bibnamefont
  {Harari}}\ and\ \bibinfo {author} {\bibfnamefont {P.}~\bibnamefont
  {Sikivie}},\ }\href {\doibase 10.1016/0370-2693(87)90032-3} {\bibfield
  {journal} {\bibinfo  {journal} {Phys. Lett. B}\ }\textbf {\bibinfo {volume}
  {195}},\ \bibinfo {pages} {361} (\bibinfo {year} {1987})}\BibitemShut
  {NoStop}%
\bibitem [{\citenamefont {Preskill}\ \emph {et~al.}(1983)\citenamefont
  {Preskill}, \citenamefont {Wise},\ and\ \citenamefont
  {Wilczek}}]{Preskill:1982cy}%
  \BibitemOpen
  \bibfield  {author} {\bibinfo {author} {\bibfnamefont {J.}~\bibnamefont
  {Preskill}}, \bibinfo {author} {\bibfnamefont {M.~B.}\ \bibnamefont {Wise}},
  \ and\ \bibinfo {author} {\bibfnamefont {F.}~\bibnamefont {Wilczek}},\ }\href
  {\doibase 10.1016/0370-2693(83)90637-8} {\bibfield  {journal} {\bibinfo
  {journal} {Phys. Lett. B}\ }\textbf {\bibinfo {volume} {120}},\ \bibinfo
  {pages} {127} (\bibinfo {year} {1983})}\BibitemShut {NoStop}%
\bibitem [{\citenamefont {Abbott}\ and\ \citenamefont
  {Sikivie}(1983)}]{Abbott:1982af}%
  \BibitemOpen
  \bibfield  {author} {\bibinfo {author} {\bibfnamefont {L.~F.}\ \bibnamefont
  {Abbott}}\ and\ \bibinfo {author} {\bibfnamefont {P.}~\bibnamefont
  {Sikivie}},\ }\href {\doibase 10.1016/0370-2693(83)90638-X} {\bibfield
  {journal} {\bibinfo  {journal} {Phys. Lett. B}\ }\textbf {\bibinfo {volume}
  {120}},\ \bibinfo {pages} {133} (\bibinfo {year} {1983})}\BibitemShut
  {NoStop}%
\bibitem [{\citenamefont {Dine}\ and\ \citenamefont
  {Fischler}(1983)}]{Dine:1982ah}%
  \BibitemOpen
  \bibfield  {author} {\bibinfo {author} {\bibfnamefont {M.}~\bibnamefont
  {Dine}}\ and\ \bibinfo {author} {\bibfnamefont {W.}~\bibnamefont
  {Fischler}},\ }\href {\doibase 10.1016/0370-2693(83)90639-1} {\bibfield
  {journal} {\bibinfo  {journal} {Phys. Lett. B}\ }\textbf {\bibinfo {volume}
  {120}},\ \bibinfo {pages} {137} (\bibinfo {year} {1983})}\BibitemShut
  {NoStop}%
\bibitem [{\citenamefont {Sikivie}\ and\ \citenamefont
  {Yang}(2009)}]{Sikivie:2009qn}%
  \BibitemOpen
  \bibfield  {author} {\bibinfo {author} {\bibfnamefont {P.}~\bibnamefont
  {Sikivie}}\ and\ \bibinfo {author} {\bibfnamefont {Q.}~\bibnamefont {Yang}},\
  }\href {\doibase 10.1103/PhysRevLett.103.111301} {\bibfield  {journal}
  {\bibinfo  {journal} {Phys. Rev. Lett.}\ }\textbf {\bibinfo {volume} {103}},\
  \bibinfo {pages} {111301} (\bibinfo {year} {2009})},\ \Eprint
  {http://arxiv.org/abs/0901.1106} {arXiv:0901.1106 [hep-ph]} \BibitemShut
  {NoStop}%
\bibitem [{\citenamefont {Erken}\ \emph {et~al.}(2012)\citenamefont {Erken},
  \citenamefont {Sikivie}, \citenamefont {Tam},\ and\ \citenamefont
  {Yang}}]{Erken:2011dz}%
  \BibitemOpen
  \bibfield  {author} {\bibinfo {author} {\bibfnamefont {O.}~\bibnamefont
  {Erken}}, \bibinfo {author} {\bibfnamefont {P.}~\bibnamefont {Sikivie}},
  \bibinfo {author} {\bibfnamefont {H.}~\bibnamefont {Tam}}, \ and\ \bibinfo
  {author} {\bibfnamefont {Q.}~\bibnamefont {Yang}},\ }\href {\doibase
  10.1103/PhysRevD.85.063520} {\bibfield  {journal} {\bibinfo  {journal} {Phys.
  Rev. D}\ }\textbf {\bibinfo {volume} {85}},\ \bibinfo {pages} {063520}
  (\bibinfo {year} {2012})},\ \Eprint {http://arxiv.org/abs/1111.1157}
  {arXiv:1111.1157 [astro-ph.CO]} \BibitemShut {NoStop}%
\bibitem [{\citenamefont {Saikawa}\ and\ \citenamefont
  {Yamaguchi}(2013)}]{Saikawa:2012uk}%
  \BibitemOpen
  \bibfield  {author} {\bibinfo {author} {\bibfnamefont {K.}~\bibnamefont
  {Saikawa}}\ and\ \bibinfo {author} {\bibfnamefont {M.}~\bibnamefont
  {Yamaguchi}},\ }\href {\doibase 10.1103/PhysRevD.87.085010} {\bibfield
  {journal} {\bibinfo  {journal} {Phys. Rev. D}\ }\textbf {\bibinfo {volume}
  {87}},\ \bibinfo {pages} {085010} (\bibinfo {year} {2013})},\ \Eprint
  {http://arxiv.org/abs/1210.7080} {arXiv:1210.7080 [hep-ph]} \BibitemShut
  {NoStop}%
\bibitem [{\citenamefont {Davidson}\ and\ \citenamefont
  {Elmer}(2013)}]{Davidson:2013aba}%
  \BibitemOpen
  \bibfield  {author} {\bibinfo {author} {\bibfnamefont {S.}~\bibnamefont
  {Davidson}}\ and\ \bibinfo {author} {\bibfnamefont {M.}~\bibnamefont
  {Elmer}},\ }\href {\doibase 10.1088/1475-7516/2013/12/034} {\bibfield
  {journal} {\bibinfo  {journal} {JCAP}\ }\textbf {\bibinfo {volume} {12}},\
  \bibinfo {pages} {034} (\bibinfo {year} {2013})},\ \Eprint
  {http://arxiv.org/abs/1307.8024} {arXiv:1307.8024 [hep-ph]} \BibitemShut
  {NoStop}%
\bibitem [{\citenamefont {Noumi}\ \emph {et~al.}(2014)\citenamefont {Noumi},
  \citenamefont {Saikawa}, \citenamefont {Sato},\ and\ \citenamefont
  {Yamaguchi}}]{Noumi:2013zga}%
  \BibitemOpen
  \bibfield  {author} {\bibinfo {author} {\bibfnamefont {T.}~\bibnamefont
  {Noumi}}, \bibinfo {author} {\bibfnamefont {K.}~\bibnamefont {Saikawa}},
  \bibinfo {author} {\bibfnamefont {R.}~\bibnamefont {Sato}}, \ and\ \bibinfo
  {author} {\bibfnamefont {M.}~\bibnamefont {Yamaguchi}},\ }\href {\doibase
  10.1103/PhysRevD.89.065012} {\bibfield  {journal} {\bibinfo  {journal} {Phys.
  Rev. D}\ }\textbf {\bibinfo {volume} {89}},\ \bibinfo {pages} {065012}
  (\bibinfo {year} {2014})},\ \Eprint {http://arxiv.org/abs/1310.0167}
  {arXiv:1310.0167 [hep-ph]} \BibitemShut {NoStop}%
\bibitem [{\citenamefont {Davidson}(2015)}]{Davidson:2014hfa}%
  \BibitemOpen
  \bibfield  {author} {\bibinfo {author} {\bibfnamefont {S.}~\bibnamefont
  {Davidson}},\ }\href {\doibase 10.1016/j.astropartphys.2014.12.007}
  {\bibfield  {journal} {\bibinfo  {journal} {Astropart. Phys.}\ }\textbf
  {\bibinfo {volume} {65}},\ \bibinfo {pages} {101} (\bibinfo {year} {2015})},\
  \Eprint {http://arxiv.org/abs/1405.1139} {arXiv:1405.1139 [hep-ph]}
  \BibitemShut {NoStop}%
\bibitem [{\citenamefont {Braaten}\ \emph {et~al.}(2016)\citenamefont
  {Braaten}, \citenamefont {Mohapatra},\ and\ \citenamefont
  {Zhang}}]{Braaten:2016kzc}%
  \BibitemOpen
  \bibfield  {author} {\bibinfo {author} {\bibfnamefont {E.}~\bibnamefont
  {Braaten}}, \bibinfo {author} {\bibfnamefont {A.}~\bibnamefont {Mohapatra}},
  \ and\ \bibinfo {author} {\bibfnamefont {H.}~\bibnamefont {Zhang}},\ }\href
  {\doibase 10.1103/PhysRevD.94.076004} {\bibfield  {journal} {\bibinfo
  {journal} {Phys. Rev. D}\ }\textbf {\bibinfo {volume} {94}},\ \bibinfo
  {pages} {076004} (\bibinfo {year} {2016})},\ \Eprint
  {http://arxiv.org/abs/1604.00669} {arXiv:1604.00669 [hep-ph]} \BibitemShut
  {NoStop}%
\bibitem [{\citenamefont {Braaten}\ \emph {et~al.}(2017)\citenamefont
  {Braaten}, \citenamefont {Mohapatra},\ and\ \citenamefont
  {Zhang}}]{Braaten:2016dlp}%
  \BibitemOpen
  \bibfield  {author} {\bibinfo {author} {\bibfnamefont {E.}~\bibnamefont
  {Braaten}}, \bibinfo {author} {\bibfnamefont {A.}~\bibnamefont {Mohapatra}},
  \ and\ \bibinfo {author} {\bibfnamefont {H.}~\bibnamefont {Zhang}},\ }\href
  {\doibase 10.1103/PhysRevD.96.031901} {\bibfield  {journal} {\bibinfo
  {journal} {Phys. Rev. D}\ }\textbf {\bibinfo {volume} {96}},\ \bibinfo
  {pages} {031901} (\bibinfo {year} {2017})},\ \Eprint
  {http://arxiv.org/abs/1609.05182} {arXiv:1609.05182 [hep-ph]} \BibitemShut
  {NoStop}%
\bibitem [{\citenamefont {Mukaida}\ \emph {et~al.}(2017)\citenamefont
  {Mukaida}, \citenamefont {Takimoto},\ and\ \citenamefont
  {Yamada}}]{Mukaida:2016hwd}%
  \BibitemOpen
  \bibfield  {author} {\bibinfo {author} {\bibfnamefont {K.}~\bibnamefont
  {Mukaida}}, \bibinfo {author} {\bibfnamefont {M.}~\bibnamefont {Takimoto}}, \
  and\ \bibinfo {author} {\bibfnamefont {M.}~\bibnamefont {Yamada}},\ }\href
  {\doibase 10.1007/JHEP03(2017)122} {\bibfield  {journal} {\bibinfo  {journal}
  {JHEP}\ }\textbf {\bibinfo {volume} {03}},\ \bibinfo {pages} {122} (\bibinfo
  {year} {2017})},\ \Eprint {http://arxiv.org/abs/1612.07750} {arXiv:1612.07750
  [hep-ph]} \BibitemShut {NoStop}%
\bibitem [{\citenamefont {Namjoo}\ \emph {et~al.}(2018)\citenamefont {Namjoo},
  \citenamefont {Guth},\ and\ \citenamefont {Kaiser}}]{Namjoo:2017nia}%
  \BibitemOpen
  \bibfield  {author} {\bibinfo {author} {\bibfnamefont {M.~H.}\ \bibnamefont
  {Namjoo}}, \bibinfo {author} {\bibfnamefont {A.~H.}\ \bibnamefont {Guth}}, \
  and\ \bibinfo {author} {\bibfnamefont {D.~I.}\ \bibnamefont {Kaiser}},\
  }\href {\doibase 10.1103/PhysRevD.98.016011} {\bibfield  {journal} {\bibinfo
  {journal} {Phys. Rev. D}\ }\textbf {\bibinfo {volume} {98}},\ \bibinfo
  {pages} {016011} (\bibinfo {year} {2018})},\ \Eprint
  {http://arxiv.org/abs/1712.00445} {arXiv:1712.00445 [hep-ph]} \BibitemShut
  {NoStop}%
\bibitem [{\citenamefont {Eby}\ \emph {et~al.}(2018)\citenamefont {Eby},
  \citenamefont {Suranyi},\ and\ \citenamefont {Wijewardhana}}]{Eby:2017teq}%
  \BibitemOpen
  \bibfield  {author} {\bibinfo {author} {\bibfnamefont {J.}~\bibnamefont
  {Eby}}, \bibinfo {author} {\bibfnamefont {P.}~\bibnamefont {Suranyi}}, \ and\
  \bibinfo {author} {\bibfnamefont {L.~C.~R.}\ \bibnamefont {Wijewardhana}},\
  }\href {\doibase 10.1088/1475-7516/2018/04/038} {\bibfield  {journal}
  {\bibinfo  {journal} {JCAP}\ }\textbf {\bibinfo {volume} {04}},\ \bibinfo
  {pages} {038} (\bibinfo {year} {2018})},\ \Eprint
  {http://arxiv.org/abs/1712.04941} {arXiv:1712.04941 [hep-ph]} \BibitemShut
  {NoStop}%
\bibitem [{\citenamefont {Braaten}\ and\ \citenamefont
  {Zhang}(2019)}]{Braaten:2019knj}%
  \BibitemOpen
  \bibfield  {author} {\bibinfo {author} {\bibfnamefont {E.}~\bibnamefont
  {Braaten}}\ and\ \bibinfo {author} {\bibfnamefont {H.}~\bibnamefont
  {Zhang}},\ }\href {\doibase 10.1103/RevModPhys.91.041002} {\bibfield
  {journal} {\bibinfo  {journal} {Rev. Mod. Phys.}\ }\textbf {\bibinfo {volume}
  {91}},\ \bibinfo {pages} {041002} (\bibinfo {year} {2019})}\BibitemShut
  {NoStop}%
\bibitem [{\citenamefont {Micahel E.~Peskin}(1995)}]{peskin}%
  \BibitemOpen
  \bibfield  {author} {\bibinfo {author} {\bibfnamefont {D.~V.~S.}\
  \bibnamefont {Micahel E.~Peskin}},\ }\href@noop {} {\emph {\bibinfo {title}
  {An Introduction to Quatum Field Theory}}}\ (\bibinfo  {publisher}
  {Addison-Wesley},\ \bibinfo {year} {1995})\BibitemShut {NoStop}%
\bibitem [{\citenamefont {Peccei}\ and\ \citenamefont
  {Quinn}(1977{\natexlab{b}})}]{Peccei:1977ur}%
  \BibitemOpen
  \bibfield  {author} {\bibinfo {author} {\bibfnamefont {R.~D.}\ \bibnamefont
  {Peccei}}\ and\ \bibinfo {author} {\bibfnamefont {H.~R.}\ \bibnamefont
  {Quinn}},\ }\href {\doibase 10.1103/PhysRevD.16.1791} {\bibfield  {journal}
  {\bibinfo  {journal} {Phys. Rev. D}\ }\textbf {\bibinfo {volume} {16}},\
  \bibinfo {pages} {1791} (\bibinfo {year} {1977}{\natexlab{b}})}\BibitemShut
  {NoStop}%
\bibitem [{\citenamefont {Marsh}(2016)}]{Marsh:2015xka}%
  \BibitemOpen
  \bibfield  {author} {\bibinfo {author} {\bibfnamefont {D.~J.~E.}\
  \bibnamefont {Marsh}},\ }\href {\doibase 10.1016/j.physrep.2016.06.005}
  {\bibfield  {journal} {\bibinfo  {journal} {Phys. Rept.}\ }\textbf {\bibinfo
  {volume} {643}},\ \bibinfo {pages} {1} (\bibinfo {year} {2016})},\ \Eprint
  {http://arxiv.org/abs/1510.07633} {arXiv:1510.07633 [astro-ph.CO]}
  \BibitemShut {NoStop}%
\bibitem [{\citenamefont {Grilli~di Cortona}\ \emph {et~al.}(2016)\citenamefont
  {Grilli~di Cortona}, \citenamefont {Hardy}, \citenamefont {Pardo~Vega},\ and\
  \citenamefont {Villadoro}}]{GrillidiCortona:2015jxo}%
  \BibitemOpen
  \bibfield  {author} {\bibinfo {author} {\bibfnamefont {G.}~\bibnamefont
  {Grilli~di Cortona}}, \bibinfo {author} {\bibfnamefont {E.}~\bibnamefont
  {Hardy}}, \bibinfo {author} {\bibfnamefont {J.}~\bibnamefont {Pardo~Vega}}, \
  and\ \bibinfo {author} {\bibfnamefont {G.}~\bibnamefont {Villadoro}},\ }\href
  {\doibase 10.1007/JHEP01(2016)034} {\bibfield  {journal} {\bibinfo  {journal}
  {JHEP}\ }\textbf {\bibinfo {volume} {01}},\ \bibinfo {pages} {034} (\bibinfo
  {year} {2016})},\ \Eprint {http://arxiv.org/abs/1511.02867} {arXiv:1511.02867
  [hep-ph]} \BibitemShut {NoStop}%
\bibitem [{\citenamefont {Di~Vecchia}\ and\ \citenamefont
  {Veneziano}(1980)}]{DiVecchia:1980yfw}%
  \BibitemOpen
  \bibfield  {author} {\bibinfo {author} {\bibfnamefont {P.}~\bibnamefont
  {Di~Vecchia}}\ and\ \bibinfo {author} {\bibfnamefont {G.}~\bibnamefont
  {Veneziano}},\ }\href {\doibase 10.1016/0550-3213(80)90370-3} {\bibfield
  {journal} {\bibinfo  {journal} {Nucl. Phys. B}\ }\textbf {\bibinfo {volume}
  {171}},\ \bibinfo {pages} {253} (\bibinfo {year} {1980})}\BibitemShut
  {NoStop}%
\bibitem [{\citenamefont {Di~Luzio}\ \emph {et~al.}(2020)\citenamefont
  {Di~Luzio}, \citenamefont {Giannotti}, \citenamefont {Nardi},\ and\
  \citenamefont {Visinelli}}]{DiLuzio:2020wdo}%
  \BibitemOpen
  \bibfield  {author} {\bibinfo {author} {\bibfnamefont {L.}~\bibnamefont
  {Di~Luzio}}, \bibinfo {author} {\bibfnamefont {M.}~\bibnamefont {Giannotti}},
  \bibinfo {author} {\bibfnamefont {E.}~\bibnamefont {Nardi}}, \ and\ \bibinfo
  {author} {\bibfnamefont {L.}~\bibnamefont {Visinelli}},\ }\href {\doibase
  10.1016/j.physrep.2020.06.002} {\bibfield  {journal} {\bibinfo  {journal}
  {Phys. Rept.}\ }\textbf {\bibinfo {volume} {870}},\ \bibinfo {pages} {1}
  (\bibinfo {year} {2020})},\ \Eprint {http://arxiv.org/abs/2003.01100}
  {arXiv:2003.01100 [hep-ph]} \BibitemShut {NoStop}%
\bibitem [{\citenamefont {Braaten}\ \emph {et~al.}(2018)\citenamefont
  {Braaten}, \citenamefont {Mohapatra},\ and\ \citenamefont
  {Zhang}}]{Braaten:2018lmj}%
  \BibitemOpen
  \bibfield  {author} {\bibinfo {author} {\bibfnamefont {E.}~\bibnamefont
  {Braaten}}, \bibinfo {author} {\bibfnamefont {A.}~\bibnamefont {Mohapatra}},
  \ and\ \bibinfo {author} {\bibfnamefont {H.}~\bibnamefont {Zhang}},\ }\href
  {\doibase 10.1103/PhysRevD.98.096012} {\bibfield  {journal} {\bibinfo
  {journal} {Phys. Rev. D}\ }\textbf {\bibinfo {volume} {98}},\ \bibinfo
  {pages} {096012} (\bibinfo {year} {2018})},\ \Eprint
  {http://arxiv.org/abs/1806.01898} {arXiv:1806.01898 [hep-ph]} \BibitemShut
  {NoStop}%
\end{thebibliography}%

\end{document}